\documentclass[10pt,american,english]{article}
\usepackage[T1]{fontenc}
\usepackage[applemac]{inputenc}
\usepackage{geometry}
\usepackage{color}
\usepackage{array}
\usepackage{float}
\usepackage{textcomp}
\usepackage{url}
\usepackage{multirow}
\usepackage{amsmath}
\usepackage{amssymb}
\usepackage{graphicx}
\usepackage{setspace}
\makeatletter

\providecommand{\tabularnewline}{\\}
\newcommand{\lyxdot}{.}

\@ifundefined{date}{}{\date{}}
\date{}
\usepackage[svgnames]{xcolor}

\usepackage{colortbl}

\makeatother

\usepackage{babel}
\begin{document}

\title{The Social Cost of Carbon with \\
Economic and Climate Risks\vspace{-1.6cm}
\thanks{We thank Kenneth Arrow, Buz Brock, Varadarajan V. Chari, Jesus Fernandez-Villaverde,
Larry Goulder, Lars Peter Hansen, Tom Hertel, Larry Karp, Tim Lenton,
Robert Litterman, Alena Miftakhova, Karl Schmedders, Christian Traeger,
Rick van der Ploeg, and Ole Wilms for comments on earlier versions
of this paper. We thank the anonymous referees for their helpful comments.
We are grateful for comments from audiences at \textcolor{black}{the
2014 Minnesota Conference on Economic Models of Climate Change, the
2014 Stanford Institute for Theoretical Economics, and the 2012 IIES
Conference on Climate and the Economy}\textcolor{red}{.} Cai acknowledges
support from the Hoover Institution and the National Science Foundation
grant SES-0951576. Financial support for Lontzek was provided by the
Zürcher Universitätsverein, the University of Zurich, and the Ecosciencia
Foundation. This research is part of the Blue Waters sustained-petascale
computing project, which is supported by the National Science Foundation
(awards OCI-0725070 and ACI-1238993) and the State of Illinois. Blue
Waters is a joint effort of the University of Illinois at Urbana-Champaign
and its National Center for Supercomputing Applications. This research
was also supported in part by NIH through resources provided by the
Computation Institute and the Biological Sciences Division of the
University of Chicago and Argonne National Laboratory, under grant
1S10OD018495-01. We also thank the\textcolor{black}{{} HTCondor team
of the University of Wisconsin-Madison }for their support. The earlier
versions of this paper include ``The Social Cost of Stochastic and
Irreversible Climate Change'' (NBER working paper 18704), ``DSICE:
A Dynamic Stochastic Integrated Model of Climate and Economy'' (RDCEP
working paper 12-02), and ``Tipping Points in a Dynamic Stochastic
IAM'' (RDCEP working paper 12-03).%
}}
\maketitle
\begin{abstract}
\begin{singlespace}
There is great uncertainty about future climate conditions and the
appropriate policies for managing interactions between the climate
and the economy. We develop a multidimensional computational model
to examine how uncertainties and risks in the economic and climate
systems affect the social cost of carbon (SCC)---that is, the present
value of the marginal damage to economic output caused by carbon emissions.
The SCC is substantially increased by economic and climate risks at
both current and future times. Furthermore, the SCC is itself a stochastic
process with significant variation; for example, the basic elements
of risk incorporated into our model cause the SCC in 2100 to be, with
significant probability, ten times what it would be without those
risks. We have only imprecise information about what parameter values
are best for approximating reality. To deal with this parametric uncertainty
we perform extensive uncertainty quantification and show that these
findings are robust for a wide range of alternative specifications.
More generally, this work shows that large-scale computing can enable
economists to examine substantially more complex and realistic models
for the purposes of policy analysis. \end{singlespace}

\end{abstract}
\begin{singlespace}
\textbf{Key words:} Climate policy, social cost of carbon, climate
tipping process, Epstein--Zin preferences, stochastic growth, long-run
risk
\end{singlespace}

\vspace{0.3cm}

\begin{singlespace}
\begin{minipage}[t]{1\columnwidth}%
\begin{singlespace}
Yongyang Cai

Becker Friedman Institute at the University of Chicago, and Hoover
Institution
\end{singlespace}

Stanford, CA 94305, USA

\begin{singlespace}
yycai@stanford.edu

\vspace{0.25cm}

Kenneth L. Judd

Hoover Institution and NBER
\end{singlespace}

Stanford, CA 94305, USA

\begin{singlespace}
kennethjudd@mac.com

\vspace{0.25cm}

Thomas S. Lontzek
\end{singlespace}

Department of Business Administration, University of Zurich

8044 Zurich, Switzerland

\begin{singlespace}
thomas.lontzek@business.uzh.ch\end{singlespace}
\end{minipage}

\vspace{0.3cm}

\end{singlespace}

\newpage{}

\begin{doublespace}

\section{Introduction}
\end{doublespace}

Global warming has been recognized as a growing potential threat to
economic well-being. This concern has lead to an increasing number
of national and international discussions on how to respond to this
threat. Determining which policies should be implemented will require
merging\textcolor{black}{{} quantitative assessments of the likely economic
impacts of carbon emissions with models of how the economic and climate
systems interact; this is the purpose of}\textit{\textcolor{black}{{}
integrated assessment models}}\textcolor{black}{{} (IAMs). This paper
expands the scope of IAMs by adding uncertainties and risks to a canonical
model of the economic and climate systems, and shows that such risks
significantly raise the optimal level of carbon emission mitigation.}

\textcolor{black}{The impact of carbon emissions on society is measured
by the social cost of carbon (SCC), defined as the marginal economic
loss caused by an extra metric ton of atmospheric carbon. The Intergovernmental
Panel on Climate Change (IPCC), whose reviews summarize scientific
studies on climate change, reports that estimates of the SCC vary
across studies with an average estimate of \$43 per ton of carbon
(Yohe et al. 2007).}%
\footnote{\textcolor{black}{We denote all monetary units in United States dollars.
The social cost of carbon is sometimes measured in units of carbon
dioxide. To convert the social-cost-of-carbon values presented in
this study to units of carbon dioxide divide by 44/12.}%
}\textcolor{black}{{} The }Interagency Working Group on Social Cost of
Carbon (IWG)---a joint effort involving several United States federal
agencies---came to similar conclusions in a report\textcolor{black}{{}
based on often-used integrated assessment models (IWG 2010). Most
IAM models assume myopic expectations. The DICE (Dynamic Integrated
Climate and Economy) model of Nordhaus (2008) is one of the few forward-looking
integrated assessment models}%
\footnote{\textcolor{black}{Most integrated assessment models are static and
only a few, such as those of Nordhaus (2008), Manne and Richels (2005),
and Nordhaus and Yang (1996), are based on dynamic models of agent
decision making.}%
}\textcolor{black}{{} and} suggests a social cost of carbon of \$35 per
ton of carbon. This rough consensus has been challenged by Pindyck
(2013) and IPCC (2014), who argue that all of these estimates are
limited because they come from IAMs that ignore the considerable risk
and uncertainty in both the economic and the climate system, and their
interactions. Those models also assume that people are far less risk
averse than indicated by economic observations of the price of risk. 

\textcolor{black}{This study presents Dynamic Stochastic Integration
of Climate and the Economy (DSICE), a computational, dynamic, stochastic
general equilibrium framework for studying global models of both the
economy and the climate. We apply it to the specific issue of how
the social cost of carbon depends on stochastic features of both the
climate and the economy when we apply empirically plausible specifications
for the willingness to pay to reduce economic risk. The examples we
study demonstrate the flexibility of the DSICE framework.}

We first examine how economic risks affect the social cost of carbon.
Specifically, we assume that factor productivity growth displays long-run
risk as modeled by Bansal and Yaron (2004) and Beeler and Campbell
(2011). We calibrate the stochastic productivity growth process to
match observed moments of the growth rates for per capita consumption.
We combine this with recursive utility specifications of dynamic preferences
that use parameter values consistent with observations about how much
people are willing to pay to reduce consumption risk. This version
of our model demonstrates that realistic specifications of risks in
the economic system will imply substantially greater social costs
of carbon; for example, t\textcolor{black}{he 2005 SCC in our benchmark
parameter case is \$61 per ton of carbon, 65 percent larger than the
SCC in the absence of productivity shocks.}

\textcolor{black}{Current empirical analyses do not give us precise
estimates of critical parameters. Therefore, following the response
surface methods from the uncertainty quantification literature (see
Oberkampf and Roy (2010) for a comprehensive discussion of verification,
validation, and uncertainty quantification (VVUQ)), we also examine
a large set of empirically plausible values for key parameters, such
as the risk aversion parameter and the elasticity of inter-temporal
substitution. We find that the 2005 SCC ranges from \$35 to \$115
per ton of carbon over the parameter values we consider. These results
demonstrate that we should be skeptical of studies that rely }on a
single parameterization, but the application of uncertainty quantification
methods supports the qualitative claim that economic risks significantly
increase the SCC.

\textcolor{black}{Of equal interest and greater novelty are our results
on the dynamics of the future SCC. Stochastic factor productivity
creates riskiness in future output and carbon emissions, which in
turn makes the SCC a random process. Conventional IAMs assume no riskiness
and produce a deterministic path for the future SCC. A common interpretation
of results in deterministic models is that a quantity's path represents
its expectation in more realistic models. In many cases, we find that
the mean path for the SCC is close to that implied by deterministic
models, thereby confirming the value of deterministic models for approximating
expectations. DSICE, however, can also determine the stochastic features
of the SCC process and shows that the SCC is approximately a random
walk with substantial variance. }For example, in our benchmark case
the expected SCC is \$286 in 2100\textcolor{black}{{} but with a 10\%
chance of exceeding \$700 and a 1\% chance of exceeding \$1,200. In
general, the variance of the social cost of carbon grows faster than
its mean. Any description of the future must recognize both the intrinsic
risk in any fixed model and our uncertainty about the best values
for parameters. Combining parameter uncertainty with intrinsic risk
implies a substantially larger range for the future SCC.}

The second source of uncertainty incorporated into DSICE is uncertainty
about how the climate will respond to anthropogenic emissions. \textcolor{black}{Most
integrated assessment models assume that the impact of climate on
productivity depends only on the contemporaneous temperature. This
implies a smooth, predictable and reversible pattern of damages from
global warming.} The stochastic climate feature we add is based on
the recent literature on \textit{climate tipping elements}. These
are defined as any subsystem of the Earth's system that could exhibit
critical points (\textit{tipping points}\textcolor{black}{)} where
the climate abruptly and unexpectedly jumps to a qualitatively different
state\textcolor{black}{{} (Lenton et al. 2008).} The climate literature
has identified several major tipping elements in the climate system\textcolor{black}{.}
Recent projections suggest that \textcolor{black}{the irreversible
melting of the Greenland ice sheet} could be triggered during this
century (IPCC 2014); such melting could cause the global sea level
to rise by 0.5–1.0 meter per century (Lenton et al. 2008). Joughin
et al. (2014) argue that the collapse of the west Antarctic ice sheet
is already under way. Other examples of climate tipping elements include
the\textcolor{black}{{} weakening or shutdown of the Atlantic thermohaline
circulation and the dieback of the Amazon rainforest. Current temperature
affects the likelihood of crossing a tipping point, but temperature
reversals will not reverse the tipping event. }

\textcolor{black}{Tipping points bring a new dimension to the modeling
of the climate's effect on the economy. For example, once warming
melts a glacier, that glacier will not reform even if the warming
is reversed, and damages arising from the rising sea levels will persist.
Tipping phenomena bring a qualitative new feature to climate damage
in that some of the current damage to productivity is due to past
warming. A recent review recommends that we should seriously consider
climate tipping points in order to better anticipate and prepare ourselves
for the inevitable, potentially adverse surprises they bring (National
Research Council 2013).}

DSICE is sufficiently flexible to incorporate tipping elements of
various kinds in terms of probability, duration, and impact levels.
This allows us to consider a range of beliefs about the likelihood
of warming triggering a climate tipping point,\textcolor{black}{{} and
to determine how optimal policy and the social cost of carbon depend
on the characteristics of a tipping element.} For our benchmark parameter
specification of preferences, stochastic growth, and the climate tipping
process we find that the 2005 social cost of carbon is \textcolor{black}{\$125
per ton of carbon}, but a value of \$354 is also consistent with plausible
specifications for climate tipping processes.

Some studies (e.g., Weitzman 2009) use the possibility of very high
damage caused by a very low probability catastrophic event to advocate
aggressive mitigation policies. Estimating the probability of tail
events is very difficult, a fact that reduces the force of these arguments.
Our DSICE results show that climate tipping specifications that imply
bad, but not catastrophic, events can imply a very high social cost
of carbon. Extreme disaster scenarios are not the only assumptions
that justify high social costs of carbon. 

\begin{doublespace}
\textcolor{black}{Uncertainty about damage from a climate tipping
event raises the social cost of carbon in a manner similar to the
impact of risk in }consumption-based capital asset pricing\textcolor{black}{{}
models: the social cost of carbon is proportional to the variance
of uncertainty regarding damage. }One interpretation of such variance
is that it represents our ignorance of damage from the climate tipping
process. This observation shows that DSICE could be used to determine
which scientific studies would be most cost-effective in reducing
the uncertainties faced by policymakers. This is just one example
of how the DSICE framework could be applied in future studies.
\end{doublespace}

\textcolor{black}{Our models demonstrate that any discussion of the
social cost of carbon must consider stochastic elements in both the
climate and the economic system. We also find that these elements
have nontrivial interactions. For example, in our default parameter
cases each component---when studied in isolation---sharply elevates
the social cost of carbon, but when both economic and climate risks
are included the resulting SCC process lies in between the levels
of the individual effects.} One cannot just look at these risks in
isolation and advance some ad hoc argument on how to aggregate results
across studies.

\textcolor{black}{The ability of DSICE to track the stochastic behavior
of the social cost of carbon makes it a potential tool for assessing
a variety of policy questions because uncertainties about the social
cost of carbon will affect the ranking of alternative policies.} R\&D
decisions are certainly one example of this. R\&D decisions made today
will determine the mitigation methods available in future decades.
A deterministic model would compare the expected benefits with the
expected costs, using the discount rate that is applied to all inter-temporal
decisions. \textcolor{black}{This procedure is not valid in a dynamic
stochastic context. For instance, our results show that there is a
good chance that the social cost of carbon will be so high in a few
decades that optimal policy would not only reduce carbon emissions
but would also use technologies that remove carbon from the atmosphere.
The development of such technologies could take decades to complete.
Policy discussions today about R\&D investment in developing those
technologies should not compare the expected social cost of carbon
in the future with the expected results of R\&D investments, but should
focus instead on the present value of having such technologies in
those states of the world where the SCC justifies their deployment. }

The R\&D illustration is just one example of a more general and important
point. In deterministic models, mitigation spending is a form of investment
and subjected to the same net present value criterion as all other
investments. There is a consensus that the choice of discount rate
is a major determinant of optimal policy (e.g., Stern 2007 and Nordhaus
2007). Some have argued that this is not the right way to think about
mitigation expenditures in an uncertain and risky world. Schneider
(1989), in his testimony to the Committee on Energy and Commerce in
1989, argued that investing in climate change mitigation is like \textquotedbl{}buying
insurance against the real possibility of large and potentially catastrophic
climate change.\textquotedbl{} DSICE is the first IAM modeling framework
that can capture insurance and hedging values of climate change policies.
\textcolor{black}{Further development of these ideas must be left
to future research, but basic economic intuition suggests that real
option values and insurance ideas will play significant roles in evaluating
alternative policies.}

This study also demonstrates the value of modern high-power computing
in studying basic economics questions. The versions of DSICE that
we examine are stochastic, nine-dimensional, nonlinear dynamic programming
problems. Including stochastic productivity requires the use of annual
time periods, and the long-run nature of climate change requires a
multi-century time span. The non-stationary character of the problems
makes value function iteration the only appropriate approach. The
specifications of risks make these problems among the most computationally
demanding ever solved in economics. We are able to solve them because
we use efficient multivariate methods to approximate value functions,
and reliable optimization methods to solve the Bellman equations (Cai
and Judd 2010; Cai et al. 2015). Any numerical computation has numerical
errors. Another theme of the VVUQ literature is that computational
results should be tested to verify their accuracy. Every value function
we compute passes demanding verification tests, giving us confidence
that numerical errors do not affect our economic conclusions. Some
examples required tens of thousands of core hours, and sensitivity
analysis demanded that we examine hundreds of cases to determine the
robustness of results across empirically plausible parameter values.
This study required the use of a few million core hours on Blue Waters,
a modern supercomputer. Our use of general purpose methods shows that
many economics problems with similar computational requirements can
now be solved. 

\begin{doublespace}
\textcolor{black}{The structure of this paper is as follows. Section
\ref{sec:Literature-Overview} compares our work with other studies.
We present the economic model in Section \ref{sec:A-Stochastic-IAM}
and the climate model in Section \ref{sec:Abrupt-and}. In Section
\ref{sec:The-Dynamic-Programing} we formulate the dynamic programming
problem, outline its solution method, and present a general calibration
strategy. Sections \ref{sec:Results:-Stochastic-Growth}, \ref{sec:Results-tip},
and \ref{sec:Results:-Stochastic-Growth and Tipp} present and discuss
implications for the social cost of carbon from stochastic specifications
of }factor productivity growth\textcolor{black}{{} and a climate tipping
process; first each in isolation and then combined. Section \ref{sec:Conclusion}
concludes.}
\end{doublespace}

\section{Relation to the Literature\label{sec:Literature-Overview}}

\textcolor{black}{Our work contributes to the popular debate on how
urgently policymakers should address climate change issues. Many cost-benefit
analyses of climate change, such as that of Nordhaus (2008), suggest
global climate policy should be relatively weak. Others, including
Pindyck (2013), Kopits et al. (2014), Lenton and Ciscar (2013), }and
Ackerman et al. (2013)\textcolor{black}{{} have criticized these analyses
for their unrealistic specifications. In DSICE, we describe the uncertainty
regarding economic growth and future climate impacts in more realistic
ways and support recent suggestions, such as those made in }Revesz
et al. (2014),\textcolor{black}{{} that the costs of c}arbon emissions
are underestimated.

\begin{doublespace}
DSICE follows recent developments in macroeconomic theory and data
analysis. Macroeconomists have recently argued that economic output
follows processes with persistence in growth rates, and that recursive
preferences do the best in terms of representing attitudes towards
risk. These insights have been included in a few recent papers on
global warming using simplified models. Bansal and Ochoa (2011) assume
that consumption and output are exogenous stochastic processes. DSICE
instead builds on a Ramsey-type, representative agent, stochastic
growth model. We calibrate the stochastic factor productivity growth
so that the resulting consumption process is statistically close to
empirical data. Jensen and Traeger (2014) includes endogenous output
but assumes much lower volatility than implied by empirical data.%
\footnote{See Appendix B for a detailed comparison.%
} Moreover, Jensen and Traeger (2014) assume that carbon emissions
have an immediate impact on economic productivity, whereas DSICE and
most integrated assessment models assume a time lag between emissions
and impacts. Ignoring this lag will likely overestimate the SCC and
will create an unrealistically strong correlation between short-run
fluctuations in atmospheric carbon and their economic impacts.
\end{doublespace}

DSICE also adds tipping events to the standard IAM model form. The
climate literature has identified several major tipping elements in
the climate system\textcolor{black}{. An example of a major tipping
element is} irreversible melting of the \textcolor{black}{Greenland
ice sheet}. Recent projections of future global warming suggest that
it is more likely than not that the \textcolor{black}{Greenland ice
sheet} tipping point will be triggered within this century (IPCC 2014).
Joughin et al. (2014) argue that marine west Antarctic ice sheet collapse
is already under way. Melting of the \textcolor{black}{Greenland ice
sheet (}which contains the equivalent of about seven meters of global
sea level\textcolor{black}{)} could lead to a global sea level rise
of up to 0.5–1 meter per century (Lenton et al. 2008). Other examples
of climate tipping elements include the\textcolor{black}{{} weakening
or shutdown of the Atlantic thermohaline circulation or the dieback
of the Amazon rainforest.}

\begin{doublespace}
Economic analyses of tipping elements have used a variety of specifications,
ranging from purely deterministic representations (e.g., Keller et
al. 2004; Mastrandrea and Schneider 2001; Nordhaus 2012; and Weitzman
2012) to fully stochastic specifications (e.g., Polasky et al. 2011;
Brock and Starrett 2003; and Lemoine and Traeger 2014). However, stochastic
formulations of tipping processes have not been incorporated into
major integrated assessment models such as that of Nordhaus (2008).%
\footnote{There are a few lower-dimensional, stochastic variants of Nordhaus
(2008), such as Kelly and Kolstad (1999) and Lemoine and Traeger (2014).%
} Our computational method can deal with this complexity. 
\end{doublespace}

\textcolor{black}{Our tipping element spe}cification is also the first
to address the recent critique by Kopits et al. (2014) of current
studies which assume that the full impact of a climate tipping point
is immediate\textit{ and }its level is known (see, e.g., Lemoine and
Traeger 2014). The nature of climate tipping point events is very
different from that currently being assumed in economic models. The
occurrence, or not, of the tipping point is unknown, the transition
time of the tipping process is unknown, and---furthermore---the impact
of the climate tipping is unknown (Lenton et al. 2008; National Research
Council 2013). 

\begin{doublespace}
The structure of DSICE and its calibration strategy enable us to retain
the deterministic integrated assessment model in Nordhaus's (2008)
model as a special case of DSICE, when we eliminate all shocks to
climate and the economy and adjust the preference parameters accordingly.
The model in Nordhaus (2008) is currently being used by the United
States government to design climate policy (IWG 2010) and we wish
to compare its implications for climate policy with those obtained
from our modeling approach. However, we will note that \textcolor{black}{DSICE
is not just a stochastic extension of earlier versions of Nordhaus
(2008). Our framework can be used to solve many integrated assessment
models of comparable dimensionality. }
\end{doublespace}

\textcolor{black}{The distinctive feature of DSICE is that it combines
}stochastic factor productivity growth\textcolor{black}{{} with uncertainty
about adverse, irreversible climate events, uses recursive preferences,
and enables us to examine and quantify the }impact of economic and
climate risks on the social cost of carbon. No computational framework
is infinitely powerful, but it is clear that DSICE is far less limited
by tractability concerns than are earlier integrated assessment models.
Future work will take advantage of our framework and computational
strengths to study alternative models, even ones larger than those
examined below.

\section{The Economic Model\label{sec:A-Stochastic-IAM}}

Our model's framework merges a basic dynamic stochastic general equilibrium
model with a commonly used climate model. The canonical ``Dynamic
Integrated Climate--Economy'' (DICE) model (Nordhaus, 2008) will
be a special case of our dynamic stochastic integrated assessment
framework. In particular, DSICE includes two stochastic processes
not part of earlier integrated assessment models. 

First, we include an exogenous stochastic process that affects productivity.
This productivity process could represent an autoregressive productivity
shock, or a process that implies consumption processes similar to
those described in the literature on stochastic growth with persistence. 

Second, we include a finite-state Markov process, $J_{t}$, that also
affects productivity but with transitions that are affected by temperature,
making $J_{t}$ an endogenous process representing the impact of past
climate change on current productivity. In this paper, $J_{t}$ models
tipping elements in climate dynamics (\textcolor{black}{Kriegler et
al. 2009} and Lenton et al. 2008), a feature of climate change dynamics
that has only recently has been studied by climate scientists. 

We describe the economic model in detail in this section, including
only those climate elements that directly relate to productivity.
A later section gives the details of the climate system. This approach
to exposition helps make clear the two distinct systems, climate and
economy, and their interactions. Regarding the calibration of DSICE
for numerical results, we present our calibration strategy in Section
\ref{sub:General-Calibration-Strategy} while a list of all parameters
and exogenous processes of the economic model appears in Appendices
A and B.

\subsection{Production}

The economic side of DSICE is a simple stochastic growth model where
production produces greenhouse gas emissions and productivity is affected
by the state of the climate. In this section we will model the climate
system in a general fashion that only describes the interactions between
climate and economic productivity. The next section will give the
motivation for and specification of the climate module. One advantage
of this approach is that it makes clear the limited nature of the
linkages between climate and economy and also makes clear how one
can replace our climate module with any alternative.

We assume time is discrete, with each period equal to one year. Let
$K_{t}$ be the world capital stock in trillions of dollars at time
$t$ and $L_{t}$ be the world population in millions at time $t$.
We use the exogenous population path from Nordhaus (2008)\vspace{-0.2cm}
\begin{equation}
L_{t}=6514e^{-0.035t}+8600(1-e^{-0.035t})\label{eq:l-DICE-CJL}
\end{equation}

\vspace{-0.3cm}
In the absence of any climate damage, the gross world product is described
by a Cobb--Douglas production function with
\[
f(K,L,\widetilde{A}_{t})=\widetilde{A}_{t}K^{\alpha}L^{1-\alpha},
\]
where $\alpha=0.3$ (as in Nordhaus, 2008) and $\widetilde{A}_{t}$
is productivity at time $t$. Productivity is decomposed into two
pieces: a deterministic trend $A_{t}$, and a stochastic productivity
state $\zeta_{t}$, that is to say that $\widetilde{A}_{t}\equiv\zeta_{t}A_{t}$.
The deterministic trend $A_{t}$ is taken from Nordhaus (2008) and
equals 
\begin{equation}
A_{t}=A_{0}\exp\left(\alpha_{1}(1-e^{-\alpha_{2}t})/\alpha_{2}\right),\label{eq:det-A-trend}
\end{equation}
where $\alpha_{1}$ is th\textcolor{black}{e initial growth rate and
$\alpha_{2}$ is the decline rate of the growth rate. We want to examine
how uncertainty in productivity interacts with climate change policies. }

\textcolor{black}{We use one extra state $\chi_{t}$ to help model
the stochastic productivity state $\zeta_{t}$, where $\chi_{t}$
represents the persistence of $\zeta_{t}$. More specifically, we
start with the formulation introduced in Bansal and Yaron (2004) with
the following form: 
\begin{equation}
\log\left(\zeta_{t+1}\right)=\log\left(\zeta_{t}\right)+\chi_{t}+\varrho\omega_{\zeta,t}\label{eq:zeta_process}
\end{equation}
\begin{equation}
\chi_{t+1}=r\chi_{t}+\varsigma\omega_{\chi,t}\label{eq:chi_process}
\end{equation}
Bansal and Yaron assumed that $\omega_{\zeta,t},\omega_{\chi,t}\sim i.i.d.\:\mathcal{N}(0,1)$
and $\varrho$, $r$, and $\varsigma$ are parameters. Gaussian disturbances
are, unfortunately, unbounded and would produce arbitrarily large
growth rates and output, creating an unbounded optimal growth problem
where even the existence of expected utility is unclear. Even if we
could overcome the theoretical and computational challenges, results
for the social cost of carbon could be driven by highly unlikely tail
events. An example of this is the dismal theorem of Weitzman (2009)
showing that the risk premium could be infinite for unboundedly distributed
uncertainties. We do want to avoid existence issues and excessive
dependence on extreme tail events. To this end, we construct a time-dependent,
finite-state Markov chain for $\left(\zeta_{t},\chi_{t}\right)$ that
implies conditional and unconditional moments of consumption processes
calibrated to observed market data. The Markov transition processes
are denoted $\zeta_{t+1}=g_{\zeta}(\zeta_{t},\chi_{t},\omega_{\zeta,t})$
and $\chi_{t+1}=g_{\chi}(\chi_{t},\omega_{\chi,t})$, where $\omega_{\zeta,t}$
and $\omega_{\chi,t}$ are two serially independent stochastic processes.
This approach also makes it possible to directly apply reliable numerical
methods for solving dynamic programming problems. Appendix B describes
these features in greater detail. }

DSICE assumes that output is affected by temperature. The production
function $f\left(K_{t},L_{t},\widetilde{A}_{t}\right)$ represents
output in the absence of any effects of climate on output. 

The impact of climate on output in DSICE will depend on two climate
states : global average temperature $T_{\mathrm{AT}}$, and a climate
state denoted by $J$. The climate state $J$ will model cumulative
effects of past temperatures and is represented by a finite-state
Markov chain In this paper, we parameterize $J$ to represent stages
in a climate tipping processes; therefore, we will refer to it as
the ``tipping state.'' DSICE also contains other climate states
but this version assumes that only the states $T_{\mathrm{AT},t}$
and $J_{t}$ directly affect economic decisions.. 

The function $\Omega\left(T_{\mathrm{AT},t},J_{t}\right)$ represents
the impact of climate on output, and gross world product equals
\[
\mathcal{Y}_{t}\left(K_{t},T_{\mathrm{AT},t},\zeta_{t},J_{t}\right)=\Omega\left(T_{\mathrm{AT},t},J_{t}\right)f\left(K_{t},L_{t},\zeta_{t}A_{t}\right),
\]
where 
\[
\Omega\left(T_{\mathrm{AT},t},J_{t}\right)=\frac{1-J_{t}}{1+\pi_{1}T_{\mathrm{AT},t}+\pi_{2}(T_{\mathrm{AT},t})^{2}}.
\]
As is common in the IAM literature, we will call $\Omega\left(T_{\mathrm{AT},t},J_{t}\right)$
the damage function. We will examine only cases where $\Omega\left(T_{\mathrm{AT},t},J_{t}\right)$
is bounded by unity, but that is not a requirement in our general
framework. When $J_{t}=0$, a state we will call the ``pre-tipping
state'', our damage function reduces to the damage function in Nordhaus
(2008), which is widely used in the literature. This study generalizes
the damage function to include effects of the ``tipping state''
and associated past cumulative effects. The definition of $J$ implies
$0\leq J_{t}\leq1$; therefore, the numerical value of the tipping
state $J_{t}$ equals the current damage caused by past tipping events,
so we also call $J_{t}$ the ``tipping damage level'' when $J_{t}>0$.
The stochastic dynamic structure of $J_{t}$ is specified in more
detail below in the section on the climate model.%
\footnote{The interaction between climate and output assumed here affects only
total productivity. More generally, the impact of climate could affect
the effective capital stock, the effective labor supply, or utility
and still fall within our modeling framework. %
}

The economic system affects the climate through emissions of carbon.
We assume\textcolor{black}{{} that industrial emissions are proportional
to output, with proportionality factor $\sigma_{t}$ representing
the carbon intensity of output. The social planner can mitigate (i.e.,
reduce) emissions by a factor $\mu_{t}$ with $0\leq\mu_{t}\leq1$.
The annual industrial carbon emissions (billions of metric tons of
carbon) equal}
\begin{equation}
E_{\mathrm{Ind},t}\left(K_{t},\mu_{t},\zeta_{t}\right)=\sigma_{t}(1-\mu_{t})f\left(K_{t},L_{t},\zeta_{t}A_{t}\right).\label{eq:emission-industry}
\end{equation}
 We follow Nordhaus (2008) and assume that mitigation expenditures
equal
\begin{equation}
\Psi_{t}=\theta_{1,t}\mu_{t}^{\theta_{2}}\mathcal{Y}_{t}\left(K_{t},T_{\mathrm{AT},t},\zeta_{t},J_{t}\right).\label{eq:mitigation-cost-def}
\end{equation}
World output (net of damage) is allocated across total consumption
$C_{t}$, mitigation expenditures $\Psi_{t}$, and gross capital investment
$I_{t}$, that is to say 
\begin{equation}
\mathcal{Y}_{t}=C_{t}+\Psi_{t}+I_{t},\label{eq:Invest}
\end{equation}
thus the capital stock evolves according to 
\begin{equation}
K_{t+1}=(1-\delta)K_{t}+I_{t}\label{eq:K-transition}
\end{equation}
where $\delta=0.1$ is the annual depreciation rate.

\subsection{Epstein--Zin Preferences\textcolor{red}{{} }}

The additively separable utility functions commonly used in climate-economy
models do not do well in explaining the willingness of people to pay
to avoid risk. Here instead, we use Epstein--Zin preferences (Epstein
and\textcolor{black}{{} Zin 1989).}\textcolor{red}{{} }\textcolor{black}{Let
$C_{t}$ be the stochastic consumption process. Epstein--Zin preferences
recursively define the social welfare as 
\begin{equation}
U_{t}=\left\{ \left(1-\beta\right)u(C_{t},L_{t})+\beta\left[\mathbb{E}_{t}\left\{ U_{t+1}^{1-\gamma}\right\} \right]^{\frac{1-1/\psi}{1-\gamma}}\right\} ^{\frac{1}{1-1/\psi}},\label{eq:EZ-U1}
\end{equation}
where $\mathbb{E}_{t}\mathbb{\left\{ \cdot\right\} }$ is the expectation
conditional on the states at time $t$, and $\beta$ is the discount
factor. Here, 
\[
u(C_{t},L_{t})=\frac{\left(C_{t}/L_{t}\right)^{1-1/\psi}}{1-1/\psi}L_{t}
\]
is the annual world utility function (assuming that each individual
has the same power utility function), $\psi$ is the inter-temporal
elasticity of substitution,}%
\footnote{\textcolor{black}{Here we assume $\psi>1$. When $0<\psi<1$, the
utility function $u(C_{t},L_{t})$ is negative, the formula becomes:
\[
U_{t}=-\:\left\{ -\left(1-\beta\right)u(C_{t},L_{t})+\beta\left[\mathbb{E}_{t}\left\{ \left(-U_{t+1}\right)^{1-\gamma}\mid C_{t},L_{t}\right\} \right]^{\frac{1-1/\psi}{1-\gamma}}\right\} ^{\frac{1}{1-1/\psi}}.
\]
While the standard formulation of Epstein-Zin preferences does not
divide by $1-1/\psi$ in the annual world utility function $u(C,L)$,
here we use this formulation in order that it is consistent with our
later reformulation for the Bellman equation (\ref{eq:DSICE_DP}). }%
}\textcolor{black}{{} and $\gamma$ is the risk aversi}on parameter.
Epstein--Zin preferences are flexible specifications of decision-makers'
preferences regarding uncertainty, and allow us to distinguish between
risk preference and the desire for consumption smoothing. Even though
we refer to $\gamma$ as the risk aversion parameter, the equilibrium
risk \foreignlanguage{american}{premia} will depend on interactions
between $\psi$ and $\gamma$. Epstein--Zin preferences are special
cases of Kreps--Porteus preferences (Kreps and Porteus 1978), which
were designed to model preferences over the resolution of risk. For
the special case where $\psi\gamma=1$, we have the separable utility
case used in Nordhaus (2008). 

Epstein--Zin preferences are used here because they better explain
observed equity premia. Even though climate change risks are not directly
related to equity returns, observations about the equity premia inform
us about society's willingness to pay to reduce consumption risk.%
\footnote{There may be other aspects of climate change that affect social welfare,
but they are not included in DSICE, nor in the standard Nordhaus (2008)
family of models.%
} Therefore, we expect the optimal climate policy to be affected by
uncertainty regarding the economic damage arising from climate change.

\section{The Climate Model\label{sec:Abrupt-and}}

Our climate model contains three modules. The first module is the
carbon system and the second is the temperature module, both of which
are deterministic and adapted from the integrated assessment model
of Nordhaus (2008). The underlying idea is that industrial emissions
of greenhouse gases increase atmospheric carbon concentrations, causing
an increase in global average temperature, which then reduces economic
productivity as specified above. 

We also add a stochastic tipping element module which represents a
state of the climate related to past climate conditions and events.
A simple example of such a climate state is sea level. Prolonged periods
of warm atmospheric temperatures will (likely) melt ice in glaciers
(on land) which will then lead to a higher sea level, which can be
viewed as a state variable that affects economic productivity due
to, for example, flooding. The addition of the climate tipping module
and its associated uncertainties adds a novel element to integrated
assessment modeling.

We present our calibration strategy for these modules in Section \ref{sub:General-Calibration-Strategy}
while a list of all parameters and exogenous processes of the model
appears in Appendices A and D.

\subsection{Carbon Module}

We assume two sources for carb\textcolor{black}{on emissions, an industrial
source $E_{\mathrm{Ind},t}\left(K_{t},\mu_{t},\zeta_{t}\right)$ specified
above and an exogenous source, $E_{\mathrm{Land},t}$, arising from
land use. Total emission}s are denoted

\begin{equation}
\mathcal{E}_{t}\left(K_{t},\mu_{t},\zeta_{t}\right)=E_{\mathrm{Ind},t}\left(K_{t},\mu_{t},\zeta_{t}\right)+E_{\mathrm{Land},t}.\label{eq:emission-def}
\end{equation}
Carbon is distributed across three ``boxes'' representing different
carbon concentrations. The three-dimensional vector ${\bf M}_{t}=(M_{\mathrm{AT},t},M_{\mathrm{UO},t},M_{\mathrm{LO},t})^{\top}$
represents the mass of carbon concentrations in the atmosphere, and
in the upper levels of the ocean and lower levels of the ocean, respectively
(in gigatons of carbon). The carbon cycle specifies how these concentrations
evolve over time and is represented by the linear dynamical system
\[
\mathbf{M}_{t+1}=\mathbf{\Phi}_{M}\mathbf{M}_{t}+\left(\mathcal{E}_{t},0,0\right)^{\top},
\]
where $\mathbf{\Phi}_{M}$ is the matrix
\begin{equation}
\mathbf{\Phi}_{M}=\left[\begin{array}{ccc}
1-\phi_{12} & \phi_{21} & 0\\
\phi_{12} & 1-\phi_{21}-\phi_{23} & \phi_{32}\\
0 & \phi_{23} & 1-\phi_{32}
\end{array}\right].\label{eq:carbon-tran-matrix}
\end{equation}
The coefficients of the matrix $\mathbf{\Phi}_{M}$ have natural interpretations.
The coefficient $\phi_{ij}$ is the rate at which carbon diffuses
from level $i$ to level $j$, for $i,j\in$ \{atmosphere, upper ocean,
lower ocean\}. Since this is a closed system except for the emission
input $\mathcal{E}_{t}$, the column sums of $\mathbf{\Phi}_{M}$
must be unity.

\subsection{Temperature Module}

The temperature module models two temperatures in the atmosphere and
the ocean, measured in degrees Celsius. That system is represented
by the vector ${\bf T}_{t}=(T_{\mathrm{AT},t},T_{\mathrm{OC},t})^{\top}$,
and evolves according to 
\begin{equation}
\mathbf{T}_{t+1}=\mathbf{\Phi}_{T}^{\mathrm{}}\mathbf{T}_{t}+\left(\xi_{1}\mathcal{F}_{t}\left(M_{\mathrm{AT},t}\right),0\right)^{\top},\label{eq:temp-tran-eq}
\end{equation}
where the heat diffusion process between ocean and air is represented
by the matrix 
\begin{equation}
\mathbf{\Phi}_{T}=\left[\begin{array}{cc}
1-\varphi_{21}-\xi_{2} & \varphi_{21}\\
\varphi_{12} & 1-\varphi_{12}
\end{array}\right].\label{eq:temp-tran-matrix}
\end{equation}
The coefficient $\varphi_{ij}$ is the heat diffusion rate from level
$i$ to level $j$, for $i,j\in$ \{atmosphere, ocean\}, and $\xi_{2}$
is the rate of atmospheric temperature change by infrared radiation
to space (Schneider and Thompson 1981). Atmospheric temperature is
affected by exogenous external forcing, $F_{\mathrm{EX},t}$, and
by interactions between radiation and \textcolor{black}{carbon} in
the atmosphere. Total radiative forcing at $t$ is 
\begin{equation}
\mathcal{F}_{t}\left(M_{\mathrm{AT},t}\right)=\eta\log_{2}\left(M_{\mathrm{AT},t}/M_{\mathrm{AT}}^{*}\right)+F_{\mathrm{EX},t},\label{eq:forcing-def}
\end{equation}
where $M_{\mathrm{AT}}^{*}$ is the preindustrial atmospheric carbon
concentration and $\eta$ is the radiative forcing parameter.

\subsection{Tipping Element Module\label{sub:Tipping-Points-Module}}

We next describe the dynamics of $J_{t}$, the climate state representing
the effects of past temperatures on some aspect of the climate that
is not captured by the current temperatures and carbon states. We
choose a Markov chain for $J_{t}$ so that the changes in $J_{t}$
model a tipping with damage as a function of that state. At each time
$t$, $J_{t}$ is one of the finite number of states in the set denoted
$\left\{ \mathcal{J}_{1},\mathcal{J}_{2},...,\mathcal{J}_{n_{\negthinspace J}}\right\} $
for some positive integer $n_{\negthinspace J}$. The transition probabilities
depend on climate states in a general way, but this study considers
tipping elements only. With this focus, we will use terminology appropriate
for climate tipping elements. 

At the initial time, the state of the tipping element is represented
by $J_{t}=0$. A common approach is to assume that a tipping point
is defined by a definite (but possibly unknown to the planner) threshold,
the crossing of which leads immediately to abrupt change. We instead
assume that a tipping point is a probabilistic function of climate
conditions. Specifically, we follow the common assumption that warming
alone causes tipping (IPCC 2014; Smith et al. 2009), but assume that
if $J_{t}=0$, the probability that tipping does not occur in the
year $t$ equals 
\begin{equation}
p_{1,1,t}=\exp\left\{ -\lambda\max\left\{ 0,\, T_{\mathrm{AT},t}-\underline{T_{\mathrm{AT}}}\right\} \right\} ,\label{eq:probTrigger}
\end{equation}
where $\lambda$ is the (linear) hazard rate parameter and $\underline{T_{\mathrm{AT}}}$
is the temperature for which $p_{1,1,t}=1$.

One common approach to modeling tipping events is to assume a critical
(but perhaps unknown) temperature threshold such that the tipping
event happens if temperature reaches the threshold (see Keller et
al. 2004). With economic uncertainty, temperatures can fall or rise.
If temperature were to cross a threshold, the tipping event would
immediately happen even if the temperature quickly fell below the
threshold. We prefer a smoother representation of how climate subsystems
respond to world average temperature, such as accomplished with our
specification in (\ref{eq:probTrigger}).

Once tipping has occurred, there will be more transitions of $J_{t}$.
This study assumes irreversibility, implying that all positive probability
transitions increase $J_{t}$.\textcolor{black}{{} Integrated assessment
models that include a tipping element typically assume that all impacts
are realized immediately (e.g., }Lemoine\textcolor{black}{{} and Traeger
2014). In our framework, that is equivalent to assuming there are
only two climate states related to tipping: }$\left\{ \mathcal{J}_{1},\mathcal{J}_{2}\right\} $.\textcolor{black}{{}
However, climate scientists do not regard this as a realistic description
of the tipping elements they consider. Lenton et al. (2008) characterize
the transition scales for various tipping elements, arguing, for example,
that the loss of Arctic summer sea-ice could be complete after about
ten years, but that the melting of the Greenland ice sheet would,
once it began, continue for more than three hundred years before reaching
its long-run state. In DSICE we present more appropriate specifications
for the Markov chain states} $\left\{ \mathcal{J}_{1},\mathcal{J}_{2},...,\mathcal{J}_{n_{\negthinspace J}}\right\} $
and the transition probabilities, and can\textcolor{black}{{} examine
the more realistic and general tipping elements preferred by climate
scientists}.%
\footnote{Lenton and Ciscar (2013) suggest that tipping elements might exhibit
domino effects in the sense that several tipping elements could be
sequential and the tipping of one might trigger a whole cascade of
additional tipping elements. The flexibility of the Markov process
approach for $J_{t}$ makes possible analysis of multiple tipping
elements, but that is left for future studies.%
} The applications of DSICE presented below will examine a variety
of specifications for the details of the Markov process that represents
the effects of tipping elements.

\section{The Dynamic Programming Problem\label{sec:The-Dynamic-Programing}}

We formulate the nine-dimensional, social planner's dynamic optimization
problem as a dynamic programming problem. The nine states include
six continuous state variables (the capital stock $K$, the three-dimensional
carbon system $\mathbf{M}$, and the two-dimensional temperature vector
$\mathbf{T}$) and three discrete state variables (the climate shock
$J$, the stochastic productivity state $\zeta$, and the persistence
of its growth rate, $\chi$). Let $\mathbf{S}\equiv\left(K,\mathbf{M},\mathbf{T},\zeta,\chi,J\right)$
denote the nine-dimensional state variable vector and let $\mathbf{S}^{+}$
denote its next period's state vector.

The Epstein--Zin utility definition expressed utility in terms of
consumption. We make a nonlinear change of variables%
\footnote{That is, $V_{t}\left(\mathbf{S}\right)=\left[U_{t}\left(\mathbf{S}\right)\right]^{1-\frac{1}{\psi}}/(1-\beta)$.%
} and express the Bellman equation in terms of utils, $\left(U_{t}\right)^{1-\frac{1}{\psi}}$
. The Bellman equation then becomes
\begin{eqnarray}
V_{t}\left(\mathbf{S}\right)=\max_{C,\mu} &  & u(C_{t},L_{t})+\beta\left[\mathbb{E}_{t}\left\{ \left(V_{t+1}\left(\mathbf{S}^{+}\right)\right)^{\frac{1-\gamma}{1-1/\psi}}\right\} \right]^{\frac{1-1/\psi}{1-\gamma}},\nonumber \\
\text{s.t.} &  & K^{+}=(1-\delta)K+\mathcal{Y}_{t}(K,T_{\mathrm{AT}},\zeta,J)-C_{t}-\Psi_{t},\nonumber \\
 &  & \mathbf{M}^{+}=\mathbf{\Phi}_{M}\mathbf{M}+\left(\mathcal{E}_{t}\left(K,\mu,\zeta\right),0,0\right)^{\top},\nonumber \\
 &  & \mathbf{T}^{+}=\mathbf{\Phi}_{T}\mathbf{T}+\left(\xi_{1}\mathcal{F}_{t}\left(M_{\mathrm{AT}}\right),0\right)^{\top},\nonumber \\
 &  & \zeta^{+}=g_{\zeta}(\zeta,\chi,\omega_{\zeta}),\nonumber \\
 &  & \chi^{+}=g_{\chi}(\chi,\omega_{\chi}),\nonumber \\
 &  & J^{+}=g_{J}(J,\mathbf{T},\omega_{J}),\label{eq:DSICE_DP}
\end{eqnarray}
for $t=0,1,\ldots,599$, and any $\psi>1$.%
\footnote{When $0<\psi<1$, the objective function of the optimization problem
is 
\[
u(C_{t},L_{t})-\beta\left[\mathbb{E}_{t}\left\{ \left(-V_{t+1}\left(\mathbf{S}^{+}\right)\right)^{\frac{1-\gamma}{1-1/\psi}}\right\} \right]^{\frac{1-1/\psi}{1-\gamma}}.
\]
} The terminal value function $V_{600}$ is given in Appendix E. In
the model, consumption $C$ and emission control rate $\mu$ are two
control variables.

\subsection{The Social Cost of Carbon }

In this paper, we follow the jargon of the climate literature which
interprets the social cost of carbon as a marginal concept--that is
to say, the monetized economic loss caused by an increase in atmospheric
carbon by one metric ton. In our model, the social cost of carbon
is the marginal cost of atmospheric carbon expressed in terms of the
numeraire good, which can be either consumption or capital as there
are no adjustment costs. We define the social cost of carbon (SCC)
to be the marginal rate of substitution between atmospheric carbon
concentration and capital, as in 
\begin{equation}
\mathrm{SCC}_{t}=-1000\left(\partial V_{t}/\partial M_{\mathrm{AT},t}\right)/\left(\partial V_{t}/\partial K_{t}\right).\label{eq:SCC-formula}
\end{equation}
It will be important to remember that the social cost of carbon is
a relative shadow price--that is to say, a ratio of two marginal utilities,
and does not express the total social cost of climate damage.%
\footnote{Because $K$ is measured in trillions of dollars and $M_{\mathrm{AT}}$
is measured in billions of tons of carbon, the 1,000 factor is needed
to express the social cost of carbon in units of dollars per ton of
carbon.%
} For example, as we change economic and/or climate risks, the social
cost of carbon may go up or down because that change in risks will
affect both the marginal value of carbon and the marginal cost of
consumption (or, equivalently, the marginal value of investment).
DSICE is a general equilibrium model where the results arise from
the assumptions about tastes and technology as well as their equilibrium
interactions.

Often, the social cost of carbon and the term \textit{carbon tax}
are used interchangeably. The optimal carbon tax is the tax on carbon
that would equate private and social costs of carbon. In our model
we also examine the optimal carbon tax, which is the Pigovian tax
policy because the externality from \textcolor{black}{carbon emissions}
can be directly dealt with by a carbon tax and because there are no
other market imperfections. The social planner in our model chooses
mitigation $\mu_{t}$, which is equivalent to choosing a carbon tax
equal to $1000\theta_{1,t}\theta_{2}\mu_{t}^{\theta_{2}-1}/\sigma_{t}$
in units of dollars per ton of carbon. If $\mu_{t}<1$, the carbon
tax equals the social cost of carbon. However, if $\mu_{t}=1$ then
the carbon tax only equals that level which will drive emissions to
zero, and may be far less than the social cost of carbon. In such
cases, mitigation policies have reached their limit of effectiveness.
Alternative policies may reduce carbon concentrations directly, as
would carbon removal and storage technologies, or reduce temperature
directly, as would some solar geoengineering technologies. We do not
explicitly include those technologies in our model but our social
cost of carbon numbers will identify equilibrium paths along which
the social cost of carbon is so high that these more direct technologies
may be competitive. We leave a quantitative analysis of those issues
to future studies.

\subsection{General Calibration Strategy\label{sub:General-Calibration-Strategy}}

Our calibration strategy has the purpose of enabling us to contrast
how a stochastic representation of both climate and the economy along
with plausible preferences will affect estimates of the social cost
of carbon. For example, the United States government uses the integrated
assessment model in Nordhaus (2008) as one of three (deterministic)
models to calculate the social cost of carbon (IWG 2010). The United
States government study thus neglects any effects of decision making
under uncertainty on the optimal social cost of carbon. 

For the deterministic part of DSICE we use the general model structure
of Nordhaus (2008) and some of its parameters. This approach allows
us to retain the Nordhaus (2008) model as a special case of our model
when we eliminate all shocks to climate and the economy and adjust
the preference parameters accordingly. We will compare all our results
to that special case. In addition, we calibrate preferences, the long-run
risk specification of stochastic growth and the stochastic climate
tipping process, and the parameters governing the dynamics of the
three carbon states and the two temperature states. We next summarize
our calibration strategy, and provide additional technical details
in Appendices B, C, and D.

\subsubsection*{Parameters of the Deterministic Part of the Economic Model}

The parameters describing the deterministic part of the economic model
are taken directly from Nordhaus (2008). This includes the specification
of the production function as well as exogenous processes for world
population and the deterministic trend in productivity. Furthermore,
we retain the Nordhaus (2008) specification for the carbon intensity
of output and deterministic damage from climate change.

\subsubsection*{Parameters of the Stochastic Growth Process}

The parameters in the productivity process $(\zeta,\chi)$ were chosen
so that the solution of the stochastic growth benchmark, in the absence
of any impact of climate, produces a consumption growth process whose
properties, in terms of long-run variances and conditional covariances,
are similar to those of U.S. data on per capita consumption growth.%
\footnote{We are grateful to Ravi Bansal for providing us with the annual per
capita data on real consumption used in Bansal et al. (2012) and Beeler
and Campbell (2011) and obtained from the Bureau of Economic Analysis
website.%
} For this purpose we use a version of our model without stochastic
climate impact because the consumption data we use comes from the
20th century when the damage to productivity, caused by climate changes
was much smaller than today. In addition, the calibration of the stochastic
growth process requires a careful formulation of the Markov chains
for the productivity shock $\zeta_{t}$ and the rate of its growth
persistence $\chi_{t}$. Markov chains with only a few states cannot
represent the kind of persistence properties observed in Bansal and
Yaron (2004). After examining various possibilities, we chose $n_{\zeta}=91$
values of $\zeta_{t}$ and $n_{\chi}=19$ values of $\chi_{t}$ at
each time $t$; the time dependence is required due to the fact that
the variance of consumption levels grows over time. In Appendix B
we show that the statistics of simulation paths of our consumption
growth from our calibrated parameters are close to those of the empirical
data.

\subsubsection*{Parameters of the Deterministic Parts of Climate and Temperature
Modules}

The climate and temperature modules of DSICE are adapted from Nordhaus
(2008) with the basic idea that for a given emission scenario the
five-dimensional module (two dimensions for climate and three for
the carbon cycle) produces paths for carbon concentrations and temperature
levels which approximately match those of large, heavily-dimensional
climate models. We have also used Nordhaus (2008) as the source for
specifying the exogenous processes of emissions arising from biological
processes and external radiative forcing. \textcolor{black}{Nevertheless,
Cai et al. (2012a) pointed out that the natural interpretation of
the computer code in Nordhaus (2008) has future carbon concentrations
causing temperature increases today, and its ten-year time step formulation
produces results significantly different than results from shorter
time steps. For the analyses in the present paper, we had to recalibrate
the Nordhaus (2008) parameters governing the dynamics of the three
carbon states and the two temperature states to fit into our dynamic
programming framework with annual time steps. More precisely, using
a benchmark emissions path we use the model in Nordhaus (2008) to
generate time series data of ten-year time frequency for the three
carbon states and the two temperature states.} We then calibrate the
parameters of the dynamics of our climate and carbon states so that
our model, with the same benchmark emissions path, will produce paths
for these five state variables that agree (at every tenth year) with
the decadal data from Nordhaus (2008). The mathematical details of
this calibration method are presented in Appendix C.

\subsubsection*{Parameters of the Stochastic Climate Tipping Process}

For the implementation of the representative stochastic tipping point
process we need to specify parameter values for the likelihood of
tipping and the expected duration of the tipping process as well as
the mean and variance of the post-tipping impacts. 

\begin{doublespace}
The key parameters for the tipping element in our examples will be
given values that the climate science literature considers plausible.
\textcolor{black}{Unfortunately, there is no consensus for these values.
For the likelihood of triggering a tipping point process, Kriegler
et al. (2009) state that there is a substantial lack of knowledge
about the underlying physical processes of climate tipping elements.
So far expert elicitation studies have been conducted to assess the
character of those processes. While the subjective probabilities implied
by these opinions vary greatly, they do represent the range of beliefs
in the climate science literature. Any policy maker would also be
presented with the same high level of uncertainty. We cannot determine
which probability assessments are correct, but we use them to construct
a map from the subjective beliefs of experts to the implications for
policy choices and the social cost of carbon. We use the Lenton (2010)
summary of the findings from Kriegler et al. (2008) and other expert
elicitation studies to calibrate the likelihood of triggering the
tipping point process.}

\textcolor{black}{We also need to choose values for climate damage
associated with our tipping element. One theme of this study is the
uncertainty in that damage and the rate at which post-tipping damage
increases, reflecting climate scientists' descriptions of tipping
point events. We treat the trigger of the tipping event as well as
its duration as a stochastic process. Tipping point events have potentially
very large impacts on the economy (see IPCC 2014; Smith et al. 2009)
but there is great uncertainty about their magnitude. There are few
studies that attempt to estimate climate damage, so we rely on the
range of views represented in the literature. Stern (2007) reviews
existing models which include the risk of tipping point events and
estimates impacts at the order of 5–10 percent of gross world product.
Nordhaus (2008) assumes that catastrophic damage can amount up to
30 percent of gross world product, and Hope (2011) calibrates damage
from tipping to include the range 5–25 percent, with a central level
of 15 percent of gross world product.\vspace{-0.5cm}
}
\end{doublespace}

\subsubsection*{Preference Parameters}

The data are not definitive on the correct values for the inter-temporal
elasticity of substitution \textcolor{black}{$\psi$, }and the risk
aversion parameter \textcolor{black}{$\gamma$}. \textcolor{black}{Bansal
and Yaron (2004) combine consumption data and asset returns to argue
that $\psi$ is between 0.5 and 1.5, and $\gamma$ is from $7.5$
to 10. Bansal and Ochoa (2011) use $\psi=1.5$ and $\gamma=10$. Vissing-Jørgensen
and Attanasio (2003) find $\gamma$ between 5 and 10 and $\psi>1$.
Vissing-Jørgensen (2002) and Campbell and Cochrane (1999) find evidence
of $\psi<1$. Barro (2009) uses $\psi=2$ and $\gamma=4$, and Pindyck
and Wang (2013) use $\psi=1.5$ and $\gamma=3.066$. The DICE model
of Nordhaus (2008) is deterministic and its utility function is equivalent
to $\psi=0.5$ in our Epstein-Zin utility function. The absence of
uncertainty in DICE implies that Epstein-Zin preferences do not depend
on $\gamma$, and utility is time separable. }

\textcolor{black}{Due to the lack of precise knowledge about preferences,
we solve DSICE for a broad range of values for $\gamma$, $0.5\leq\gamma\leq20$,
and for $\psi$, $0.5\leq\psi\leq2.0$, and examine how the social
cost of carbon depends on risk preferences. In our benchmark parameter
specification we follow Bansal and Yaron (2004) and assume} $\psi=1.5$
and $\gamma=10$. \textcolor{black}{However, due to the lack of precise
knowledge of preferences, we will solve our model for a broad range
of values covering $0.5\leq\gamma\leq20$ for the risk aversion parameter
and $0.5\leq\psi\leq2.0$ }for the inter-temporal elasticity of substitution\textcolor{black}{.
We will examine how the social cost of carbon depends on risk preferences.}

\subsection{Numerical Solution Method}

We solve the nine-dimensional problem specified in (\ref{eq:DSICE_DP})
using value function iteration. Three state variables, $(\zeta,\chi,J)$,
are discretized and the stochastic shocks are modeled as transitions
of finite-state Markov chains. The productivity process states, $(\zeta,\chi)$,
use Markov chains that have enough states to ensure that the resulting
consumption processes match the conditional variance and autocorrelation
observed in consumption data, and $J$ is calibrated to represent
processes discussed in the climate literature. At each discrete point
in $(\zeta,\chi,J)$ space, the value function over the six continuous
states, $\left(K,\mathbf{M},\mathbf{T}\right)$, is approximated by
multivariate orthogonal polynomials. The range of each continuous
state variable is chosen so that all simulation paths stay in that
range. This is a large problem but the use of parallel programming
methods and hardware makes this tractable. See Appendices B and E
of this paper and Cai et al. (2015) for a more extended discussion
of the mathematical and computational details.

\subsection{A Verification Test of Code}

One theme of the VVUQ literature (\textcolor{black}{e.g., Oberkampf
and Roy 2010)}) is the value of tests that check the correctness of
the computer code. One common test is to apply the code to special
cases where we know the solution. If all uncertainty is removed, then
our model reduces to a deterministic optimal control problem that
can be solved by nonlinear programming methods. We compare these optimal
control solutions to our value function iteration results to see if
the value functions imply an optimal path equal to the nonlinear programming
results. Our tests show that paths implied by the value functions
have at least three-digit accuracy, often significantly more. See
Appendix F for more details.

\subsection{Presentation of Results}

\textcolor{black}{Our specifications of a stochastic component in
factor productivity growth and a stochastic climate are each novel
contributions to the economics of climate change. Therefore, we first
study the implications for climate policy of each component in isolation
before we investigate their interactive implications. In Sections
6, 7, and 8 respectively we analyze the implications for climate policy
of only stochastic growth, only stochastic climate, and stochastic
growth and stochastic climate combined. }

\textcolor{black}{We compare our stochastic examples with one deterministic
benchmark example with CRRA utility using $\psi=0.5$ as in Nordhaus
(2008). The DICE family of models is based on the continuous-time
differential equation system in Schneider and Thompson (1981). Based
on the DICE model of Nordhaus (2008), Cai et al. (2012b) analyze alternative
time steps and find that one-year step size gives excellent solutions
to the continuous-time system. Thus, our deterministic benchmark example
is chosen to be the one in Cai et al. (2012b) with annual time steps.
DSICE will also use one year time periods in all examples in this
paper. }

In each of the following three sections we define a benchmark parameter
specification and show the distribution of optimal dynamic paths for
the social cost of carbon and other variables. We also perform a sensitivity
analysis on some parameters and report in tables how today's optimal
level of the social cost of carbon and other variables is affected
by different parameter choices.

The simulation of optimal paths is performed as follows: At time $t=0$
we specify the levels of the six continuous states based on today's
observed levels of the capital stock, the three masses of carbon in
the atmosphere, the upper ocean, and the lower ocean, and the two
temperature levels in the atmosphere and the ocean. We also assume
that today's stochastic productivity state is at its observed mean
with zero persistence of its growth rate.%
\footnote{The values of the nine state variables at the initial time $\left(K_{0},\mathbf{M}_{0},\mathbf{T}_{0},\zeta_{0},\chi_{0},J_{0}\right)$
are given in Appendix A.\textcolor{black}{{} We use 2005 as the first
year in order to be comparable with Nordhaus (2008) and similar studies.
Our experience indicates that using, e.g., 2010 as the start date
will not change the qualitative results.}%
} For the climate, we also assume that a tipping point has not yet
been triggered. After initializing the state space, we use the value
function to compute current optimal decisions (i.e., at $t=0$). From
the combination of optimal social decisions and the realization of
current-period shocks we obtain next-year levels of the state variables
(i.e., at $t=1$). We continue this simulation process until the terminal
time. In our benchmark cases, we simulate 10,000 such paths to examine
the distributions of states, decisions, and in particular the social
cost of carbon.

\section{The Social Cost of Carbon with (only) Stochastic Growth \label{sec:Results:-Stochastic-Growth}}

This section analyzes the impact of stochastic growth and risk preferences
on the social cost of carbon and other features of the combined climate
and economic system. We do not incorporate the risk of a climate tipping
point in any of the model runs in this section. We first describe
a benchmark example with parameter specifications that produce consumption
processes matching historical data. We will call this our \textit{stochastic
growth benchmark}. This stochastic growth benchmark allows us to exposit
key features of the resulting dynamic processes, such as consumption,
output, productivity, climate states, and the social cost of carbon.
We then perform a sensitivity analysis for empirically plausible alternative
preference parameters, focusing on how preference assumptions affect
the social cost of carbon at the initial time.

\subsection{The Stochastic Growth Benchmark }

In this stochastic growth benchmark example we assume Epstein--Zin
preferences with $\psi=1.5$ and $\gamma=10$ and characterize the
stochastic growth process by assuming $\varrho=0.035$, $r=0.775$,
and $\varsigma=0.008$. As the first step we present Table \ref{tab:Stat-LRR_EconVar}
listing the mean and standard deviation of variables such as per capita
output growth $g_{y,t}$, the social cost of carbon $\mathrm{SCC}_{t}$
(on $\log_{10}$ scale), and per capita consumption growth $g_{c,t}$
at years 2020, 2050, and 2100, from our 10,000 simulation paths of
the solution to the dynamic programming problem. We also perform a
lag-1 linear autoregression analysis after detrending. That is, for
a time series $x_{t}$ (which could represent any variable in Table
\ref{tab:Stat-LRR_EconVar}), we assume 
\begin{equation}
x_{t+1}-\overline{x}_{t+1}=\Lambda(x_{t}-\overline{x}_{t})+\epsilon_{t},\label{eq:AR-def}
\end{equation}
where $\overline{x}_{t}$ is the mean of $x_{t}$ from the 10,000
simulation points at time $t$. For each simulation path, by fitting
with its first 100 years data, we get an estimate of $\Lambda$, and
the standard deviation of the lag-1 autoregression residuals, denoted
$\sigma(\epsilon)$, which is also known as the one-period-ahead conditional
standard deviation. In total, we obtain 10,000 estimates of $\Lambda$
and $\sigma(\epsilon)$. Table \ref{tab:Stat-LRR_EconVar} also reports
their mean and standard errors. 

Table \ref{tab:Stat-LRR_EconVar} tells us that both the mean and
the standard deviation of $\log_{10}(\mathrm{SCC}_{t})$ are expanding
over time. However, the standard deviation of $\log_{10}(\mathrm{SCC}_{t})$
between 2020 and 2100 more than triples, thus increasing much faster
than the growth rate of its mean. Moreover, the mean of $\Lambda$
for $\log_{10}(\mathrm{SCC}_{t})$ is larger than 1, implying that
$\log_{10}(\mathrm{SCC}_{t})$ is non-stationary over time. Similarly,
the ratio of abatement expenditure to gross output, $\Psi_{t}/\mathcal{Y}_{t}$,
is also shown to be non-stationary, with expanding mean and standard
deviation over time. 

Table \ref{tab:Stat-LRR_EconVar} also shows that the mean and standard
deviation of the other four variables \textcolor{black}{(i.e., per
capita output growth $g_{y,t}$,} per capita consumption growth $g_{c,t}$,
ratio of consumption to gross output $C_{t}/\mathcal{Y}_{t}$, and
ratio of capital investment to gross output \textcolor{black}{$I_{t}/\mathcal{Y}_{t}$})
are nearly independent of time. Moreover, for each of these four variables,
the means of their $\Lambda$ are less than 1 with small standard
errors such that their 95 percent confidence levels are also below
1. This finding suggests that the four variables are stationary and
have a mean-reverting property. 

\begin{table}[H]
\begin{centering}
\textcolor{black}{\small{}}%
\begin{tabular}{c|cccccc}
\hline 
 & $g_{y,t}$ & \textbf{\textcolor{black}{\small{}$C_{t}/\mathcal{Y}_{t}$}} & \textbf{\textcolor{black}{\small{}$I_{t}/\mathcal{Y}_{t}$}} & \textbf{\textcolor{black}{\small{}$\Psi_{t}/\mathcal{Y}_{t}$}} & $\log_{10}(\mathrm{SCC}_{t})$ & \textbf{\textcolor{black}{\small{}$g_{c,t}$}}\tabularnewline
\hline 
\textcolor{black}{\small{}mean at 2020 } & \textcolor{black}{\small{}0.013} & \textcolor{black}{\small{}0.697} & \textcolor{black}{\small{}0.302} & \textcolor{black}{\small{}$9.0(-4)$} & \textcolor{black}{\small{}1.924} & \textcolor{black}{\small{}0.014}\tabularnewline
\textcolor{black}{\small{}mean at 2050} & \textcolor{black}{\small{}0.013} & \textcolor{black}{\small{}0.705} & \textcolor{black}{\small{}0.293} & \textcolor{black}{\small{}$2.1(-3)$} & \textcolor{black}{\small{}2.153} & \textcolor{black}{\small{}0.013}\tabularnewline
\textcolor{black}{\small{}mean at 2100} & \textcolor{black}{\small{}0.012} & \textcolor{black}{\small{}0.704} & \textcolor{black}{\small{}0.290} & \textcolor{black}{\small{}$6.1(-3)$} & \textcolor{black}{\small{}2.457} & \textcolor{black}{\small{}0.012}\tabularnewline
\hline 
\textcolor{black}{\small{}standard deviation at 2020} & \textcolor{black}{\small{}0.038} & \textcolor{black}{\small{}0.027} & \textcolor{black}{\small{}0.026} & \textcolor{black}{\small{}$3.2(-4)$} & \textcolor{black}{\small{}0.087} & \textcolor{black}{\small{}0.024}\tabularnewline
\textcolor{black}{\small{}standard deviation at 2050} & \textcolor{black}{\small{}0.039} & \textcolor{black}{\small{}0.028} & \textcolor{black}{\small{}0.027} & \textcolor{black}{\small{}$1.5(-3)$} & \textcolor{black}{\small{}0.184} & \textcolor{black}{\small{}0.024}\tabularnewline
\textcolor{black}{\small{}standard deviation at 2100} & \textcolor{black}{\small{}0.039} & \textcolor{black}{\small{}0.029} & \textcolor{black}{\small{}0.027} & \textcolor{black}{\small{}$5.9(-3)$} & \textcolor{black}{\small{}0.279} & \textcolor{black}{\small{}0.025}\tabularnewline
\hline 
\textcolor{black}{\small{}mean of $\Lambda$ } & \textcolor{black}{\small{}0.184} & \textcolor{black}{\small{}0.854} & \textcolor{black}{\small{}0.847} & \textcolor{black}{\small{}1.008} & \textcolor{black}{\small{}1.003} & \textcolor{black}{\small{}0.458}\tabularnewline
\textcolor{black}{\small{}standard error of $\Lambda$} & \textcolor{black}{\small{}0.117} & \textcolor{black}{\small{}0.057} & \textcolor{black}{\small{}0.056} & \textcolor{black}{\small{}0.077} & \textcolor{black}{\small{}0.015} & \textcolor{black}{\small{}0.135}\tabularnewline
\hline 
\textcolor{black}{\small{}mean of $\sigma(\epsilon)$ } & \textcolor{black}{\small{}0.037} & \textcolor{black}{\small{}0.014} & \textcolor{black}{\small{}0.013} & \textcolor{black}{\small{}$2.3(-4)$} & \textcolor{black}{\small{}0.013} & \textcolor{black}{\small{}0.021}\tabularnewline
\textcolor{black}{\small{}standard error of $\sigma(\epsilon)$} & \textcolor{black}{\small{}0.003} & \textcolor{black}{\small{}0.001} & \textcolor{black}{\small{}0.001} & \textcolor{black}{\small{}$2.7(-4)$} & \textcolor{black}{\small{}0.001} & \textcolor{black}{\small{}0.002}\tabularnewline
\hline 
\end{tabular}
\par\end{centering}{\small \par}

\protect\caption{\textcolor{black}{\small{}\label{tab:Stat-LRR_EconVar}Statistics
from 10,000 simulation paths ($a(-n)$ means $a\times10^{-n}$) for
$g_{y,t}$ (per capita output growth), }\textbf{\textcolor{black}{\small{}$C_{t}/\mathcal{Y}_{t}$}}\textcolor{black}{\small{}
(ratio of consumption to gross output), }\textbf{\textcolor{black}{\small{}$I_{t}/\mathcal{Y}_{t}$}}\textcolor{black}{\small{}
(ratio of capital investment to gross output), }\textbf{\textcolor{black}{\small{}$\Psi_{t}/\mathcal{Y}_{t}$}}\textcolor{black}{\small{}
(ratio of abatement expenditure to gross output), $\log_{10}(\mathrm{SCC}_{t})$
(social cost of carbon }on $\log_{10}$ scale\textcolor{black}{\small{}),
and }\textbf{\textcolor{black}{\small{}$g_{c,t}$}}\textcolor{black}{\small{}
(per capita consumption growth). Note that our statistics for consumption
growth from this optimal control with climate impacts are very close
to those without climate impacts, which we use for calibration and
describe in Appendix B. This implies that these statistics are also
close to those from observed data from 1930 to 2008 for the United
States---the mean, standard deviation, $\Lambda_{2}$, and $\sigma(\epsilon)$
of the empirical data are 0.019, 0.022, 0.46, and 0.018 respectively,
well covered by the 90\% confidence intervals of the statistics from
our simulation; see Table \ref{tab:Calib-prod-shock} in Appendix
B. }}
\end{table}

\begin{figure}[H]
\noindent \centering{}\includegraphics[scale=0.75]{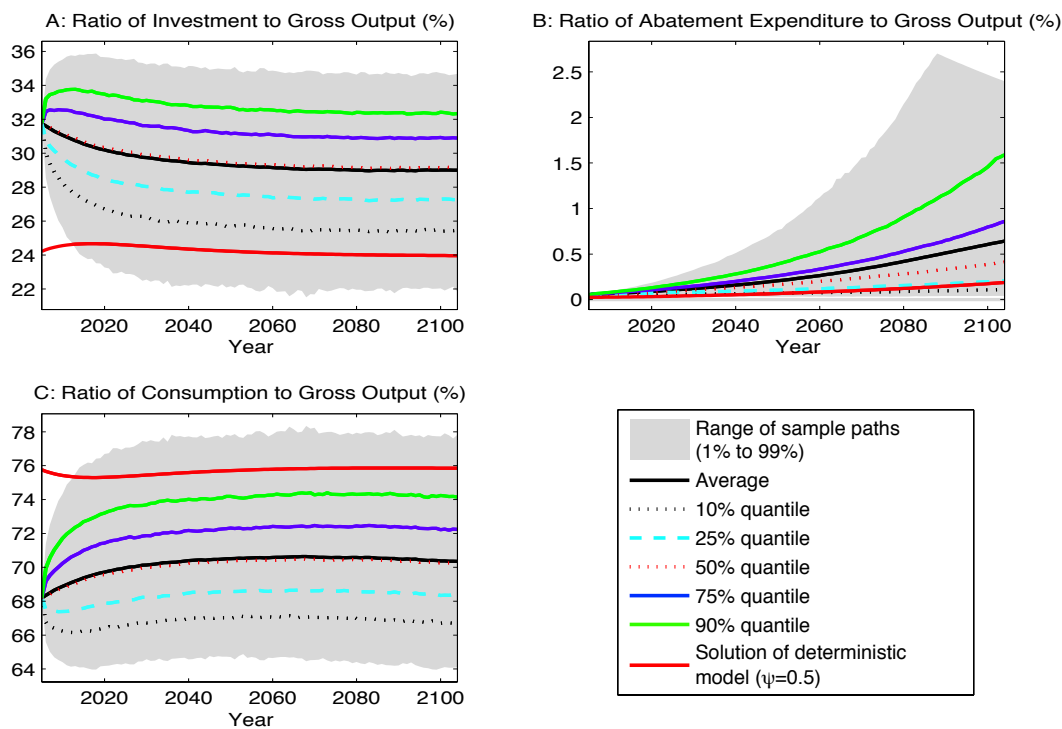}\textcolor{black}{\small{}\protect\caption{\textcolor{black}{\small{}Simulation results of the stochastic growth
benchmark---ratios to gross world output \label{fig:sim-res-LRR-ratio-1}}}
}
\end{figure}

In comparison to a model with purely deterministic growth, our model
implies a lower ratio of consumption to gross world output and higher
ratios of investment in capital accumulation and abatement. Figure
\ref{fig:sim-res-LRR-ratio-1} presents the details, displaying the
optimal dynamic distribution of the three ratios for the first 100
years with various quantiles of 10,000 simulations of the solution
to the dynamic programming problem (\ref{eq:DSICE_DP}). 

For example, the black dotted lines represent the 10 percent quantiles
for each ratio. Similarly, the cyan dashed lines, the red dotted lines,
the blue solid lines and the green solid lines represent the 25 percent,
50 percent, 75 percent, and 90 percent quantiles at each time respectively.
The black solid lines represent the sample mean path. As explained
earlier, a special case of DSICE (i.e., all variances are zero and
$\psi=0.5$) makes it comparable with the deterministic model of Nordhaus
(2008). We denote this special deterministic case by the red solid
lines. The lower (upper) edge of the gray areas represent the 1 percent
(99 percent) quantile; the gray areas represent the 98 percent probability
range of each ratio.%
\footnote{We will use the same graphical exposition for all plots describing
stochastic processes.%
} 

From Figure \ref{fig:sim-res-LRR-ratio-1} we see that \textcolor{black}{with
more than 90 percent probability $I_{t}/\mathcal{Y}_{t}$} will be
greater than in the case of a purely deterministic model (the red
solid line is below the black dotted line). Furthermore, we find that
at the initial time \textcolor{black}{$I_{t}/\mathcal{Y}_{t}$} is
at 32 percent, about 8 percent higher than under the deterministic
growth assumption and the expected difference is about 5 percent toward
the end of this century. Overall, the assumption of stochastic factor
productivity growth with persistence leads to a significant expected
increase in capital investments and thus to a precautionary buildup
of the capital stock.%
\footnote{The simulation paths of gross world output, capital, and per capita
consumption growth are shown in Figure \ref{fig:sim-LRR-econ} in
Appendix G.%
} Also, \textcolor{black}{throughout the 10,000 simulations of our
model we found $I_{t}/\mathcal{Y}_{t}$ to range roughly between 22
and 33 percent, expanding the distributional results reported in Table
}\ref{tab:Stat-LRR_EconVar}. 

\medskip{}

Panel C presents quite the opposite statistical picture for $C_{t}/\mathcal{Y}_{t}$\textcolor{black}{.
We see that with more than 90 percent probability }$C_{t}/\mathcal{Y}_{t}$\textcolor{black}{{}
will be lower than in a purely deterministic model and that toward
the end of this century }$C_{t}/\mathcal{Y}_{t}$\textcolor{black}{{}
appears to stabilize at about 70 percent; a reduction of about 6 percent
over the deterministic model. Overall, this reduction is not fully
offset by higher capital investments and, as panel B indicates, that
difference is allocated to expenditures on abatement of emissions.
We find that the latter, which is denoted by }$\Psi_{t}/\mathcal{Y}_{t}$\textcolor{black}{,
is generally quite low and does not exceed 0.2 percent in this century
in the deterministic case. Yet, as the black solid line in panel B
indicates, there is a 50 percent probability that the expenditures
on emissions abatement will be at least three times higher by the
year 2100 when growth is modeled stochastically. Furthermore, with
more than 20 percent probability more than 1 percent of gross world
output should be devoted to mitigation by the year 2100. }

The optimal allocation of gross world output is a portfolio choice
problem where abatement expenditures are a form of investment. Thus,
savings are split into investing in the capital stock or reducing
the capital stock. \textcolor{black}{In sum, Figure }\ref{fig:sim-res-LRR-ratio-1}
indicates that the inclusion of stochastic growth with persistence
will have significant impacts on the optimal allocation of gross world
output and that deterministic specifications and, most likely, even
certainty equivalent formulations will fail to account for these impacts.

We have investigated the dynamics of the economic variables, now we
study the relation between them. Table \ref{tab:Correlation} reports
the correlation matrices of the growth rates of five economic variables:
gross world output, consumption, capital investment, abatement expenditure,
and the social cost of carbon, in 2020 and 2100. We see that all reported
correlation numbers are almost the same in 2020 and 2100. Moreover,
the growth of abatement expenditure is almost independent of all the
other four variables, the growth of both consumption and capital investment
are highly correlated with the growth of gross output, and the growth
of the social cost of carbon is also highly correlated with the growth
of consumption. 

\begin{table}[H]
\begin{centering}
{\small{}}%
\begin{tabular}{c|rrrrr|rrrrr}
\hline 
 & \multicolumn{5}{c|}{{\small{}Correlation at 2020}} & \multicolumn{5}{c}{{\small{}Correlation at 2100}}\tabularnewline
\hline 
 & {\small{}$g_{\mathcal{Y},t}$} & {\small{}$g_{\mathcal{C},t}$} & {\small{}$g_{I,t}$} & {\small{}$g_{\mathcal{\varPsi},t}$} & {\small{}$g_{\mathrm{SCC},t}$} & {\small{}$g_{\mathcal{Y},t}$} & {\small{}$g_{\mathcal{C},t}$} & {\small{}$g_{I,t}$} & {\small{}$g_{\mathcal{\varPsi},t}$} & {\small{}$g_{\mathrm{SCC},t}$}\tabularnewline
\hline 
{\small{}$g_{\mathcal{Y},t}$} & {\small{}1} & {\small{}0.90} & {\small{}0.95} & {\small{}-0.02} & {\small{}0.77} & {\small{}1} & {\small{}0.90} & {\small{}0.94} & {\small{}-0.00} & {\small{}0.78}\tabularnewline
{\small{}$g_{\mathcal{C},t}$} & {\small{}0.90} & {\small{}1} & {\small{}0.71} & {\small{}-0.02} & {\small{}0.91} & {\small{}0.90} & {\small{}1} & {\small{}0.70} & {\small{}-0.00} & {\small{}0.91}\tabularnewline
{\small{}$g_{I,t}$} & {\small{}0.95} & {\small{}0.71} & {\small{}1} & {\small{}-0.02} & {\small{}0.58} & {\small{}0.94} & {\small{}0.70} & {\small{}1} & {\small{}0.00} & {\small{}0.57}\tabularnewline
{\small{}$g_{\mathcal{\varPsi},t}$} & {\small{}-0.02} & {\small{}-0.02} & {\small{}-0.02} & {\small{}1} & {\small{}-0.01} & {\small{}-0.00} & {\small{}-0.00} & {\small{}0.00} & {\small{}1} & {\small{}0.00}\tabularnewline
{\small{}$g_{\mathrm{SCC},t}$} & {\small{}0.77} & {\small{}0.91} & {\small{}0.58} & {\small{}-0.01} & {\small{}1} & {\small{}0.78} & {\small{}0.91} & {\small{}0.57} & {\small{}0.00} & {\small{}1}\tabularnewline
\hline 
\end{tabular}
\par\end{centering}{\small \par}

\protect\caption{{\small{}\label{tab:Correlation}Correlation between the growth rates
of gross world output $\mathcal{Y}_{t}$, consumption $C_{t}$, investment
$I_{t}$, abatement expenditure $\Psi_{t}$, and the social cost of
carbon $\mathrm{SCC}_{t}$ }}
\end{table}

We next study how the stochastic growth specification affects the
dynamics of the social cost of carbon, the carbon tax, the emission
control rate, and the two most important climate states: atmospheric
carbon concentration and surface temperature. 

\begin{figure}[H]
\noindent \begin{centering}
\includegraphics[scale=0.55]{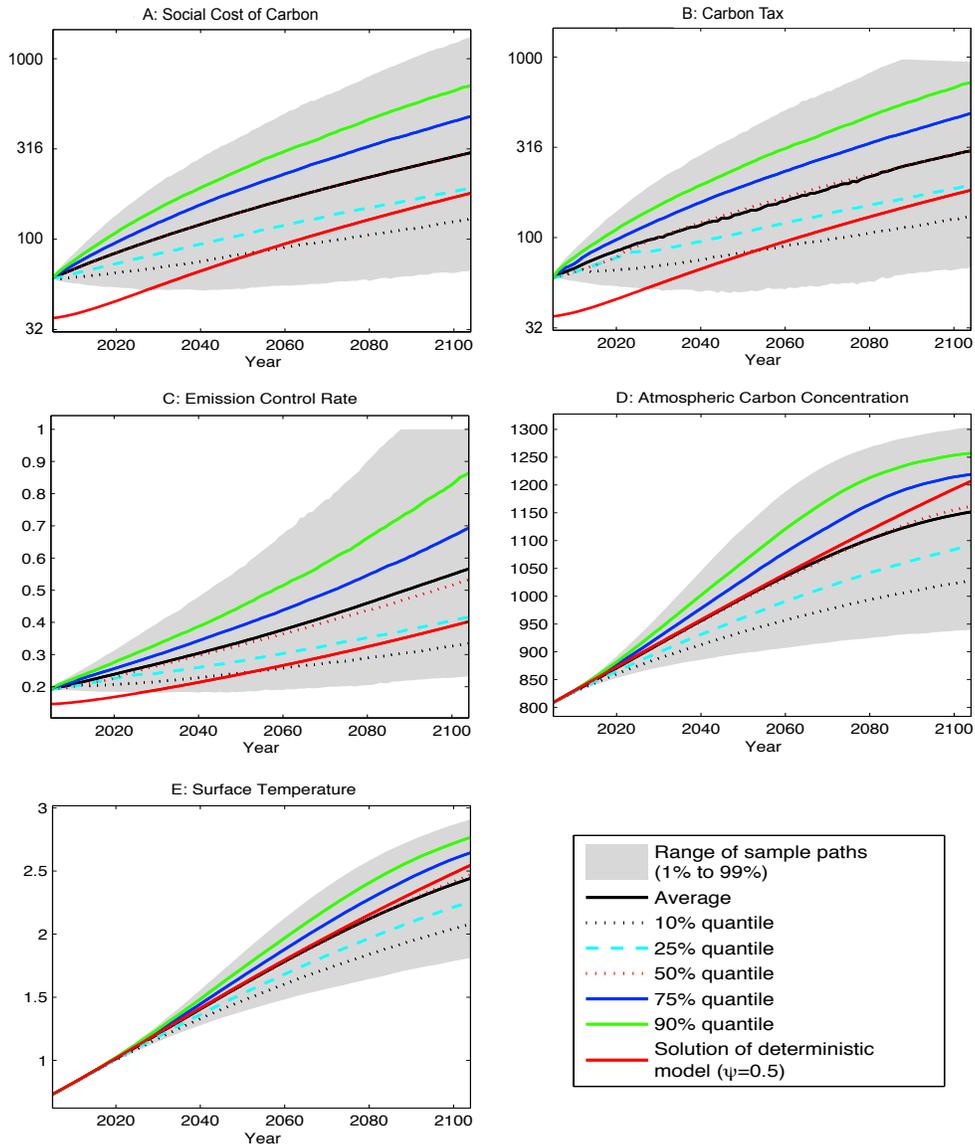}
\par\end{centering}

\protect\caption{\textcolor{black}{\small{}Simulation results of the stochastic growth
benchmark---climate system and policies \label{fig:sim-res-LRR-climate}}}
\end{figure}

We first consider panel C in Figure \ref{fig:sim-res-LRR-climate}
which shows the distribution of the rate of emission control. We observe
that with 90 percent probability, the stochastic growth representation
implies a higher emission control rate until the middle of this century
and that by 2100 the probability is reduced to 75 percent. The generally
much higher emission control rate is directly linked to the increased
share of abatement expenditures to gross world output. The latter
occurs because the prospects of much higher growth translate to expectations
of increasing gross emissions, which in turn increase the atmospheric
carbon stock and ultimately enhance global warming with its exponentially
increasing damage to gross world output.

Both panels D and E indicate that on average the stochastic growth
representation will lead to a slightly lower accumulation of carbon
in the atmosphere and also have a small cooling effect on temperature.
However, mostly due to the uncertainty about economic growth over
the present century, we see that the range of atmospheric temperature
in 2100 is about 1 degree Celsius, suggesting a large uncertainty
about the extent of global warming and its impacts on the world economy
and the environment. In fact, at the 2009 Copenhagen climate convention
most climate scientists agreed that keeping changes in temperature
levels below 2 degrees Celsius is necessary to prevent climate change
from having dangerous impacts.

As explained earlier, the monetized impacts of climate change are
expressed by the social cost of carbon which we depict in panel A
(on log$_{10}$ scale). \textcolor{black}{The optimal initial social
cost of carbon is \$61 per ton of carbon, about 65 percent higher
than in a purely deterministic economy, such as that assumed in Nordhaus
(2008), whose study is currently being used for the design of climate
policy in the United States, and to which our model compares when
we eliminate the uncertainty about growth. }Furthermore, the range
resulting from the 1 percent and 99 percent quantiles of the simulation
increases substantially over time. Here, the social cost of carbon
in 2100 varies from \textcolor{black}{\$65 per ton of carbon} to \textcolor{black}{\$1,200
per ton of carbon} and even the 10 percent and 90 percent quantiles
in 2100 show a range of \textcolor{black}{\$125 per ton of carbon}
to \textcolor{black}{\$660 per ton of carbon}.

A major observation of our analysis is that by the year 2100 about
3 percent of our model runs produce a carbon tax (panel B) much lower
than the social cost of carbon. This coincides with the simulation
paths for which the emission control rate hits its upper bound of
100 percent, implying that it is optimal to have zero emissions. We
obtain the social cost of carbon from the optimization framework of
a social planner, and a (Pigovian) carbon tax policy in this case
could be implemented to equate private and social costs of carbon
in the absence of other market imperfections, and thus achieve the
first best policy. When the emission control rate is at its limit,
the carbon tax needs only to be large enough to eliminate all emissions,
but as our results show it can be far less than the social cost of
carbon. The gap between the carbon tax and the social cost of carbon
can be large, with the largest carbon tax in 2100 less than \textcolor{black}{\$1,000
per ton of carbon} while the largest social cost of carbon is \textcolor{black}{\$2,000
per ton of carbon} among the 10,000 simulations.

A direct implication of this finding for climate policy is that with
more than 3 percent probability, mitigation policies will reach the
limit of their effectiveness and since the social cost of carbon will
be much larger than the carbon tax can internalize, alternative policies
such as carbon removal and storage or solar geoengineering technologies
may become competitive. 

Our findings point to one very important fact: there is great uncertainty
about all aspects of the combined economic and climate system. For
many variables, the mean value at each point in time is close to the
solution of the purely deterministic model. Tracking the mean is all
one can ask of any deterministic model, and in that sense deterministic
models can be successful. However, there is great uncertainty about
the future value of each key variable. This fact is of particular
importance for understanding the social cost of carbon. The social
cost of carbon is the marginal cost of extra carbon in terms of wealth,
making it the marginal rate of substitution between mitigation expenditures
and investment expenditures in physical capital. At the margin, those
two uses of savings have different impacts on future economic variables,
making allocation decisions between mitigation and investment essentially
a portfolio choice problem; a large social cost of carbon represents
the amount of investment in new capital that one is willing to sacrifice
to reduce carbon emissions by a gigaton.

\subsection{Uncertainty Quantification for Preference Parameters\label{sub:LRR-SCC}}

Empirical work suggests plausible values for $\psi$ and $\gamma$,
but the data do not give us precise values for the key parameters.
A basic method in the uncertainty quantification literature is to
recompute the social cost of carbon over a range of parameter choices
that reflect the range of economic analyses. We examine different
values of $\psi$ and $\gamma$ to determine the sensitivity of the
social cost of carbon to alternative preference specifications. Each
example will differ from the stochastic growth benchmark only in the
preference specification, as we will always use the benchmark stochastic
growth productivity process. Thus, the following results are comparative
dynamics only for changes in $\psi$ and $\gamma$.

Our sensitivity analysis will look only at the social cost of carbon
at the initial time. Simulations show that the dynamic stochastic
process for the social cost of carbon is qualitatively similar to
the stochastic growth benchmark in all cases. The social cost of carbon
at the initial time can be thought of as the initial value for the
social cost of carbon process, which is volatile for any choice of
preference parameters. Table \ref{tab:SCC-LRR} lists the initial
optimal social cost of carbon from our model under stochastic growth
assuming the values of the elasticity of inter-temporal substitution
to be $\psi=0.5$, 0.75, 1.25, 1.5, and 2.0, and the risk aversion
parameter to be $\gamma=0.5,$ 2, 6, 10, and 20. Recall that, in our
benchmark example with $\psi=1.5$ and $\gamma=10$, the optimal initial
social cost of carbon is \textcolor{black}{\$61 per ton of carbon}.
Table \ref{tab:SCC-LRR} indicates that for our benchmark case with
$\gamma=10$ the social cost of carbon is almost invariant to alternative
specifications of $\psi$, while keeping $\psi=1.5$, the social costs
of carbon at the initial time will be smaller for higher levels of
$\gamma$. 

\begin{table}[H]
\begin{centering}
\textcolor{black}{\small{}}%
\begin{tabular}{c|c|rrrrr}
\hline 
\multirow{2}{*}{\textcolor{black}{\small{}$\psi$}} & \textcolor{black}{\small{}Deterministic } & \multicolumn{5}{c}{\textcolor{black}{\small{}$\gamma$}}\tabularnewline
\cline{3-7} 
 & \textcolor{black}{\small{}Growth Case} & \textcolor{black}{\small{}0.5} & \textcolor{black}{\small{} 2} & \textcolor{black}{\small{}6} & \textcolor{black}{\small{}10} & \textcolor{black}{\small{}20}\tabularnewline
\hline 
\textcolor{black}{\small{}$0.5$} & \textcolor{black}{\small{}37} & \textcolor{black}{\small{}35} & \textcolor{black}{\small{}39} & \textcolor{black}{\small{}52} & \textcolor{black}{\small{}61} & \textcolor{black}{\small{}69}\tabularnewline
\textcolor{black}{\small{}$0.75$} & \textcolor{black}{\small{}54} & \textcolor{black}{\small{}53} & \textcolor{black}{\small{}55} & \textcolor{black}{\small{}58} & \textcolor{black}{\small{}60} & \textcolor{black}{\small{}62}\tabularnewline
\textcolor{black}{\small{}$1.25$} & \textcolor{black}{\small{}82} & \textcolor{black}{\small{}83} & \textcolor{black}{\small{}77} & \textcolor{black}{\small{}65} & \textcolor{black}{\small{}61} & \textcolor{black}{\small{}56}\tabularnewline
\textcolor{black}{\small{}$1.5$} & \textcolor{black}{\small{}94} & \textcolor{black}{\small{}95} & \textcolor{black}{\small{}85} & \textcolor{black}{\small{}68} & \textcolor{black}{\small{}61} & \textcolor{black}{\small{}55}\tabularnewline
\textcolor{black}{\small{}$2.0$} & \textcolor{black}{\small{}111 } & \textcolor{black}{\small{}115} & \textcolor{black}{\small{}97} & \textcolor{black}{\small{}71} & \textcolor{black}{\small{}62} & \textcolor{black}{\small{}54}\tabularnewline
\hline 
\end{tabular}
\par\end{centering}{\small \par}

\textcolor{black}{\small{}\protect\caption{\textcolor{black}{\small{}Initial social cost of carbon (\$ per ton
of carbon) under stochastic growth\label{tab:SCC-LRR}}}
}
\end{table}

More generally, Table \ref{tab:SCC-LRR} shows that the social cost
of carbon is sensitive to the preference parameters, ranging from
\$35 (in the case of $\psi=0.5$ and $\gamma=0.5$) to \$115 (in the
case of $\psi=2$ and $\gamma=0.5$). We see that when the inter-temporal
elasticity of substitution ($\psi$) is less than unity, a higher
$\gamma$ will imply a higher social cost of carbon. However, when
the inter-temporal elasticity of substitution is larger than 1, the
social cost of carbon is decreasing in $\gamma$. Moreover, when $\gamma\leq6$
a higher inter-temporal elasticity of substitution implies a higher
social cost of carbon, but for $\gamma=20$ the opposite is the case.
An increase in the inter-temporal elasticity of substitution reduces
the preference for consumption-smoothing but that tells us little
about the social cost of carbon. The social cost of carbon is the
value of mitigation expenditures relative to the value of capital
investment, both of which are chosen. Our results show that the interplay
between the inter-temporal elasticity of substitution and the risk
aversion parameter in our model with stochastic growth is nontrivial
for the social cost of carbon. 

From Table \ref{tab:SCC-LRR}, we see also that when $\gamma\geq2$,
a higher volatility of uncertain economic growth implies a lower initial
social cost of carbon for cases with $\psi>1$, or implies an higher
initial social cost of carbon for cases with $\psi>1$ (as the volatility
of the deterministic case is zero). The outcomes of general equilibrium
come from a mix of income and price effects, making it difficult to
arrive at simple explanations. Our sensitivity analysis enables us
to examine quantitatively the impact of alternative parametric assumptions.

Tables \ref{tab:invest-LRR} and \ref{tab:abate-LRR} report the sensitivity
of the initial ratios of capital investment to gross world output
($I_{t}/\mathcal{Y}_{t}$) and abatement expenditure to gross world
output ($\Psi_{t}/\mathcal{Y}_{t}$) respectively, for the combinations
of $\gamma$ and the inter-temporal elasticity of substitution. 

For example, when $\psi=1.5$ and $\gamma=10$, initial $I_{t}/\mathcal{Y}_{t}$
is 0.317 and $\Psi_{t}/\mathcal{Y}_{t}$ is $5.5\times10^{-4}$. From
Table \ref{tab:invest-LRR} we see that capital investment is increasing
in $\gamma$. Table \ref{tab:abate-LRR} displays the same pattern
for abatement expenditure as that displayed for the social cost of
carbon in Table \ref{tab:SCC-LRR}: When the inter-temporal elasticity
of substitution is less than 1, a higher $\gamma$ will imply a higher
ratio; when the inter-temporal elasticity of substitution is larger
than 1, the ratio is decreasing in $\gamma$; and a higher $\psi$
does not necessarily imply a higher or lower ratio of abatement expenditure
to gross world output. 

\begin{table}[H]
\begin{centering}
\textcolor{black}{\small{}}%
\begin{tabular}{c|c|ccccc}
\hline 
\multirow{2}{*}{\textcolor{black}{\small{}$\psi$}} & \textcolor{black}{\small{}Deterministic } & \multicolumn{5}{c}{\textcolor{black}{\small{}$\gamma$}}\tabularnewline
\cline{3-7} 
 & \textcolor{black}{\small{}Growth Case} & \textcolor{black}{\small{}0.5} & \textcolor{black}{\small{}2} & \textcolor{black}{\small{}6} & \textcolor{black}{\small{}10} & \textcolor{black}{\small{}20}\tabularnewline
\hline 
\textcolor{black}{\small{}$0.5$} & \textcolor{black}{\small{}0.242} & \textcolor{black}{\small{}0.240} & \textcolor{black}{\small{}0.249} & \textcolor{black}{\small{}0.272} & \textcolor{black}{\small{}0.291} & \textcolor{black}{\small{}0.321}\tabularnewline
\textcolor{black}{\small{}$0.75$} & \textcolor{black}{\small{}0.259} & \textcolor{black}{\small{}0.257} & \textcolor{black}{\small{}0.265} & \textcolor{black}{\small{}0.285} & \textcolor{black}{\small{}0.299} & \textcolor{black}{\small{}0.318}\tabularnewline
\textcolor{black}{\small{}$1.25$} & \textcolor{black}{\small{}0.278} & \textcolor{black}{\small{}0.276} & \textcolor{black}{\small{}0.286} & \textcolor{black}{\small{}0.303} & \textcolor{black}{\small{}0.312} & \textcolor{black}{\small{}0.322}\tabularnewline
\textcolor{black}{\small{}$1.5$} & \textcolor{black}{\small{}0.284} & \textcolor{black}{\small{}0.283} & \textcolor{black}{\small{}0.293} & \textcolor{black}{\small{}0.310} & \textcolor{black}{\small{}0.317} & \textcolor{black}{\small{}0.324}\tabularnewline
\textcolor{black}{\small{}$2.0$} & \textcolor{black}{\small{}0.293} & \textcolor{black}{\small{}0.293} & \textcolor{black}{\small{}0.305} & \textcolor{black}{\small{}0.322} & \textcolor{black}{\small{}0.328} & \textcolor{black}{\small{}0.331}\tabularnewline
\hline 
\end{tabular}
\par\end{centering}{\small \par}

\textcolor{black}{\small{}\protect\caption{\textcolor{black}{\small{}Initial ratio of capital investment to gross
world output under stochastic growth\label{tab:invest-LRR}}}
}
\end{table}

\begin{table}[H]
\begin{centering}
\textcolor{black}{\small{}}%
\begin{tabular}{c|c|ccccc}
\hline 
\multirow{2}{*}{\textcolor{black}{\small{}$\psi$}} & \textcolor{black}{\small{}Deterministic } & \multicolumn{5}{c}{\textcolor{black}{\small{}$\gamma$}}\tabularnewline
\cline{3-7} 
 & \textcolor{black}{\small{}Growth} & \textcolor{black}{\small{}0.5} & \textcolor{black}{\small{}2} & \textcolor{black}{\small{}6} & \textcolor{black}{\small{}10} & \textcolor{black}{\small{}20}\tabularnewline
\hline 
\textcolor{black}{\small{}$0.5$} & \textcolor{black}{\small{}$2.6(-4)$} & \textcolor{black}{\small{}$2.4(-4)$} & \textcolor{black}{\small{}$2.9(-4)$} & \textcolor{black}{\small{}$4.5(-4)$} & \textcolor{black}{\small{}$5.7(-4)$} & \textcolor{black}{\small{}$7.1(-4)$}\tabularnewline
\textcolor{black}{\small{}$0.75$} & \textcolor{black}{\small{}$4.7(-4)$} & \textcolor{black}{\small{}$4.7(-4)$} & \textcolor{black}{\small{}$4.9(-4)$} & \textcolor{black}{\small{}$5.5(-4)$} & \textcolor{black}{\small{}$5.7(-4)$} & \textcolor{black}{\small{}$5.9(-4)$}\tabularnewline
\textcolor{black}{\small{}$1.25$} & \textcolor{black}{\small{}$9.1(-4)$} & \textcolor{black}{\small{}$9.1(-4)$} & \textcolor{black}{\small{}$8.4(-4)$} & \textcolor{black}{\small{}$6.5(-4)$} & \textcolor{black}{\small{}$5.5(-4)$} & \textcolor{black}{\small{}$5.0(-4)$}\tabularnewline
\textcolor{black}{\small{}$1.5$} & \textcolor{black}{\small{}$1.1(-3)$} & \textcolor{black}{\small{}$1.1(-3)$} & \textcolor{black}{\small{}$9.8(-4)$} & \textcolor{black}{\small{}$6.8(-4)$} & \textcolor{black}{\small{}$5.5(-4)$} & \textcolor{black}{\small{}$4.8(-4)$}\tabularnewline
\textcolor{black}{\small{}$2.0$} & \textcolor{black}{\small{}$1.4(-3)$} & \textcolor{black}{\small{}$1.6(-3)$} & \textcolor{black}{\small{}$1.2(-3)$} & \textcolor{black}{\small{}$7.3(-4)$} & \textcolor{black}{\small{}$5.9(-4)$} & \textcolor{black}{\small{}$4.7(-4)$}\tabularnewline
\hline 
\end{tabular}
\par\end{centering}{\small \par}

\textcolor{black}{\small{}\protect\caption{\textcolor{black}{\small{}Initial ratio of abatement expenditure to
gross world output under stochastic growth\label{tab:abate-LRR}.
Note that $a(-n)$ represents $a\times10^{-n}$}}
}
\end{table}

We also study the effects of alternative preference specifications
on the dynamics (level and distribution) of the social cost of carbon
and per capita consumption growth. Table \ref{tab:LRR_4extremeCases}
lists these effects for only the extreme parameter cases of our simulations,
which are $\psi=0.5$ or 2, and $\gamma=0.5$ or 20. First, we verify
the statistical results of per capita consumption over the large range
of preference parameters.\textcolor{black}{{} Its mean growth rate is
quite stable over this century, }ranging from 1.1 percent to 1.4 percent
per year, close to that of the benchmark case shown in Table \ref{tab:Stat-LRR_EconVar}.
We also report the other statistics used in Table \ref{tab:Stat-LRR_EconVar}
such as the statistics of $\Lambda$ and $\sigma(\epsilon)$ for the
lag-1 autoregression (\ref{eq:AR-def}). These numbers show that a
smaller $\psi$ will produce a larger volatility for the per capita
consumption growth $g_{c}$. \textcolor{black}{Furthermore, the volatility
of $g_{c}$ is quite invariant to different values of the risk aversion
parameter. }In addition, all the cases show the mean-reverting property
of the consumption growth: the 95\% confidence level of $\Lambda$
is always below one (for lower $\psi$ levels, $g_{c}$ has a smaller
$\Lambda$, implying a faster reverting rate), and its mean and standard
deviation are almost independent of time for each reported $(\psi,\gamma)$
combination. 

Regarding the sensitivity of the social costs of carbon, we recall
from Figure \ref{fig:sim-res-LRR-climate} of our benchmark parameter
case that the social cost of carbon is highly volatile. Here, we show
that this high volatility is also persistent when we assume alternative
preference specifications. More precisely, the mean and---in particular---the
standard deviation of \textcolor{black}{$\log_{10}(\mathrm{SCC})$
are increasing over time. Furthermore, the mean of }$\Lambda$ is
not less than 1, implying non-stationarity of \textcolor{black}{$\log_{10}(\mathrm{SCC})$. }

\begin{table}[H]
\begin{centering}
\textcolor{black}{\small{}}%
\begin{tabular}{c|cc|cc|cc|cc}
\hline 
 & \multicolumn{4}{c|}{\textcolor{black}{\small{}$\log_{10}(\mathrm{SCC})$}} & \multicolumn{4}{c}{\textcolor{black}{\small{}$g_{c}$}}\tabularnewline
\hline 
\textcolor{black}{\small{}$\psi$} & \multicolumn{2}{c|}{\textcolor{black}{\small{}$0.5$}} & \multicolumn{2}{c|}{\textcolor{black}{\small{}$2$}} & \multicolumn{2}{c|}{\textcolor{black}{\small{}$0.5$}} & \multicolumn{2}{c}{\textcolor{black}{\small{}$2$}}\tabularnewline
\hline 
\textcolor{black}{\small{}$\gamma$} & \textcolor{black}{\small{}$0.5$} & \textcolor{black}{\small{}$20$} & \textcolor{black}{\small{}$0.5$} & \textcolor{black}{\small{}$20$} & \textcolor{black}{\small{}$0.5$} & \textcolor{black}{\small{}$20$} & \textcolor{black}{\small{}$0.5$} & \textcolor{black}{\small{}$20$}\tabularnewline
\hline 
\textcolor{black}{\small{}mean at 2020 } & \textcolor{black}{\small{}1.632} & \textcolor{black}{\small{}2.045} & \textcolor{black}{\small{}2.171} & \textcolor{black}{\small{}1.858} & \textcolor{black}{\small{}0.011} & \textcolor{black}{\small{}0.014} & \textcolor{black}{\small{}0.013} & \textcolor{black}{\small{}0.014}\tabularnewline
\textcolor{black}{\small{}mean at 2050} & \textcolor{black}{\small{}1.878} & \textcolor{black}{\small{}2.369} & \textcolor{black}{\small{}2.372} & \textcolor{black}{\small{}2.069} & \textcolor{black}{\small{}0.013} & \textcolor{black}{\small{}0.013} & \textcolor{black}{\small{}0.013} & \textcolor{black}{\small{}0.013}\tabularnewline
\textcolor{black}{\small{}mean at 2100} & \textcolor{black}{\small{}2.211} & \textcolor{black}{\small{}2.763} & \textcolor{black}{\small{}2.656} & \textcolor{black}{\small{}2.347} & \textcolor{black}{\small{}0.012} & \textcolor{black}{\small{}0.012} & \textcolor{black}{\small{}0.012} & \textcolor{black}{\small{}0.012}\tabularnewline
\hline 
\textcolor{black}{\small{}standard deviation at 2020} & \textcolor{black}{\small{}0.068} & \textcolor{black}{\small{}0.113} & \textcolor{black}{\small{}0.097} & \textcolor{black}{\small{}0.074} & \textcolor{black}{\small{}0.036} & \textcolor{black}{\small{}0.035} & \textcolor{black}{\small{}0.023} & \textcolor{black}{\small{}0.021}\tabularnewline
\textcolor{black}{\small{}standard deviation at 2050} & \textcolor{black}{\small{}0.181} & \textcolor{black}{\small{}0.251} & \textcolor{black}{\small{}0.189} & \textcolor{black}{\small{}0.161} & \textcolor{black}{\small{}0.037} & \textcolor{black}{\small{}0.037} & \textcolor{black}{\small{}0.023} & \textcolor{black}{\small{}0.022}\tabularnewline
\textcolor{black}{\small{}standard deviation at 2100} & \textcolor{black}{\small{}0.283} & \textcolor{black}{\small{}0.370} & \textcolor{black}{\small{}0.282} & \textcolor{black}{\small{}0.249} & \textcolor{black}{\small{}0.039} & \textcolor{black}{\small{}0.039} & \textcolor{black}{\small{}0.025} & \textcolor{black}{\small{}0.023}\tabularnewline
\hline 
\textcolor{black}{\small{}mean of $\Lambda$} & \textcolor{black}{\small{}1.008} & \textcolor{black}{\small{}1.005} & \textcolor{black}{\small{}1.000} & \textcolor{black}{\small{}1.005} & \textcolor{black}{\small{}0.100} & \textcolor{black}{\small{}0.091} & \textcolor{black}{\small{}0.584} & \textcolor{black}{\small{}0.708}\tabularnewline
\textcolor{black}{\small{}standard error of $\Lambda$} & \textcolor{black}{\small{}0.012} & \textcolor{black}{\small{}0.014} & \textcolor{black}{\small{}0.017} & \textcolor{black}{\small{}0.014} & \textcolor{black}{\small{}0.116} & \textcolor{black}{\small{}0.118} & \textcolor{black}{\small{}0.148} & \textcolor{black}{\small{}0.101}\tabularnewline
\hline 
\textcolor{black}{\small{}mean of $\sigma(\epsilon)$} & \textcolor{black}{\small{}0.009} & \textcolor{black}{\small{}0.016} & \textcolor{black}{\small{}0.017} & \textcolor{black}{\small{}0.011} & \textcolor{black}{\small{}0.036} & \textcolor{black}{\small{}0.036} & \textcolor{black}{\small{}0.018} & \textcolor{black}{\small{}0.015}\tabularnewline
\textcolor{black}{\small{}standard error of $\sigma(\epsilon)$} & \textcolor{black}{\small{}0.001} & \textcolor{black}{\small{}0.001} & \textcolor{black}{\small{}0.001} & \textcolor{black}{\small{}0.001} & \textcolor{black}{\small{}0.003} & \textcolor{black}{\small{}0.003} & \textcolor{black}{\small{}0.003} & \textcolor{black}{\small{}0.002}\tabularnewline
\hline 
\end{tabular}
\par\end{centering}{\small \par}

\protect\caption{\textcolor{black}{\small{}\label{tab:LRR_4extremeCases}Statistics
of the social cost of carbon (}on log$_{10}$ scale\textcolor{black}{\small{})
and per capita consumption growth $g_{c}$ for four extreme cases
of preference parameters} }
\end{table}

\subsection{\textcolor{black}{\normalsize{}Impact of Growth Parameters }}

\textcolor{black}{The deterministic productivity trend $A_{t}$ has
a growth rate $\alpha_{1}\exp(-\alpha_{2}t$) at time $t$ from Equation
(\ref{eq:det-A-trend}). It has two important parameters: the initial
growth rate, $\alpha_{1}$, and the decline rate of the growth of
the productivity trend, $\alpha_{2}$, with their default values 0.0092
and 0.001 respectively. We next carry out a sensitivity analysis on
them by letting $\alpha_{1}=0$ (the growth of the productivity trend
is 0, i.e., $A_{t}\equiv A_{0}$) or $\alpha_{2}=0$ (the growth of
the productivity trend is constant, i.e., $A_{t}=A_{0}\exp\left(\alpha_{1}t\right)$
with $\alpha_{1}=0.0092$).}

\textcolor{black}{Table \ref{tab:impact-growth-param} lists the simulation
statistics of the social cost of carbon and per capita consumption
growth for the two cases described above, with all remaining parameters
of the stochastic growth benchmark case unchanged (we also list the
statistics for the stochastic growth benchmark case in which $A_{t}=0.0092e^{-0.001t}$).
When $A_{t}$ is a constant (i.e., $\alpha_{1}=0$), the initial-time
social cost of carbon is only \$39 per ton of carbon ($\log_{10}(\mathrm{SCC})=1.591$),
much less than the \$61 per ton of carbon of the stochastic growth
benchmark case. From Table \ref{tab:impact-growth-param} we find
that the values of} $\alpha_{1}$ and $\alpha_{2}$ (different deterministic
productivity trends) change the means of consumption growth $g_{c}$,
but have little impact on the standard deviation, $\Lambda$, and
$\sigma(\epsilon)$. Nevertheless, we still observe the mean-reverting
property of consumption growth. In contrast, the different deterministic
productivity trends do change both the mean and the standard deviation
of $\log_{10}(\mathrm{SCC}_{t})$. A smaller productivity trend $A_{t}$
leads to a smaller mean and also a smaller standard deviation of $\log_{10}(\mathrm{SCC}_{t})$.
The non-stationarity of \textcolor{black}{$\log_{10}(\mathrm{SCC})$
still exists for both cases as their means of $\Lambda$ are still
larger than 1.}

\begin{table}[H]
\begin{centering}
\textcolor{black}{\small{}}%
\begin{tabular}{c|ccc|ccc}
\hline 
 & \multicolumn{3}{c|}{\textcolor{black}{\small{}$\log_{10}(\mathrm{SCC})$}} & \multicolumn{3}{c}{\textcolor{black}{\small{}$g_{c}$}}\tabularnewline
\hline 
\noalign{\vskip\doublerulesep}
growth of $A_{t}$ & 0 & 0.0092 & $0.0092e^{-0.001t}$ & 0 & 0.0092 & $0.0092e^{-0.001t}$\tabularnewline
\hline 
\textcolor{black}{\small{}initial-time solution} & \textcolor{black}{\small{}1.591} & \textcolor{black}{\small{}1.792} & 1.785 & \textcolor{black}{\small{}—} & \textcolor{black}{\small{}—} & \textcolor{black}{\small{}—}\tabularnewline
\hline 
\textcolor{black}{\small{}mean at 2020 } & \textcolor{black}{\small{}1.662} & \textcolor{black}{\small{}1.934} & \textcolor{black}{\small{}1.924} & \textcolor{black}{\small{}~0.002} & \textcolor{black}{\small{}0.014} & \textcolor{black}{\small{}0.014}\tabularnewline
\textcolor{black}{\small{}mean at 2050} & \textcolor{black}{\small{}1.714} & \textcolor{black}{\small{}2.167} & \textcolor{black}{\small{}2.153} & \textcolor{black}{\small{}~0.000} & \textcolor{black}{\small{}0.013} & \textcolor{black}{\small{}0.013}\tabularnewline
\textcolor{black}{\small{}mean at 2100} & \textcolor{black}{\small{}1.703} & \textcolor{black}{\small{}2.490} & \textcolor{black}{\small{}2.457} & \textcolor{black}{\small{}-0.000} & \textcolor{black}{\small{}0.013} & \textcolor{black}{\small{}0.012}\tabularnewline
\hline 
\textcolor{black}{\small{}standard deviation at 2020} & \textcolor{black}{\small{}0.096} & \textcolor{black}{\small{}0.086} & \textcolor{black}{\small{}0.087} & \textcolor{black}{\small{}~0.024} & \textcolor{black}{\small{}0.024} & \textcolor{black}{\small{}0.024}\tabularnewline
\textcolor{black}{\small{}standard deviation at 2050} & \textcolor{black}{\small{}0.215} & \textcolor{black}{\small{}0.183} & \textcolor{black}{\small{}0.184} & \textcolor{black}{\small{}~0.025} & \textcolor{black}{\small{}0.025} & \textcolor{black}{\small{}0.024}\tabularnewline
\textcolor{black}{\small{}standard deviation at 2100} & \textcolor{black}{\small{}0.393} & \textcolor{black}{\small{}0.278} & \textcolor{black}{\small{}0.279} & \textcolor{black}{\small{}~0.025} & \textcolor{black}{\small{}0.025} & \textcolor{black}{\small{}0.025}\tabularnewline
\hline 
\textcolor{black}{\small{}mean of $\Lambda$} & \textcolor{black}{\small{}1.005} & \textcolor{black}{\small{}1.003} & \textcolor{black}{\small{}1.003} & \textcolor{black}{\small{}~0.431} & \textcolor{black}{\small{}0.458} & \textcolor{black}{\small{}0.458}\tabularnewline
\textcolor{black}{\small{}standard error of $\Lambda$} & \textcolor{black}{\small{}0.016} & \textcolor{black}{\small{}0.015} & \textcolor{black}{\small{}0.015} & \textcolor{black}{\small{}~0.149} & \textcolor{black}{\small{}0.134} & \textcolor{black}{\small{}0.135}\tabularnewline
\hline 
\textcolor{black}{\small{}mean of $\sigma(\epsilon)$} & \textcolor{black}{\small{}0.017} & \textcolor{black}{\small{}0.013} & \textcolor{black}{\small{}0.013} & \textcolor{black}{\small{}~0.021} & \textcolor{black}{\small{}0.021} & \textcolor{black}{\small{}0.021}\tabularnewline
\textcolor{black}{\small{}standard error of $\sigma(\epsilon)$} & \textcolor{black}{\small{}0.001} & \textcolor{black}{\small{}0.001} & \textcolor{black}{\small{}0.001} & \textcolor{black}{\small{}~0.002} & \textcolor{black}{\small{}0.002} & \textcolor{black}{\small{}0.002}\tabularnewline
\hline 
\end{tabular}
\par\end{centering}{\small \par}

\protect\caption{\textcolor{black}{\small{}\label{tab:impact-growth-param}Statistics
of the social cost of carbon }(on $\log_{10}$ scale)\textcolor{black}{\small{}
and per capita consumption growth $g_{c}$ for cases with different
rates of growth in the deterministic productivity trend}}
\end{table}

To better understand the mechanism driving this result, recall Equation
(\ref{eq:SCC-formula}) for the social cost of carbon and the fact
that the social cost of carbon is itself a ratio of the negative shadow
price of carbon and the shadow price of capital. Thus, any change
in the risk structure will affect both shadow prices. Here, a lower
expected growth rate (as the trend is set to zero) will reduce the
expectation of the level of the capital stock when compared to our
stochastic growth benchmark, since the stochastic factor productivity
growth process is the same in both cases. At the same time, since
the economy now grows without a positive trend of about 0.92 percent
per year, gross emissions, which are proportional to gross world output,
will also grow less by that amount. Consequently, less carbon will
accumulate in the atmosphere at each point in time. A careful comparison
of these effects is difficult in general equilibrium, but we note
here that the value function is concave in capital and convex in the
carbon stock, leading to the combined effect of reducing the social
cost of carbon.

In the $\alpha_{1}=0$ case there is only a stochastic component of
growth. As a consequence, the expectation of the mean of $g_{c}$
is close to zero over the 21st century, while the standard deviation
of $g_{c}$ and its residual is close to that of the stochastic growth
benchmark case with deterministic trend. 

For the case $\alpha_{2}=0$, the initial-time social cost of carbon
is \textcolor{black}{\$62 per ton of carbon} ($\log_{10}(\mathrm{SCC})=1.792$),
only slightly larger than the stochastic growth benchmark. A model
version with $\alpha_{2}=0$ actually implies a higher deterministic
trend as the reduction in the trend is set to zero. As a consequence,
the opposite effects of those of the\textcolor{red}{{} }$\alpha_{1}=0$
case are observed. The expected social cost of carbon is also higher
throughout the rest of this century, while the standard deviation
of the social cost of carbon and its residuals is similar to that
of the stochastic benchmark case. Similar insights are obtained from
studying the per capita consumption growth rate of the $\alpha_{2}=0$
case.

\section{Social Cost of Carbon with (only) Stochastic Climate Tipping\label{sec:Results-tip}}

We next study how a tipping element in the climate system may affect
the social cost of carbon in the absence of any economic uncertainty.
First, we present a Markov chain specification of a representative
climate tipping element. Based on our calibration, we then set up
a benchmark parameter specification and study optimal climate policy;
we call this our \textit{climate tipping benchmark}. In a final step
we present an extensive multidimensional sensitivity analysis with
broad ranges of each of the parameters. \textcolor{black}{In light
of the numbers provi}ded in the few studies available, we conclude
that the impact from potential tipping point events should be carefully
assessed. Appendix D presents a more detailed derivation of our parameter
choices.

\subsection{A Markov Chain Specification of the Climate Tipping Process}

As described in Section \ref{sub:Tipping-Points-Module}, we have
a pre-tipping stage where $J_{t}=0$ until the climate tipping process
begins, followed by several stages of increasing damage. We assume
that the probability function of staying in the pre-tipping stage
is $p_{1,1,t}$, given by the formula (\ref{eq:probTrigger}). The
hazard rate parameter $\lambda$ together with the temperature process
determines the duration of the pre-tipping stage. 

We assume that the long-run post-tipping damage level, denoted $\mathcal{J}_{\infty}$,
is uncertain. We assume that we do not know $\mathcal{J}_{\infty}$
until the climate tipping process is triggered. This is a stark simplification,
but it does allow us to distinguish the expected duration of the post-tipping
process, $\overline{\mathcal{D}}$, which is always known, from the
uncertainty about the ultimate damage level. The fact that the uncertainty
about $\mathcal{J}_{\infty}$ is resolved when the climate tipping
process is triggered allows us to make statements about the relative
impact of the hazard rate of the tipping point event, the expected
duration of the post-tipping process, and the mean and variance of
$\mathcal{J}_{\infty}$. The unconditional mean of $\mathcal{J}_{\infty}$
is denoted $\overline{\mathcal{J}}_{\infty}$, and the variance of
$\mathcal{J}_{\infty}$ is $q\overline{\mathcal{J}}_{\infty}^{2}$,
where $q$ is called the ``mean squared-variance ratio''. The ratio
$q$ is analogous to the square of the Sharpe ratio, a concept used
in portfolio theory, and arises naturally in our discussion of results. 

In our examples, we assume that there are three possible post-tipping
damage levels in the long run. For each long-run damage level, the
tipping process moves through five more stages, the fifth being the
final state, and each transition occurs at the same fixed rate. Thus,
if $\overline{\mathcal{D}}$ is the expected duration of the post-tipping
process, then each transient stage has the expected duration $\overline{D}_{i}=\overline{\mathcal{D}}/4$
and all transitions have an exponential distribution, implying $p_{i,i+1,t}=\exp\left(-4/\overline{\mathcal{D}}\right)$.
Experimentation indicated that five tipping stages is adequate to
approximate processes with more states at relatively low computational
costs. These specifications imply a total of sixteen possible values
for $J_{t}$, the pre-tipping stage, and five post-tipping stages
or each of the three realizations of long-run damage. The complete
mathematical description of $J_{t}$ is contained in Appendix D.

\subsection{The Stochastic Climate Tipping Benchmark }

We choose parameter values that are roughly the average of the range
of opinions in the literature. More precisely, the climate tipping
benchmark case assumes $\lambda=0.0035$, $\overline{\mathcal{J}}_{\infty}=0.05$,
$q=0.2$, and $\mathcal{\overline{\mathcal{D}}}=50$. The choice of
$\lambda=0.0035$ implies that the conditional annual probability
of tipping increases by 0.35 percent for a warming of 1 degree Celsius.
The Epstein--Zin preference parameters in the climate tipping benchmark
case are again $\psi=1.5$ \textcolor{black}{and} $\gamma=10$. 

\begin{figure}[H]
\noindent \begin{centering}
\includegraphics[scale=0.5]{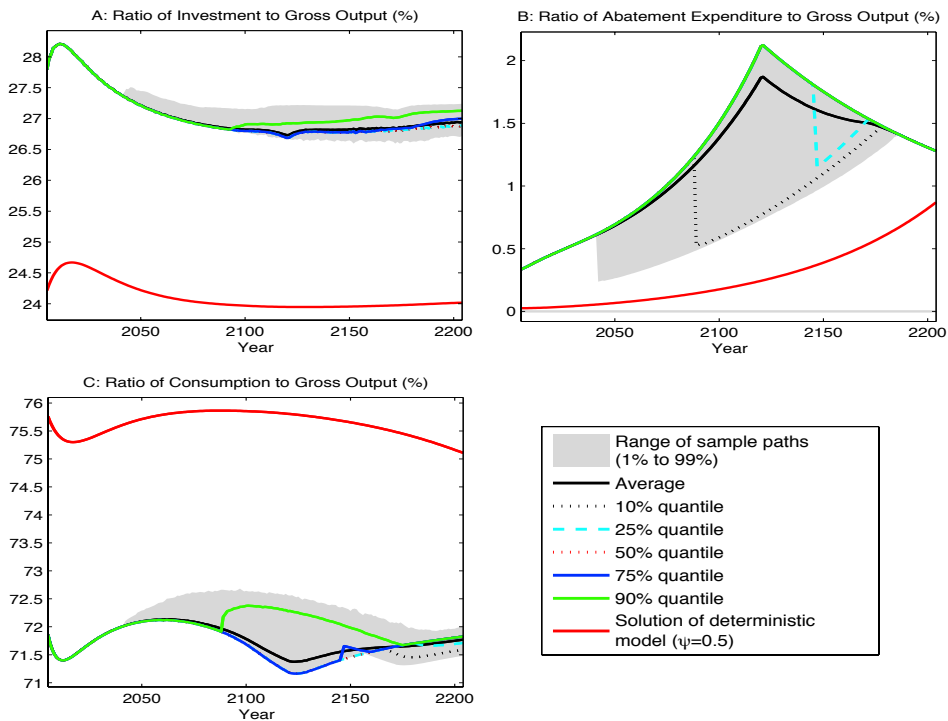}
\par\end{centering}

\protect\caption{\textcolor{black}{\small{}Simulation results for the stochastic climate
tipping benchmark---ratios to gross world output\label{fig:sim-res-tip-representative-ratio}}}
\end{figure}

We first study the effect of a possible climate tipping point on the
optimal allocation of gross world output to capital investment, consumption,
and abatement expenditures. Figure \ref{fig:sim-res-tip-representative-ratio}
shows the dynamics of the ratio of consumption to gross world output
($C_{t}/\mathcal{Y}_{t}$), \textcolor{black}{the ratio of capital
investment to gross world output ($I_{t}/\mathcal{Y}_{t}$)}, and
the ratio of abatement expenditure to gross world output ($\Psi_{t}/\mathcal{Y}_{t}$)
respectively. 

As a general finding, we first note that the absence of stochastic
growth clearly reduces the range of the distribution of all three
ratios. Now, the climate tipping point risk is the sole component
by which output can be reduced and our calibration of post-tipping
damage levels is much lower that the possible change in gross world
output due to stochastic growth with a long-run persistence.

We see that accounting for a climate tipping element implies throughout
this century a higher share of output be devoted to capital investment
(about 3 percent on average) and abatement expenditures, and a smaller
share be allocated to consumption (about 3 percent on average). Thus,
the optimal policy for addressing the tipping point risks implies
a reallocation of gross output away from consumption and toward a
precautionary buildup of the capital stock and simultaneous reductions
in emissions. 

The sign of these effects comes from the the income effect of the
threat of a tipping point, but here we also point out that the magnitude
of these changes relative to the expected value of damage from tipping
is high if one believes that future consumption should be discounted
at a market interest rate of about 5 percent. For example, \$1 in
2150 is worth about \$0.0014 in 2015 if it is discounted at a 5 percent
rate per year.

\begin{figure}[H]
\noindent \begin{centering}
\includegraphics[scale=0.54]{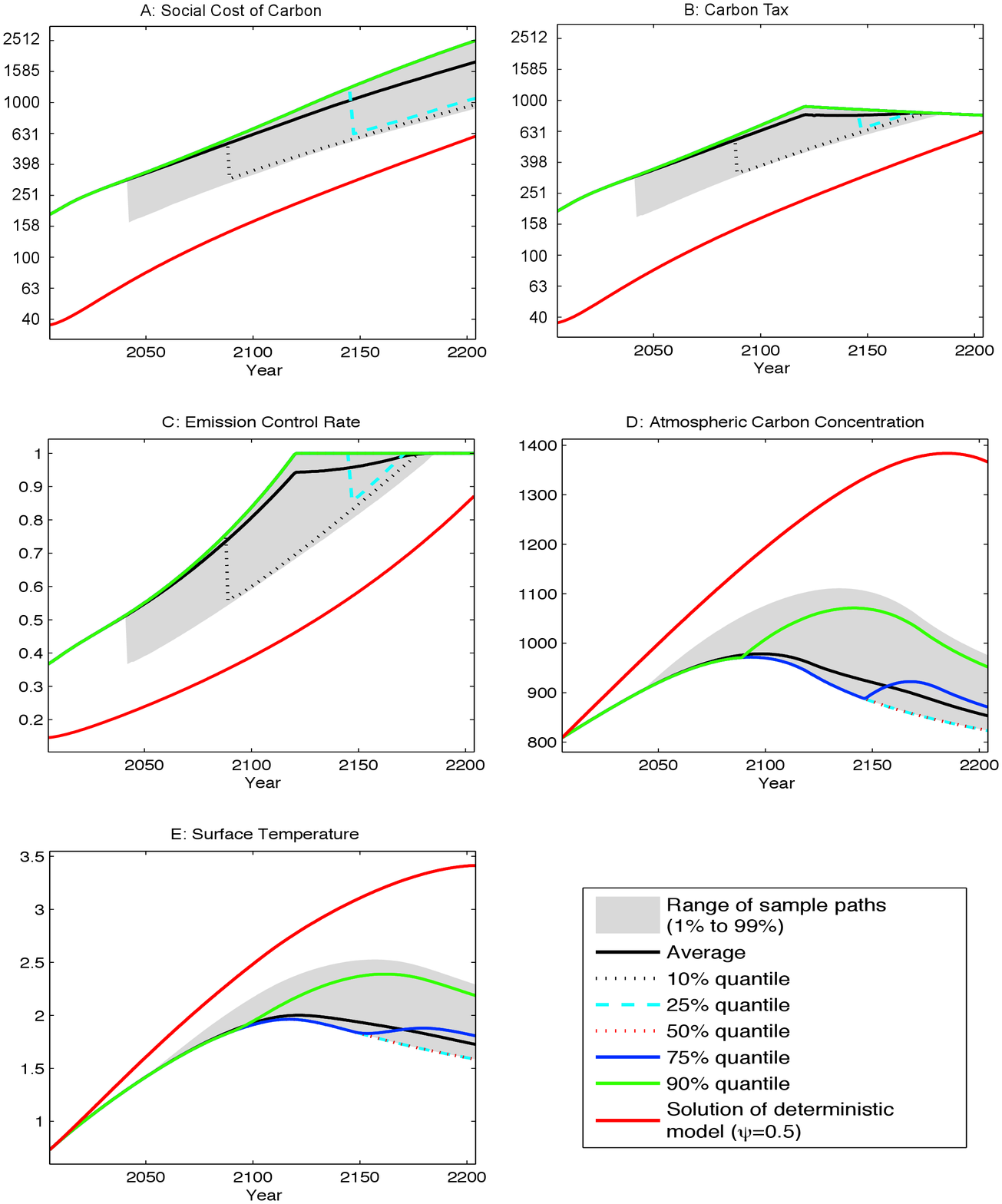}
\par\end{centering}

\protect\caption{\textcolor{black}{\small{}Simulation results for the stochastic climate
tipping benchmark---climate system and policies \label{fig:sim-res-tip-representative-climate}}}
\end{figure}

We next consider the implications of a climate tipping point on the
dynamics of the social cost of carbon, the carbon tax, the emission
control rate, and the two most important climate states (atmospheric
carbon concentration and surface temperature) respectively. Figure
\ref{fig:sim-res-tip-representative-climate} shows the results of
10,000 simulation paths over the first 200 years for these variables
and we use the same color and line conventions as in the previous
figures. 

Because of the existence of climate tipping risk and Epstein--Zin
preferences, the model version with stochastic climate tipping is
expected to result in a more intense climate policy compared to that
of a deterministic model. In fact as panel C in Figure \ref{fig:sim-res-tip-representative-climate}
shows, the optimal emission control rate follows a pattern related
to that of the abatement expenditures in the previous figure. Throughout
this century it is optimal to more than double the emission efforts
as a response to the threat of a tipping point in the climate. 

These immense emission reductions of our climate tipping benchmark
case imply a strict reduction of atmospheric carbon concentrations
(panel D) compared to that of the deterministic model. The resulting
path of surface temperature (panel E) corresponds to the temperature
paths from the lowest of the most recent emission scenarios used by
IPCC (2013), implying a peak temperature increase before 2100 of around
2 degrees Celsius and a decline afterward. 

A striking result in panel A of Figure \ref{fig:sim-res-tip-representative-climate}
is the initial social cost of carbon of \textcolor{black}{\$189 per
ton of carbon}, a large increase over the \textcolor{black}{\$38 per
ton of carbon} resulting from a model specification which ignores
the stochastic nature of the climate. To underline the significance
of this major increase in the social cost of carbon, recall our rather
conservative assumptions on the nature of the tipping point processes:
we assume an expected duration of the tipping process of 50 years,
an expected post-tipping damage of 5 percent, and a mean squared-variance
ratio of 0.2. As can be seen from the blue dashed line, these assumptions
indicate that there is a 75 percent probability that a tipping process
will not be triggered before 2150. Yet today's optimal social cost
of carbon is \textcolor{black}{\$189 per ton of carbon}, five times
the value obtained from a model run in which the climate tipping point
is ignored. This strongly suggests that analyses of climate policy
which simply ignore the potential of abrupt changes to the climate
system---as does the current United States government study (Interagency
Working Group on Social Cost of Carbon, 2010)---are significantly
underestimating the social cost of carbon. 

The dynamics of the distribution of the social costs of carbon indicate
that by 2100 the expected social cost of carbon is about \$630 per
ton of carbon, which is almost four times the \textcolor{black}{\$160
per ton of carbon} obtained from a deterministic model. Furthermore,
the expected social cost of carbon in year 2200 is about \$\textcolor{black}{1,700
per ton of carbon} but our simulations produce a range from about
\$\textcolor{black}{1,000 per ton of carbon} to \textcolor{black}{\$2,500
per ton of carbon.}

Related to the analysis in the previous section, we also note here
that by the year 2125 some of our 10,000 simulated paths will produce
a carbon tax which is less than the social cost of carbon, indicating
that the opportunities to address the climate externality are exhausted
as mitigation may not be larger than 100 percent. In fact, it appears
that, with a slightly higher than 75 percent probability, mitigation
policies will reach their limits of effectiveness by 2125 and alternative
carbon management options might be useful.

\subsection{Uncertainty Quantification for the Climate Tipping Process\label{sub:Sensitivity-tip}}

We next examine how the initial social cost of carbon is affected
by parameter uncertainty in the climate tipping process by recomputing
the social cost of carbon over a range of parameter choices that reflect
scientific opinions. We examine the six-dimensional collection (2,430
cases) of parameter values defined by the tensor product of the following
finite sets: \vspace{-0.5cm}

\[
\lambda\in\left\{ 0.0025,0.0035,0.0045\right\} ,\;\;\;\;\;\;\;\overline{\mathcal{J}}_{\infty}\in\left\{ 0.025,0.05,0.10\right\} ,\;\;\;\;\;\;\; q\in\{0,0.2,0.4\},
\]

\vspace{-1cm}

\[
\psi\in\{0.5,0.75,1.25,1.5,2.0\},\;\;\;\;\;\;\;\gamma\in\{0.5,2,6,10,14,20\},\;\;\;\;\;\;\;\overline{\mathcal{D}}\in\{5,50,200\}.
\]
We compute the social cost of carbon at the initial time for all the
2,430 cases and Table \ref{tab:SCC-tip} presents the initial social
cost of carbon for some of the representative cases. For example,
when $\lambda=0.0035$, $\overline{\mathcal{J}}_{\infty}=0.05$, $q=0.2$,
and $\mathcal{\overline{\mathcal{D}}}=50$ (i.e., the last row in
Table \ref{tab:SCC-tip}) the initial social cost of carbon is \$189
per ton for $\psi=1.5$ and $\gamma=10$. The value of the social
cost of carbon with the climate tipping process is always greater
than in the deterministic case in which the climate tipping process
is ignored. This is expected since the tipping element increases future
possible damage.

Table \ref{tab:SCC-tip} also shows that the social cost of carbon
is larger for a higher inter-temporal elasticity of substitution $\psi$,
and a higher value of the risk aversion parameter $\gamma$. Furthermore,
we observe that the initial-time social cost of carbon increases with
higher (present-discounted) expected damages from the climate tipping
process, which can be caused by a higher mean damage level ($\overline{\mathcal{J}}_{\infty}$),
a higher hazard rate parameter ($\lambda$), a shorter expected duration
of the tipping process ($\overline{\mathcal{D}}$) , or by a higher
mean squared-variance ratio of the expected damage level ($q$). As
mentioned earlier, our specification of a climate tipping point in
an economic growth model is unique by the standards of how climate
scientists view the nature of climate tipping points (e.g., Lenton
and Ciscar 2013). In the nomenclature of our model, previous studies
often assume $\overline{\mathcal{J}}_{\infty}>0.15$ (sometimes even
0.3), \textcolor{black}{$\mathcal{\overline{\mathcal{D}}=}1$, and
the implied $\lambda$ is much higher than 0.0045. The insights obtained
using models with such extreme and in some case wrong assumptions
should be assessed with special care. }

\begin{table}[H]
\begin{centering}
\textcolor{black}{\small{}}%
\begin{tabular}{c|c|c|c|ccc|ccc}
\hline 
 &  &  &  & \multicolumn{6}{c}{\textcolor{black}{\small{}Social cost of carbon ($\mathrm{SCC}$)}}\tabularnewline
\cline{5-10} 
\textcolor{black}{\small{}$\lambda$} & \textcolor{black}{\small{}$\overline{\mathcal{J}}_{\infty}$} & \textcolor{black}{\small{}$\mathcal{\overline{\mathcal{D}}}$} & \textcolor{black}{\small{}$q$} & \multicolumn{3}{c|}{\textcolor{black}{\small{}$\psi=0.5$}} & \multicolumn{3}{c}{\textcolor{black}{\small{}$\psi=1.5$}}\tabularnewline
\cline{5-10} 
 &  &  &  & \textcolor{black}{\small{}$\gamma=0.5$} & \textcolor{black}{\small{}$\gamma=10$} & \textcolor{black}{\small{}$\gamma=20$} & \textcolor{black}{\small{}$\gamma=0.5$} & \textcolor{black}{\small{}$\gamma=10$} & \textcolor{black}{\small{}$\gamma=20$}\tabularnewline
\hline 
\textcolor{black}{\small{}0.0025} & \textcolor{black}{\small{}0.025} & \textcolor{black}{\small{}5} & \textcolor{black}{\small{}0} & \textcolor{black}{\small{}43} & \textcolor{black}{\small{}44} & \textcolor{black}{\small{}45} & \textcolor{black}{\small{}128 } & \textcolor{black}{\small{}131 } & \textcolor{black}{\small{}135 }\tabularnewline
 &  &  & \textcolor{black}{\small{}0.4} & \textcolor{black}{\small{}43} & \textcolor{black}{\small{}45} & \textcolor{black}{\small{}46} & \textcolor{black}{\small{}128 } & \textcolor{black}{\small{}134 } & \textcolor{black}{\small{}142 }\tabularnewline
\cline{3-10} 
 &  & \textcolor{black}{\small{}200} & \textcolor{black}{\small{}0} & \textcolor{black}{\small{}39} & \textcolor{black}{\small{}39} & \textcolor{black}{\small{}39} & \textcolor{black}{\small{}110 } & \textcolor{black}{\small{}111 } & \textcolor{black}{\small{}112 }\tabularnewline
 &  &  & \textcolor{black}{\small{}0.4} & \textcolor{black}{\small{}39} & \textcolor{black}{\small{}39} & \textcolor{black}{\small{}39} & \textcolor{black}{\small{}110 } & \textcolor{black}{\small{}112 } & \textcolor{black}{\small{}114 }\tabularnewline
\cline{2-10} 
 & \textcolor{black}{\small{}0.10} & \textcolor{black}{\small{}5} & \textcolor{black}{\small{}0} & \textcolor{black}{\small{}65} & \textcolor{black}{\small{}83} & \textcolor{black}{\small{}110 } & \textcolor{black}{\small{}260 } & \textcolor{black}{\small{}364 } & \textcolor{black}{\small{}482 }\tabularnewline
 &  &  & \textcolor{black}{\small{}0.4} & \textcolor{black}{\small{}65} & \textcolor{black}{\small{}103 } & \textcolor{black}{\small{}194 } & \textcolor{black}{\small{}261 } & \textcolor{black}{\small{}467 } & \textcolor{black}{\small{}722 }\tabularnewline
\cline{3-10} 
 &  & \textcolor{black}{\small{}200} & \textcolor{black}{\small{}0} & \textcolor{black}{\small{}47} & \textcolor{black}{\small{}50} & \textcolor{black}{\small{}54} & \textcolor{black}{\small{}170 } & \textcolor{black}{\small{}195 } & \textcolor{black}{\small{}230 }\tabularnewline
 &  &  & \textcolor{black}{\small{}0.4} & \textcolor{black}{\small{}47} & \textcolor{black}{\small{}52} & \textcolor{black}{\small{}62} & \textcolor{black}{\small{}171 } & \textcolor{black}{\small{}224 } & \textcolor{black}{\small{}343 }\tabularnewline
\hline 
\textcolor{black}{\small{}0.0045} & \textcolor{black}{\small{}0.025} & \textcolor{black}{\small{}5} & \textcolor{black}{\small{}0} & \textcolor{black}{\small{}47} & \textcolor{black}{\small{}49} & \textcolor{black}{\small{}50} & \textcolor{black}{\small{}147 } & \textcolor{black}{\small{}150 } & \textcolor{black}{\small{}154 }\tabularnewline
 &  &  & \textcolor{black}{\small{}0.4} & \textcolor{black}{\small{}47} & \textcolor{black}{\small{}50} & \textcolor{black}{\small{}52} & \textcolor{black}{\small{}147 } & \textcolor{black}{\small{}155 } & \textcolor{black}{\small{}164 }\tabularnewline
\cline{3-10} 
 &  & \textcolor{black}{\small{}200} & \textcolor{black}{\small{}0} & \textcolor{black}{\small{}40} & \textcolor{black}{\small{}41} & \textcolor{black}{\small{}41} & \textcolor{black}{\small{}119 } & \textcolor{black}{\small{}120 } & \textcolor{black}{\small{}121 }\tabularnewline
 &  &  & \textcolor{black}{\small{}0.4} & \textcolor{black}{\small{}40} & \textcolor{black}{\small{}41} & \textcolor{black}{\small{}41} & \textcolor{black}{\small{}119 } & \textcolor{black}{\small{}121 } & \textcolor{black}{\small{}124 }\tabularnewline
\cline{2-10} 
 & \textcolor{black}{\small{}0.10} & \textcolor{black}{\small{}5} & \textcolor{black}{\small{}0} & \textcolor{black}{\small{}85} & \textcolor{black}{\small{}109} & \textcolor{black}{\small{}143 } & \textcolor{black}{\small{}369 } & \textcolor{black}{\small{}480 } & \textcolor{black}{\small{}584 }\tabularnewline
 &  &  & \textcolor{black}{\small{}0.4} & \textcolor{black}{\small{}85} & \textcolor{black}{\small{}140 } & \textcolor{black}{\small{}252 } & \textcolor{black}{\small{}370 } & \textcolor{black}{\small{}586 } & \textcolor{black}{\small{}817 }\tabularnewline
\cline{3-10} 
 &  & \textcolor{black}{\small{}200} & \textcolor{black}{\small{}0} & \textcolor{black}{\small{}54} & \textcolor{black}{\small{}58} & \textcolor{black}{\small{}63} & \textcolor{black}{\small{}222 } & \textcolor{black}{\small{}259 } & \textcolor{black}{\small{}305 }\tabularnewline
 &  &  & \textcolor{black}{\small{}0.4} & \textcolor{black}{\small{}54} & \textcolor{black}{\small{}62} & \textcolor{black}{\small{}76} & \textcolor{black}{\small{}223 } & \textcolor{black}{\small{}306 } & \textcolor{black}{\small{}445 }\tabularnewline
\hline 
\textcolor{black}{\small{}0.0035} & \textcolor{black}{\small{}0.05} & \textcolor{black}{\small{}50} & \textcolor{black}{\small{}0.2} & \textcolor{black}{\small{}48} & \textcolor{black}{\small{}52} & \textcolor{black}{\small{}55} & \textcolor{black}{\small{}171 } & \textcolor{black}{\small{}189 } & \textcolor{black}{\small{}216 }\tabularnewline
\hline 
\end{tabular}
\par\end{centering}{\small \par}

\protect\caption{\textcolor{black}{\small{}\label{tab:SCC-tip}Initial social cost
of carbon (\$ per ton of carbon) with stochastic climate tipping}}
\end{table}

Here, the increase in the social cost of carbon relative to $q$ is
very minor when $\gamma$ or $\overline{\mathcal{J}}_{\infty}$ is
small (a small $\overline{\mathcal{J}}_{\infty}$ implies a much smaller
variance of the uncertain, final, post-tipping damage level which
is equal to $q\overline{\mathcal{J}}_{\infty}^{2}$), particularly
for large expected durations. However, it becomes more visible for
larger values of $\gamma$. Therefore, the effect of rising uncertainty
about future damage from abrupt climate change is amplified with higher
values of the risk aversion parameter. In models assuming separable
preferences, it is only the mean of the uncertain damage, and not
its variance, that affects the social cost of carbon. 

Keeping our climate tipping benchmark preference parameters, the most
pessimistic description of the tipping point process will imply today's
optimal social cost of carbon to be \textcolor{black}{\$568 per ton
of carbon}, while increasing the risk parameter to 20 would further
increase today's social cost of carbon to \textcolor{black}{\$817
per ton of carbon}.

\begin{figure}[H]
\centering{}\includegraphics[width=0.5\textwidth,height=0.3\textheight]{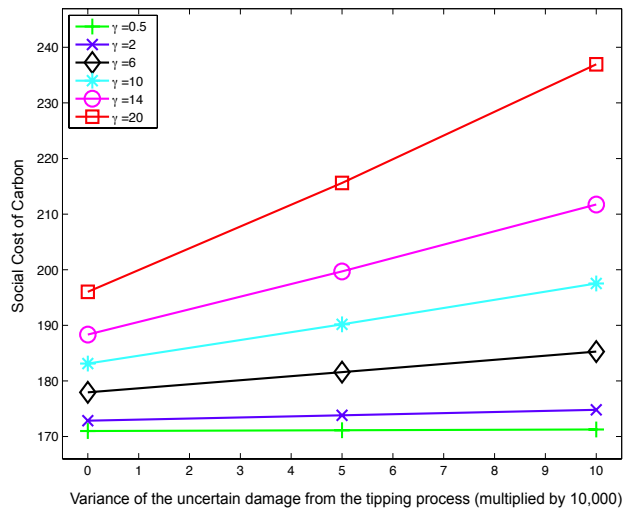}\protect\caption{\textcolor{black}{\small{}Sensitivity of the social cost of carbon
to the risk-aversion parameter and uncertainty regarding post-tipping
damage \label{fig:Effect-of-RiskAversionVolDamage}}}
\end{figure}

In addition, from the results of all the 2,430 cases, we found that
there exists one common pattern: the initial social cost of carbon
is linear in $q$. Figure \ref{fig:Effect-of-RiskAversionVolDamage}
shows the numbers for the social cost of carbon for various values
of $\gamma$ and $q$, when $\psi=1.5$, $\lambda=0.0035$, $\overline{\mathcal{J}}_{\infty}=0.05$,
and $\overline{\mathcal{D}}=50$.%
\footnote{ In Figure \ref{fig:Effect-of-RiskAversionVolDamage}, the horizontal
axis is the variance of the uncertain damage level at the final absorbing
stage, namely $q\overline{\mathcal{J}}_{\infty}^{2}$, and it is scaled
by 10,000. %
} Other cases have the same qualitative pattern, so we omit them here.
In Figure \ref{fig:Effect-of-RiskAversionVolDamage}, the blue line,
the black line, the red line, and the green line represent the case
of $\gamma=0.5$, 2, 6, 10, 14, and 20 respectively. We see that all
these lines are straight, and that a higher $\gamma$ implies a greater
slope, meaning that it is more sensitive to the variance of the uncertain
damage level. 

This is not surprising since it fits into the logic of the basic consumption-based
capital asset pricing model (Lucas 1978) which tells us that the price
of risk is related to its covariance with the aggregate endowment.
Since the magnitude of the damage is proportional to output, the damage
is strongly related to output; in fact, in this case the climate damage
is the only stochastic element of output conditional on the tipping
event. Therefore, the correlation is unity and the social cost of
carbon has a price of risk component which is linear in the variance
of the uncertain damage level.

Note that the social cost of carbon increases as the variance of the
uncertain damage level increases. One interpretation of variance is
that it represents our ignorance of the consequences of an unfolding
tipping process. With that interpretation, the horizontal difference
represents the decline in the social cost of carbon that would result
if we carried out more scientific research and reduced the uncertainty
regarding the post-tipping damage level. This observation shows that
our model could be used to identify the value of reducing uncertainty
and to indicate which kinds of scientific studies would be the most
valuable to pursue. This is a point left for further development in
future studies.

\section{Social Cost of Carbon with Stochastic Growth \& Climate Tipping\label{sec:Results:-Stochastic-Growth and Tipp}}

The previous two sections have examined the impacts on the social
cost of carbon from stochastic growth and stochastic climate tipping,
both in isolation. The real world system includes both uncertainties
and this section presents the results of our model in the presence
of long-run risk in both economic growth and in the climate tipping
process. The optimal policy will now have to balance the need to delay
the triggering of the tipping point process with the accumulation
of additional capital in the face of stochastic growth, and with the
desire to smooth our consumption patterns. We study a \textit{stochastic
growth and climate tipping benchmark }case of parameter specification
and carry out a sensitivity analysis.

\subsection{The Stochastic Growth \& Climate Tipping Benchmark\label{sub:benchmark-LRR-tip}}

\begin{figure}[H]
\centering{}\includegraphics[width=0.8\textwidth]{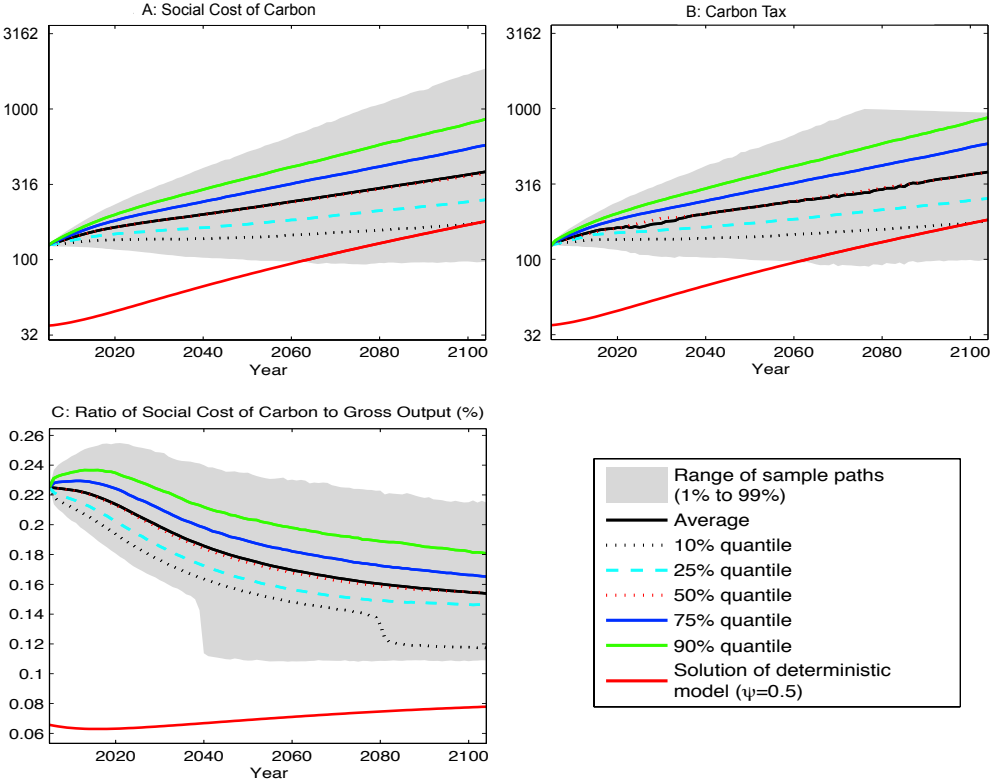}\protect\caption{\textcolor{black}{\small{}Simulation results for stochastic growth
and climate tipping benchmark\label{fig:sim-res-LRR-tip}}}
\end{figure}

We use $\psi=1.5$, $\gamma=10$, $\lambda=0.0035$, $\overline{\mathcal{J}}_{\infty}=0.05$,
$q=0.2$, and $\mathcal{\overline{\mathcal{D}}}=50$ for the stochastic
growth and climate tipping benchmark. Figure \ref{fig:sim-res-LRR-tip}
shows the results of 10,000 simulation paths over the first 100 years
for dynamics of the social cost of carbon, the carbon tax, and the
ratio of social cost of carbon to gross world output respectively.
Other variables such as capital, consumption and its growth, atmospheric
carbon concentration, and surface temperature have pictures visually
similar to the corresponding pictures in Figures \ref{fig:sim-res-LRR-climate}
and \ref{fig:sim-res-tip-representative-climate}, so we omit them.
We use the same line and color types as in previous figures. 

We first study the social cost of carbon. Its initial-time level is
\textcolor{black}{\$125 per ton of carbon} and at 2100 the average
(or expected) social cost of carbon is around \$\textcolor{black}{400
per ton of carbon}. Thus, the path of the expected social cost of
carbon falls between its paths obtained from our analyses of each
risk component in isolation. At the initial time the social cost of
carbon of \textcolor{black}{\$125 per ton of carbon} is even exactly
the average of the numbers obtained from the two cases of the previous
sections (\textcolor{black}{\$61 per ton of carbon} and \textcolor{black}{\$189
per ton of carbon}). Compared to a deterministic model, which would
ignore both risk components and have $\psi=0.5$, we find that the
initial social cost of carbon increases by a factor of 3.2 and that
with 90 percent probability the social cost of carbon will be significantly
higher throughout this century. 

The presence of both stochastic growth and climate tipping risk also
increases the variance of the future social cost of carbon relative
to the separate stochastic growth and climate tipping benchmarks.
For example, the social cost of carbon in 2100 ranges from \textcolor{black}{\$100
per ton of carbon} (the 1 percent quantile) to \textcolor{black}{\$1,700
per ton of carbon} (the 99 percent quantile). The carbon tax, which
we present in panel B, is also more likely to hit its upper bound
after 2072 than in either of the single risk benchmarks. The combination
of these risks implies that there is a probability of about 7 percent
that mitigation policies will have reached the limit of their effectiveness
by 2100.

\textcolor{black}{Some recent research has argued for simple rules
of thumb for the social cost of carbon. In particular, }Golosov et
al. (2014) set up a dynamic, forward-looking climate--economy model
with logarithmic utility and full capital depreciation to argue that
the optimal social cost of carbon is proportional to output. Barrage
(2014) shows that the benchmark in Golosov et al. (2014) implies that
the ratio of the social cost of carbon to decadal gross world output
is $8\times10^{-5}$ and constant over time with constant productivity
growth, but that it increases over time for the productivity process
in Nordhaus (2008), approaching $8\times10^{-5}$ from below. Direct
comparisons with deterministic models like those analyzed in Barrage
(2014) are difficult. Nevertheless, for our stochastic growth and
climate tipping benchmark case, we compute the ratio of the social
cost of carbon to gross world output ($\mathrm{SCC}_{t}/\mathcal{Y}_{t}$),
which we show\textcolor{black}{{} }in panel C in Figure \ref{fig:sim-res-LRR-tip}.%
\footnote{More precisely, we report the ratio $\Gamma_{t}/1000/\mathcal{Y}_{t}$
in order to have the same units as those used in Golosov et al. (2014). %
}

First, we note that when compared to the deterministic model version,
$\mathrm{SCC}_{t}/\mathcal{Y}_{t}$ is about three times larger at
the initial time while at 2100 it is expected to be about twice as
large. Second, we find that the expected $\mathrm{SCC}_{t}/\mathcal{Y}_{t}$
is decreasing over time by about 50 percent, and is thus not constant.
Third, and most importantly, we find that the ratio of the social
cost of carbon to gross world output is not close to any simple path,
but is rather a stochastic process varying over the interval {[}0.00108,
0.00215{]} at year 2100 and over an even larger interval for the second
half of this century. These results indicate that when there are multiple
uncertainties, such as in the economic and climate systems, incorporating
them into an integrated assessment model might lead to substantially
different qualitative and quantitative results. In this example case,
we find no useful, robust, dynamic relation between the social cost
of carbon and gross world output.

\subsection{Sensitivity of the Stochastic Growth \& Climate Tipping Benchmark}

We compute the sensitivity of the stochastic growth and climate tipping
benchmark case to several parameters. Table \ref{tab:SCC-LRR-tip}
lists the initial social cost of carbon for selected combinations
of the parameter values for sensitivity analysis. These parameters
are the hazard rate $\lambda$, the post-tipping damage level $\overline{\mathcal{J}}_{\infty}$,
the mean duration time of the tipping process $\mathcal{\overline{\mathcal{D}}}$,
the mean squared-variance ratio $q$, the elasticity of inter-temporal
substitution $\psi$, the risk aversion parameter $\gamma$, and the
rate of persistence in the long-run risk of stochastic growth $r$.

\begin{table}[H]
\begin{centering}
\textcolor{black}{\small{}}%
\begin{tabular}{c|c|c|c|c|cc|cc}
\hline 
 &  &  &  &  & \multicolumn{4}{c}{\textcolor{black}{\small{}Social cost of carbon ($\mathrm{SCC}$)}}\tabularnewline
\cline{6-9} 
\textcolor{black}{\small{}$\lambda$} & \textcolor{black}{\small{}$\overline{\mathcal{J}}_{\infty}$} & \textcolor{black}{\small{}$\mathcal{\overline{\mathcal{D}}}$} & \textcolor{black}{\small{}$q$} & \textcolor{black}{\small{}$r$} & \multicolumn{2}{c|}{\textcolor{black}{\small{}$\psi=0.5$}} & \multicolumn{2}{c}{\textcolor{black}{\small{}$\psi=1.5$}}\tabularnewline
\cline{6-9} 
 &  &  &  &  & \textcolor{black}{\small{}$\gamma=2$} & \textcolor{black}{\small{}$\gamma=10$} & \textcolor{black}{\small{}$\gamma=2$} & \textcolor{black}{\small{}$\gamma=10$}\tabularnewline
\hline 
\textcolor{black}{\small{}0.0035} & \textcolor{black}{\small{}0.05} & \textcolor{black}{\small{}50} & \textcolor{black}{\small{}0} & \textcolor{black}{\small{}0.775} & \textcolor{black}{\small{}55} & \textcolor{black}{\small{}112 } & \textcolor{black}{\small{}155 } & \textcolor{black}{\small{}116 }\tabularnewline
\textcolor{black}{\small{}0.0035} & \textcolor{black}{\small{}0.10} & \textcolor{black}{\small{}50} & \textcolor{black}{\small{}0} & \textcolor{black}{\small{}0.775} & \textcolor{black}{\small{}73} & \textcolor{black}{\small{}189 } & \textcolor{black}{\small{}236 } & \textcolor{black}{\small{}191 }\tabularnewline
\textcolor{black}{\small{}0.0045} & \textcolor{black}{\small{}0.05} & \textcolor{black}{\small{}50} & \textcolor{black}{\small{}0} & \textcolor{black}{\small{}0.775} & \textcolor{black}{\small{}59} & \textcolor{black}{\small{}123 } & \textcolor{black}{\small{}170 } & \textcolor{black}{\small{}128 }\tabularnewline
\textcolor{black}{\small{}0.0035} & \textcolor{black}{\small{}0.05} & \textcolor{black}{\small{}5} & \textcolor{black}{\small{}0} & \textcolor{black}{\small{}0.775} & \textcolor{black}{\small{}63} & \textcolor{black}{\small{}133 } & \textcolor{black}{\small{}175 } & \textcolor{black}{\small{}136 }\tabularnewline
\textcolor{black}{\small{}0.0045} & \textcolor{black}{\small{}0.10} & \textcolor{black}{\small{}5} & \textcolor{black}{\small{}0} & \textcolor{black}{\small{}0.775} & \textcolor{black}{\small{}102} & \textcolor{black}{\small{}293} & \textcolor{black}{\small{}318} & \textcolor{black}{\small{}274}\tabularnewline
\textcolor{black}{\small{}0.0045} & \textcolor{black}{\small{}0.10} & \textcolor{black}{\small{}5} & \textcolor{black}{\small{}0.4} & \textcolor{black}{\small{}0.775} & \textcolor{black}{\small{}107} & \textcolor{black}{\small{}398} & \textcolor{black}{\small{}331} & \textcolor{black}{\small{}354}\tabularnewline
\textcolor{black}{\small{}0.0035} & \textcolor{black}{\small{}0.05} & \textcolor{black}{\small{}50} & \textcolor{black}{\small{}0} & \textcolor{black}{\small{}0.5} & \textcolor{black}{\small{}53} & \textcolor{black}{\small{}84} & \textcolor{black}{\small{}162 } & \textcolor{black}{\small{}133 }\tabularnewline
\hline 
\end{tabular}
\par\end{centering}{\small \par}

\protect\caption{\textcolor{black}{\small{}\label{tab:SCC-LRR-tip}Initial social cost
of carbon (\$ per ton of carbon) under stochastic growth and climate
tipping}}
\end{table}

For example, when $\psi=1.5$, $\gamma=10$, $\lambda=0.0035$, $\overline{\mathcal{J}}_{\infty}=0.05$,
$q=0$, $\mathcal{\overline{\mathcal{D}}}=50$, and $r=0.775$ (the
default values used in all of the previous examples), the initial
social cost of carbon is \$116 per ton of carbon. We compare this
number to the \$125 per ton of carbon of the stochastic growth and
climate tipping benchmark case in Subsection \ref{sub:benchmark-LRR-tip}
for which we set $q=0.2$. Thus, higher $q$ implies a higher initial
social cost of carbon, which is consistent with our observations for
cases with a stochastic climate tipping only, shown in Table \ref{tab:SCC-tip}.
This is more clearly reflected by comparing the extreme cases of the
fifth and sixth rows of Table \ref{tab:SCC-LRR-tip} (with $\lambda=0.0045$,
$\overline{\mathcal{J}}_{\infty}=10\%$, $\mathcal{\overline{\mathcal{D}}}=5$,
and $r=0.775$): for every case of $(\psi,\gamma)$, the sixth row
with $q=0.4$ has a larger social cost of carbon than the fifth row
with $q=0$. Also, the range of the initial-time social cost of carbon
is wide, from \textcolor{black}{\$53 per ton of carbon} (the case
with $\lambda=0.0035$, $\overline{\mathcal{J}}_{\infty}=0.05$, $\mathcal{\overline{\mathcal{D}}}=50$,
$r=0.5$, $\psi=0.5$, and $\gamma=2$) to \textcolor{black}{\$398
per ton of carbon} (the case with $\lambda=0.0045$, $\overline{\mathcal{J}}_{\infty}=0.10$,
$\mathcal{\overline{\mathcal{D}}}=5$, $r=0.775$, $\psi=0.5$, and
$\gamma=10$). 

Moreover, some qualitative properties found in previous examples with
only climate risk still hold. That is, a higher hazard rate parameter
$\lambda$, a higher mean damage level $\overline{\mathcal{J}}_{\infty}$,
a larger mean squared-variance ratio $q$, or a shorter expected duration
$\mathcal{\overline{\mathcal{D}}}$ will lead to a higher social cost
of carbon, although their quantitative values differ substantially. 

However, the qualitative properties of the preference parameters ($\psi$
and $\gamma$) are nontrivial now: when $\psi=0.5$, a higher $\gamma$
always implies a higher social cost of carbon, and when $\psi=1.5$,
the table shows that the effect of higher $\gamma$ on the social
cost of carbon can be positive or negative. In most cases except the
sixth row of Table \ref{tab:SCC-LRR-tip}, a higher $\gamma$ will
have a smaller social cost of carbon, while the sixth row is averse.
We also know from Table \ref{tab:SCC-tip} that if the volatility
of the economic risk goes to zero, then a higher $\gamma$ will have
a higher social cost of carbon for any $\psi$. This is partially
reflected in the last row of the table: when $\psi=1.5$ and $r=0.5$,
the difference of the social cost of carbon between $\gamma=2$ and
$\gamma=10$ is $162-133=$ \$29 per ton of carbon. This number is
less than the corresponding difference in the first data row with
$\psi=1.5$ and $r=0.775$, which is $155-116=$ \$39 per ton of carbon,
because the volatility of the economic risk with $r=0.775$ is larger
than that with $r=0.5$.

\section{\textcolor{black}{Conclusion} \label{sec:Conclusion}}

This study has presented DSICE, a computational framework for stochastic
integrated assessment of issues related to the joint evolution of
the economy and the climate. We analyzed the optimal level and dynamic
properties of the social cost of carbon and the associated optimal
carbon tax in the face of stochastic and irreversible climate change
and its interaction with economic factors including growth uncertainty
and preferences about risk. We did this in a manner that has allows
us to compare our results to the deterministic model in Nordhaus (2008),
an influential and well-known integrated assessment model. The specific
examples in this study show three basic points. 

First, we use Epstein–Zin preferences in order to specify tastes that
are more compatible with the evidence on risk aversion and the inter-temporal
elasticity of substitution. Those parametric specifications imply
significantly larger social costs of carbon. 

Second, the incorporation of long-run risk shows that for plausible
parameterizations the social cost of carbon is itself a stochastic
process with considerable uncertainty. The range of possible realizations
for the social cost of carbon can be so large that policy makers may
find it rational to pursue methods that are shown to be irrational
in standard deterministic models. Examination of parameter uncertainty
also shows that the range of plausible social cost of carbon values
is much larger than implied by other integrated assessment analyses. 

Third, climate scientists have recently argued that tipping elements
in the climate system contribute to the uncertainty regarding future
climate conditions. We incorporate tipping elements into our model,
and find that the threat of a tipping element induces significant
and immediate increases in the social cost of carbon, even for moderate
assumptions about the likelihood and impacts of the climate tipping
events. The social cost of carbon can be very high even without assuming
catastrophic climate change events, but rather by merely assuming
plausibly parameterized examples of uncertain and irreversible climate
change. We also find that tipping events with modest damage levels
in the distant future can have significant effects on the current
social cost of carbon, a finding which argues against discounting
future expected damage at high rates.

Finally, we have also shown that it is possible to solve empirically
plausible nine-dimensional models of climate and the economy that
include (i) productivity shocks of the kind studied in macroeconomics,
(ii) dynamically non-separable preferences consistent with observed
prices of risk, and (iii) stochastic tipping elements in the climate
system.\textcolor{black}{{} We find that the interaction between the
preferences and these economic and climate risks is qualitatively
nontrivial. }

\begin{singlespace}

\section*{Appendix A---Definition of Parameters }
\end{singlespace}

\begin{singlespace}
In Equations (\ref{eq:emission-industry}), (\ref{eq:mitigation-cost-def}),
(\ref{eq:emission-def}) and (\ref{eq:forcing-def}), we used $\sigma_{t}$
(carbon intensity of output.), $\theta_{1,t}$ (mitigation cost coefficient),
$E_{\mathrm{Land},t}$ (annual carbon emissions from biological processes),
and $F_{\mathrm{EX},t}$ (exogenous radiative forcing). The following
shows their definition with annual time steps, which are adapted from
the decadal formulas in Nordhaus (2008): 

\begin{equation}
\sigma_{t}=\sigma_{0}\exp\left(-0.0073(1-e^{-0.003t})/0.003\right)\label{eq:tech-factor}
\end{equation}

\begin{equation}
\theta_{1,t}=\frac{1.17\sigma_{t}\left(1+e^{-0.005t}\right)}{2\theta_{2}}\label{eq:backstop-DICE-CJL}
\end{equation}
\begin{equation}
E_{\mathrm{Land},t}=1.1e^{-0.01t}\label{eq:E-Land-CJL}
\end{equation}

\begin{equation}
F_{\mathrm{EX},t}=\begin{cases}
-0.06+0.0036t, & \mathrm{if}\; t\leq100\\
0.3, & \mathrm{otherwise}
\end{cases}\label{eq:exogenous-radiativ}
\end{equation}

The calibration of the stochastic productivity process and the climate
tipping element is discussed in Appendices B and D. The calibration
of the parameters in the climate system of the deterministic model
with the annual time steps is discussed in Appendix C. \textcolor{black}{The
values of other parameters in the carbon and }temperature systems
are the same as in Nordhaus (2008). Tables \ref{tab:Param-econ}--\ref{tab:Param-climate}
list the values and/or definition of all parameters, variables, and
symbols. 

\begin{table}[H]
\begin{singlespace}
\begin{centering}
\textcolor{black}{\small{}}%
\begin{tabular}{|l|>{\raggedright}p{100mm}|}
\hline 
\textcolor{black}{\small{}$t\in\{0,1,...,600\}$} & \textcolor{black}{\small{}time in years ($t$ represents year $t+2005$)}\tabularnewline
\textcolor{black}{\small{}$\psi\in[0.5,2]$ } & \textcolor{black}{\small{}inter-temporal elasticity of substitution
(default: 1.5)}\tabularnewline
\textcolor{black}{\small{}$\gamma\in[0.5,20]$ } & \textcolor{black}{\small{}risk aversion parameter (default: 10)}\tabularnewline
\textcolor{black}{\small{}$u$} & \textcolor{black}{\small{}utility function}\tabularnewline
\textcolor{black}{\small{}$U$} & \textcolor{black}{\small{}Epstein--Zin utility}\tabularnewline
\textcolor{black}{\small{}$V$} & \textcolor{black}{\small{}value function}\tabularnewline
\textcolor{black}{\small{}$\beta=0.985$ } & \textcolor{black}{\small{}discount factor}\tabularnewline
\textcolor{black}{\small{}$A_{t}$ } & \textcolor{black}{\small{}productivity trend at time $t$, $A_{0}=0.0272$}\tabularnewline
\textcolor{black}{\small{}$L_{t}$ } & \textcolor{black}{\small{}population at time $t$}\tabularnewline
\textcolor{black}{\small{}$K_{t}$} & \textcolor{black}{\small{}capital at time $t$ (in \$ trillions),
$K_{0}=137$}\tabularnewline
\textcolor{black}{\small{}$\Psi_{t}$} & \textcolor{black}{\small{}mitigation expenditure at time $t$}\tabularnewline
\textcolor{black}{\small{}$C_{t}$} & \textcolor{black}{\small{}consumption at time $t$}\tabularnewline
\textcolor{black}{\small{}$c_{t}$} & \textcolor{black}{\small{}per capita consumption at time $t$}\tabularnewline
\textcolor{black}{\small{}$g_{c}$} & \textcolor{black}{\small{}per capita consumption growth rate }\tabularnewline
\textcolor{black}{\small{}$I_{t}$} & \textcolor{black}{\small{}investment at time $t$}\tabularnewline
\textcolor{black}{\small{}$f$} & \textcolor{black}{\small{}production function }\tabularnewline
\textcolor{black}{\small{}$\alpha=0.3$ } & \textcolor{black}{\small{}output elasticity of capital}\tabularnewline
\textcolor{black}{\small{}$\alpha_{1}$} & \textcolor{black}{\small{}initial growth rate of the productivity
trend (default: 0.0092)}\tabularnewline
\textcolor{black}{\small{}$\alpha_{2}$} & \textcolor{black}{\small{}decline rate of the growth rate of the productivity
trend (default: 0.001)}\tabularnewline
\textcolor{black}{\small{}$\delta=0.1$ } & \textcolor{black}{\small{}annual depreciation rate }\tabularnewline
\textcolor{black}{\small{}$\mathcal{Y}_{t}$} & \textcolor{black}{\small{}gross world output at time $t$}\tabularnewline
\textcolor{black}{\small{}$\mathrm{SCC}_{t}$} & \textcolor{black}{\small{}social cost of carbon}\tabularnewline
\hline 
\end{tabular}
\par\end{centering}{\small \par}
\end{singlespace}

\protect\caption{\textcolor{black}{\small{}Parameters, variables, and symbols for the
economic system of our model\label{tab:Param-econ}}}
\end{table}

\begin{table}[H]
\begin{singlespace}
\begin{centering}
\begin{tabular}{|l|>{\raggedright}p{100mm}|}
\hline 
$\chi_{t}$  & persistence of productivity shock at time $t$, $\chi_{0}=0$\tabularnewline
$\zeta_{t}$  & stochastic productivity shock at time $t$, $\zeta_{0}=1$ \tabularnewline
$\varrho=0.035$  & productivity process parameter\tabularnewline
$r$  & productivity process parameter (default: 0.775)\tabularnewline
$\varsigma=0.008$  & productivity process parameter\tabularnewline
$\widetilde{A}_{t}=\zeta_{t}A_{t}$ & stochastic productivity at time $t$\tabularnewline
$g_{\zeta}$ & transition function of $\zeta_{t}$\tabularnewline
$g_{\chi}$ & transition function of $\chi_{t}$\tabularnewline
$\omega_{\zeta,t}$ & i.i.d. shocks in transition of $\zeta_{t}$ \tabularnewline
$\omega_{\chi,t}$ & i.i.d. shocks in transition of $\chi_{t}$ \tabularnewline
$\Lambda$ & coefficient of the lag-1 linear autoregression \tabularnewline
$\epsilon_{t}$ & residuals of the lag-1 autoregression fitting\tabularnewline
$\sigma(\epsilon)$ & one-period-ahead conditional standard deviation (standard deviation
of the lag-1 autoregression residuals $\epsilon$)\tabularnewline
\hline 
\end{tabular}
\par\end{centering}
\end{singlespace}

\protect\caption{Parameters, variables, and symbols for the stochastic growth specification\label{tab:Param-LRR}}
\end{table}

\end{singlespace}

\begin{table}[H]
\begin{singlespace}
\begin{centering}
\textcolor{black}{\small{}}%
\begin{tabular}{|l|>{\raggedright}p{100mm}|}
\hline 
\textcolor{black}{\small{}$J_{t}$} & \textcolor{black}{\small{}tipping state at time $t$. $J_{t}$ also
denotes the damage level from climate tipping. }\tabularnewline
\textcolor{black}{\small{}$\mathcal{J}_{i}$} & \textcolor{black}{\small{}possible values of $J_{t}$}\tabularnewline
\textcolor{black}{\small{}$\mathcal{J}_{\infty}$} & \textcolor{black}{\small{}uncertain long-run tipping damage level}\tabularnewline
\textcolor{black}{\small{}$p_{i,j,t}$} & \textcolor{black}{\small{}transition probability of $J_{t}$ at time
$t$}\tabularnewline
\textcolor{black}{\small{}$g_{J}$} & \multicolumn{1}{l|}{\textcolor{black}{\small{}transition function of $J_{t}$}}\tabularnewline
\textcolor{black}{\small{}$\omega_{J,t}$} & \textcolor{black}{\small{}i.i.d. shock in transition of $J_{t}$}\tabularnewline
\textcolor{black}{\small{}$\underline{T_{\mathrm{AT}}}=1$} & \textcolor{black}{\small{}surface temperature with zero probability
of tipping}\tabularnewline
\textcolor{black}{\small{}$\lambda\in[0.0025,0.0045]$ } & \textcolor{black}{\small{}hazard rate parameter}\tabularnewline
\textcolor{black}{\small{}$\overline{\mathcal{J}}_{\infty}\in[0.025,0.10]$ } & \textcolor{black}{\small{}mean tipping damage level in the long run}\tabularnewline
\textcolor{black}{\small{}$q\in[0,0.4]$ } & \textcolor{black}{\small{}mean-square-variance ratio for the uncertain
tipping damage level}\tabularnewline
\textcolor{black}{\small{}$\overline{\mathcal{D}}\in[5,200]$ } & \textcolor{black}{\small{}expected duration of the whole tipping process
in years}\tabularnewline
\hline 
\end{tabular}
\par\end{centering}{\small \par}
\end{singlespace}

\protect\caption{\textcolor{black}{\small{}Parameters, variables, and symbols for the
climate tipping element\label{tab:Param-tip}}}
\end{table}

\begin{singlespace}
\bigskip{}

\end{singlespace}

\begin{table}[H]
\begin{centering}
\textcolor{black}{\small{}}%
\begin{tabular}{|l|>{\raggedright}p{85mm}|}
\hline 
\textcolor{black}{\small{}$M_{\mathrm{AT},t}$} & \textcolor{black}{\small{}carbon concentration in atmosphere (billion
tons) at time $t$, $M_{\mathrm{AT},0}=808.9$}\tabularnewline
\textcolor{black}{\small{}$M_{\mathrm{UO},t}$} & \textcolor{black}{\small{}carbon concentration in upper ocean (billion
tons) at time $t$, $M_{\mathrm{UO},0}=1255$}\tabularnewline
\textcolor{black}{\small{}$M_{\mathrm{LO},t}$} & \textcolor{black}{\small{}carbon concentration in lower ocean (billions
tons) at time $t$, $M_{\mathrm{LO},0}=18365$}\tabularnewline
\textcolor{black}{\small{}$\mathbf{M}_{t}=\left(M_{\mathrm{AT},t},M_{\mathrm{UO},t},M_{\mathrm{LO},t}\right)^{\top}$} & \textcolor{black}{\small{}carbon concentration vector at time $t$}\tabularnewline
\textcolor{black}{\small{}$T_{\mathrm{AT},t}$} & \textcolor{black}{\small{}global average surface temperature (degrees
Celsius) at time $t$, $T_{\mathrm{AT},0}=0.7307$}\tabularnewline
\textcolor{black}{\small{}$T_{\mathrm{OC},t}$} & \textcolor{black}{\small{}global average ocean temperature (degrees
Celsius) at time $t$, $T_{\mathrm{OC},0}=0.0068$}\tabularnewline
\textcolor{black}{\small{}$\mathbf{T}_{t}=\left(T_{\mathrm{AT},t},T_{\mathrm{OC},t}\right)^{\top}$} & \textcolor{black}{\small{}temperature vector at time $t$}\tabularnewline
\textcolor{black}{\small{}$\mathbf{S}=(K,\mathbf{M},\mathbf{T},\zeta,\chi,J)$} & \textcolor{black}{\small{}nine-dimensional state vector}\tabularnewline
\textcolor{black}{\small{}$\mu_{t}$} & \textcolor{black}{\small{}emission control rate at time $t$}\tabularnewline
\textcolor{black}{\small{}$\Omega_{t}$} & \textcolor{black}{\small{}damage factor function at time $t$}\tabularnewline
\textcolor{black}{\small{}$\mathcal{F}_{t}$} & \textcolor{black}{\small{}radiative forcing at time $t$}\tabularnewline
\textcolor{black}{\small{}$\mathcal{E}_{t}$} & \textcolor{black}{\small{}emission at time $t$}\tabularnewline
\textcolor{black}{\small{}$\pi_{1}=0$ } & \textcolor{black}{\small{}damage factor parameter}\tabularnewline
\textcolor{black}{\small{}$\pi_{2}=0.0028388$ } & \textcolor{black}{\small{}damage factor parameter}\tabularnewline
\textcolor{black}{\small{}$\theta_{2}=2.8$ } & \textcolor{black}{\small{}mitigation cost parameter}\tabularnewline
\textcolor{black}{\small{}$\sigma_{0}=0.13418$ } & \textcolor{black}{\small{}initial technology factor }\tabularnewline
\textcolor{black}{\small{}$\sigma_{t}$ } & \textcolor{black}{\small{}technology factor at time $t$}\tabularnewline
\textcolor{black}{\small{}$\Phi_{\mathrm{M}}$} & \textcolor{black}{\small{}transition matrix of carbon cycle}\tabularnewline
\textcolor{black}{\small{}$\Phi_{\mathrm{T}}$} & \textcolor{black}{\small{}transition matrix of temperature system}\tabularnewline
\textcolor{black}{\small{}$\phi_{12}=0.019$ } & \textcolor{black}{\small{}rate of carbon flux: atmosphere to upper
ocean}\tabularnewline
\textcolor{black}{\small{}$\phi_{23}=0.0054$ } & \textcolor{black}{\small{}rate of carbon flux: upper ocean to lower
ocean}\tabularnewline
\textcolor{black}{\small{}$\phi_{21}=0.01$} & \textcolor{black}{\small{}rate of carbon flux: upper ocean to atmosphere}\tabularnewline
\textcolor{black}{\small{}$\phi_{32}=0.00034$} & \textcolor{black}{\small{}rate of carbon flux: lower ocean to upper
ocean}\tabularnewline
\textcolor{black}{\small{}$\xi_{1}=0.037$ } & \textcolor{black}{\small{}temperature transition parameter}\tabularnewline
\textcolor{black}{\small{}$\xi_{2}=0.047$ } & \textcolor{black}{\small{}rate of atmospheric temperature decrease
due to infrared radiation to space}\tabularnewline
\textcolor{black}{\small{}$\varphi_{12}=0.01$ } & \textcolor{black}{\small{}rate of heat diffusion: atmosphere to ocean}\tabularnewline
\textcolor{black}{\small{}$\varphi_{21}=0.0048$} & \textcolor{black}{\small{}rate of heat diffusion: ocean to atmosphere}\tabularnewline
\textcolor{black}{\small{}$\eta=3.8$ } & \textcolor{black}{\small{}radiative forcing parameter }\tabularnewline
\textcolor{black}{\small{}$\theta_{1,t}$ } & \textcolor{black}{\small{}adjusted cost for backstop at time $t$}\tabularnewline
\textcolor{black}{\small{}$F_{\mathrm{EX},t}$} & \textcolor{black}{\small{}exogenous radiative forcing in year $t$}\tabularnewline
$E_{\mathrm{Ind},t}$ & industrial carbon emissions \textcolor{black}{\small{}(billions tons)}
in year $t$ \tabularnewline
\textcolor{black}{\small{}$E_{\mathrm{Land},t}$} & \textcolor{black}{\small{}carbon emissions (billions tons) from land
use in year $t$}\tabularnewline
\textcolor{black}{\small{}$M_{\mathrm{AT}}^{*}=596.4$ } & \textcolor{black}{\small{}preindustrial atmospheric carbon concentration}\tabularnewline
\hline 
\end{tabular}
\par\end{centering}{\small \par}

\protect\caption{\textcolor{black}{\small{}Parameters, variables, and symbols in the
temperature and carbon systems of our model\label{tab:Param-climate}}}
\end{table}

\begin{singlespace}

\section*{Appendix B---Calibration of the Productivity Process}
\end{singlespace}

\begin{singlespace}
We construct a time-dependent, finite-state Markov chain for the productivity
process $\left(\zeta_{t},\chi_{t}\right)$, which depends on three
parameters: $\varrho$, $r$, and $\varsigma$. Our calibration target
is to choose these three parameters so that we match the conditional
and unconditional moments of endogenous consumption growth with the
statistics from observed annual market data. The overall strategy
used basic solution and simulation methods. The consumption data was
that from 1930 to 2008. For each choice of $\varrho$, $r$, and $\varsigma$
and each choice of the sizes of the Markov chains, $n_{\zeta}$ and
$n_{\chi}$ we tested, we solved our model assuming damage from climate
is zero (i.e., $J_{t}=\pi_{1}=\pi_{2}=0$). We also use our default
case preference parameters $\psi=1.5$ and $\gamma=10$. Once we obtained
a solution, we computed 10,000 simulations of the consumption process
over the first century and computed the conditional and unconditional
moments of the per capita consumption growth paths. 

We now present the details. Our construction of the Markov chain is
adapted from the following processes%
\footnote{To avoid confusion, we use $\widehat{\zeta}_{t}$ and $\widehat{\chi}_{t}$
to represent the continuous random variables at each time $t$, while
$\zeta_{t}$ and $\chi_{t}$ represent the Markov chains. %
} 
\[
\log\left(\hat{\zeta}_{t+1}\right)=\log\left(\hat{\zeta}_{t}\right)+\hat{\chi}_{t}+\varrho\hat{\omega}_{\zeta,t},
\]

\[
\hat{\chi}_{t+1}=r\hat{\chi}_{t}+\varsigma\hat{\omega}_{\chi,t},
\]
where $\hat{\omega}_{\zeta,t}$, $\hat{\omega}_{\chi,t}\sim i.i.d.\:\mathcal{N}(0,1)$.
We know that $\widehat{\chi}_{t}$ is a normal random variable with
mean $0$. Denote $\Upsilon_{t}\equiv\mathrm{Var}\left\{ \widehat{\chi}_{t}\right\} $,
we have $\Upsilon_{t+1}=r^{2}\Upsilon_{t}+\varsigma^{2}$, then from
recursive iteration, we get 
\begin{equation}
\Upsilon_{t}=\frac{\varsigma^{2}\left(1-r^{2t}\right)}{1-r^{2}}\label{eq:var-chi}
\end{equation}
for $t>0$. We also know that $\log\left(\widehat{\zeta}_{t}\right)$
is a normal random variable with mean 0, and 
\[
\log\left(\widehat{\zeta}_{t}\right)=\sum_{s=1}^{t-1}\widehat{\chi}_{t}+\varrho\sum_{s=0}^{t-1}\hat{\omega}_{\zeta,s}.
\]
Denote $\Delta_{t}\equiv\mathrm{Var}\left\{ \log\left(\widehat{\zeta}_{t}\right)\right\} $.
Since $\mathbb{E}\left\{ \widehat{\chi}_{t}\widehat{\chi}_{s}\right\} =r^{\left|t-s\right|}\Upsilon_{\min\{t,s\}}$,
we have 

\begin{equation}
\Delta_{t}=\mathbb{E}\left\{ \left(\log\left(\widehat{\zeta}_{t}\right)\right)^{2}\right\} =\left(\sum_{s=1}^{t-1}\Upsilon_{s}+2\sum_{\tau=2}^{t-1}\sum_{s=1}^{\tau-1}r^{\tau-s}\Upsilon_{s}\right)+\varrho^{2}t,\label{eq:var-logZeta}
\end{equation}
for $t>0$. 

Now we use the the values of $\Upsilon_{t}$ in (\ref{eq:var-chi})
and $\Delta_{t}$ in (\ref{eq:var-logZeta}) to define the values
of our Markov chains $\left(\zeta_{t},\chi_{t}\right)$. We choose
the values of $\chi_{t}$ as $\left\{ \chi_{t,j}:\, j=1,...,n_{\chi}\right\} $
where $\chi_{t,j}$ are equally spaced in $\left[-3\sqrt{\Upsilon_{t}},3\sqrt{\Upsilon_{t}}\right]$
(the probability that the continuously distributed random variable
$\widehat{\chi}_{t}$ is in $\left[-3\sqrt{\Upsilon_{t}},3\sqrt{\Upsilon_{t}}\right]$
is 99.7 percent), at each time $t$. Similarly, we choose the values
of $\zeta_{t}$ as $\left\{ \zeta_{t,i}:\, i=1,...,n_{\zeta}\right\} $
so that $\log\left(\zeta_{t,i}\right)$ are equally spaced in $\left[-3\sqrt{\Delta_{t}},3\sqrt{\Delta_{t}}\right]$
(the probability that the continuously distributed random variable
$\log\left(\widehat{\zeta}_{t}\right)$ is in $\left[-3\sqrt{\Delta_{t}},3\sqrt{\Delta_{t}}\right]$
is 99.7 percent). 

Next we set the transition probability matrices for our Markov chains
$\left(\zeta_{t},\chi_{t}\right)$. For the paired values $\left\{ \left(\zeta_{t,i},\chi_{t,j}\right):i=1,...,n_{\zeta},j=1,...,n_{\chi}\right\} $
($\zeta_{0,i}\equiv1$ and $\chi_{0,j}\equiv0$), we apply the method
in Tauchen (1986) to define the transition probability from $\chi_{t,j}$
to $\chi_{t+1,j'}$: 
\begin{eqnarray*}
\Pr\left\{ \chi_{t+1,j'}\mid\chi_{t,j}\right\}  & = & \Phi\left(\frac{1}{\varsigma}\left(\frac{\chi_{t+1,j'}+\chi_{t+1,j'+1}}{2}-r\chi_{t,j}\right)\right)\\
 &  & -\Phi\left(\frac{1}{\varsigma}\left(\frac{\chi_{t+1,j'-1}+\chi_{t+1,j'}}{2}-r\chi_{t,j}\right)\right),
\end{eqnarray*}
for $j'=2,...,n_{\chi}-1$, and 
\[
\Pr\left\{ \chi_{t+1,1}\mid\chi_{t,j}\right\} =\Phi\left(\frac{1}{\varsigma}\left(\frac{\chi_{t+1,1}+\chi_{t+1,2}}{2}-r\chi_{t,j}\right)\right),
\]

\[
\Pr\left\{ \chi_{t+1,n}\mid\chi_{t,j}\right\} =1-\Phi\left(\frac{1}{\varsigma}\left(\frac{\chi_{t+1,n-1}+\chi_{t+1,n}}{2}-r\chi_{t,j}\right)\right),
\]
where $\Phi\left(\cdot\right)$ is the cumulative normal distribution
function, for any $j=1,...,n_{\chi}$. Similarly, the transition probability
from $\left(\zeta_{t,i},\chi_{t,j}\right)$ to $\zeta_{t+1,i'}$ is
defined as 
\begin{eqnarray*}
\Pr\left\{ \zeta_{t+1,i'}\mid\left(\zeta_{t,i},\chi_{t,j}\right)\right\}  & = & \Phi\left(\frac{1}{\varrho}\left(\frac{\log\left(\zeta_{t+1,i'}\right)+\log\left(\zeta_{t+1,i'+1}\right)}{2}-\left(\log\left(\zeta_{t,i}\right)+\chi_{t,j}\right)\right)\right)\\
 &  & -\Phi\left(\frac{1}{\varrho}\left(\frac{\log\left(\zeta_{t+1,i'}\right)+\log\left(\zeta_{t+1,i'+1}\right)}{2}-\left(\log\left(\zeta_{t,i}\right)+\chi_{t,j}\right)\right)\right),
\end{eqnarray*}
for $i'=2,...,n_{\zeta}-1$, and 
\[
\Pr\left\{ \zeta_{t+1,1}\mid\left(\zeta_{t,i},\chi_{t,j}\right)\right\} =\Phi\left(\frac{1}{\varrho}\left(\frac{\log\left(\zeta_{t+1,1}\right)+\log\left(\zeta_{t+1,2}\right)}{2}-\left(\log\left(\zeta_{t,i}\right)+\chi_{t,j}\right)\right)\right),
\]

\[
\Pr\left\{ \zeta_{t+1,n}\mid\left(\zeta_{t,i},\chi_{t,j}\right)\right\} =1-\Phi\left(\frac{1}{\varrho}\left(\frac{\log\left(\zeta_{t+1,n-1}\right)+\log\left(\zeta_{t+1,n}\right)}{2}-\left(\log\left(\zeta_{t,i}\right)+\chi_{t,j}\right)\right)\right),
\]
for any $i=1,...,n_{\zeta},j=1,...,n_{\chi}$. 

We examine several possibilities for the size of the Markov chains,
and find that $n_{\zeta}=91$ and $n_{\chi}=19$ are enough for a
good approximation (i.e., distributions of solutions of our stochastic
growth benchmark example in the first 100 years are almost invariant
to a higher $n_{\zeta}$ or $n_{\chi}$) with the calibrated values
at $\varrho=0.035$, $r=0.775$, and $\varsigma=0.008$. Moreover,
with $n_{\zeta}=91$ and $n_{\chi}=19$, our transition probability
matrix is not nearly degenerated%
\footnote{\begin{singlespace}
A nearly degenerated probability matrix means that at one state, the
maximum of its nonzero probabilities to next nodes is more than 99.9\%.\end{singlespace}
} at all times in our examples.

We compare the performance of our model to match the statistics of
market data with that of Jensen and Traeger (2014) who specifies a
model for labor productivity growth. Because the logarithm of the
total factor productivity is equal to $(1-\alpha)$ times the logarithm
of the labor productivity, where $\alpha=0.3$ is the output elasticity
of capital, the process implied in Jensen and Traeger (2014) is equivalent
to the following total factor productivity process:

\[
\log\left(\tilde{\zeta}_{t+1}\right)=\log\left(\tilde{\zeta}_{t}\right)+\tilde{\chi}_{t}+0.0133\tilde{\omega}_{\zeta,t},
\]
\[
\tilde{\chi}_{t+1}=0.5\tilde{\chi}_{t}+0.0133\tilde{\omega}_{\chi,t}.
\]
Table \ref{tab:Calib-prod-shock} provides the comparison of the statistics
of the observed annual consumption data, with the simulation statistics
of our model ( with $\varrho=0.035$, $r=0.775$, and $\varsigma=0.008$),
and with the simulation statistics of our model when using the Jensen
and Traeger (2014) parameters ($\varrho=0.0133$, $r=0.5$, and $\varsigma=0.0133$).
For each simulation path of the per capita consumption growth, $g_{c}$,
we compute its standard deviation $\sigma(g_{c})$ and order-$j$
autocorrelations (i.e., the correlation between $g_{c,t+j}$ and $g_{c,t}$).
The medians of $\sigma(g_{c})$ and the order-1 autocorrelation from
10,000 simulation paths of our model are 0.023 and 0.43, close to
the corresponding numbers from the annual market data, and the order-2
autocorrelation of the data is also inside its 90 percent corresponding
confidence interval from our simulations.%
\footnote{Since we use the productivity trend of Nordhaus (2008) with an initial-time
growth rate of 0.92 percent (and decreasing afterwards), our median
of $\mathbb{E}\left(g_{c}\right)$ is smaller than the one of the
market data of the last century, but our confidence interval still
contains the historic market estimate, 0.019.%
} However, the process in Jensen and Traeger (2014) has a much smaller
median standard deviation, only 0.013, and even its confidence interval
$[0.010,0.016]$ does not contain the market estimate, 0.022. Moreover,
the 90\% confidence interval of the order-2 autocorrelation of the
simulation from the Jensen--Traeger process, $[0.27,0.70]$, does
not contain the order-2 autocorrelation of the data.

We also performed the lag-1 autoregression analysis (\ref{eq:AR-def})
for each consumption growth path $g_{c,t}$, and obtained estimates
of the coefficient $\Lambda$ and the standard deviation of the residuals
$\sigma(\epsilon)$. From Table \ref{tab:Calib-prod-shock}, we see
that the autoregression statistics with our calibrated parameter values
are also much closer to those of the data than to those with the parameters
implied by Jensen and Traeger (2014), particularly $\sigma(\epsilon)$
as its value from observed data, 0.0179, is outside of the confidence
interval from the Jensen--Traeger process, but is still within the
90 percent confidence interval from our model.

\begin{table}[H]
\begin{centering}
\textcolor{black}{\small{}}%
\begin{tabular}{c|l|lll|lll}
\hline 
 & \multicolumn{1}{c|}{\textcolor{black}{\small{}Observed Data}} & \multicolumn{3}{c|}{\textcolor{black}{\small{}Our Model }} & \multicolumn{3}{c}{\textcolor{black}{\small{}Jensen and Traeger (2014)}}\tabularnewline
\hline 
\textcolor{black}{\small{}Variable} & \textcolor{black}{\small{}Estimate} & \textcolor{black}{\small{}Median} & \textcolor{black}{\small{}5\%} & \textcolor{black}{\small{}95\%} & \textcolor{black}{\small{}Median} & \textcolor{black}{\small{}5\%} & \textcolor{black}{\small{}95\%}\tabularnewline
\hline 
\textcolor{black}{\small{}$\mathbb{E}\left(g_{c}\right)$} & \textcolor{black}{\small{}0.019} & \textcolor{black}{\small{}0.013} & \textcolor{black}{\small{}0.002} & \textcolor{black}{\small{}0.025} & \textcolor{black}{\small{}0.013} & \textcolor{black}{\small{}0.006} & \textcolor{black}{\small{}0.020}\tabularnewline
\textcolor{black}{\small{}$\sigma(g_{c})$} & \textcolor{black}{\small{}0.022} & \textcolor{black}{\small{}0.023} & \textcolor{black}{\small{}0.019} & \textcolor{black}{\small{}0.028} & \textcolor{black}{\small{}0.013} & \textcolor{black}{\small{}0.010} & \textcolor{black}{\small{}0.016}\tabularnewline
\textcolor{black}{\small{}order-1 autocorrelation} & \textcolor{black}{\small{}0.48} & \textcolor{black}{\small{}0.43} & \textcolor{black}{\small{}0.19} & \textcolor{black}{\small{}0.64} & \textcolor{black}{\small{}0.55} & \textcolor{black}{\small{}0.30} & \textcolor{black}{\small{}0.73}\tabularnewline
\textcolor{black}{\small{}order-2 autocorrelation} & \textcolor{black}{\small{}0.17} & \textcolor{black}{\small{}0.37} & \textcolor{black}{\small{}0.13} & \textcolor{black}{\small{}0.59} & \textcolor{black}{\small{}0.51} & \textcolor{black}{\small{}0.27} & \textcolor{black}{\small{}0.70}\tabularnewline
\textcolor{black}{\small{}$\Lambda$} & \textcolor{black}{\small{}0.46} & \textcolor{black}{\small{}0.48} & \textcolor{black}{\small{}0.24} & \textcolor{black}{\small{}0.68} & \textcolor{black}{\small{}0.59} & \textcolor{black}{\small{}0.36} & \textcolor{black}{\small{}0.76}\tabularnewline
\textcolor{black}{\small{}$\sigma(\epsilon)$} & \textcolor{black}{\small{}0.0179} & \textcolor{black}{\small{}0.0203} & \textcolor{black}{\small{}0.0177} & \textcolor{black}{\small{}0.023} & \textcolor{black}{\small{}0.011} & \textcolor{black}{\small{}0.009} & \textcolor{black}{\small{}0.012}\tabularnewline
\hline 
\end{tabular}
\par\end{centering}{\small \par}

\protect\caption{\textcolor{black}{\small{}Statistics of per capita consumption growth\label{tab:Calib-prod-shock}}}
\end{table}

\end{singlespace}

\begin{singlespace}

\section*{Appendix C---Calibration of the Deterministic Climate System}
\end{singlespace}

\begin{singlespace}
The climate system parameters of Nordhaus (2008) were chosen to represent
a ten-year time period discretization and match the business-as-usual
path (i.e., no emission control) in Nordhaus (2008). Analyses of economic
uncertainty and productivity shocks typically uses much shorter time
periods. Our choice of one year time periods for the economic system
requires us to choose climate system parameters, $\phi_{12}$, $\phi_{21}$,
$\phi_{23}$, $\phi_{32}$, $\xi_{1}$, $\xi_{2}$, $\varphi_{12}$,
and $\varphi_{21}$, consistent with one year periods. As in Nordhaus
(2008), $\phi_{21}=\phi_{12}\tilde{M}_{\mathrm{AT}}^{*}/\tilde{M}_{\mathrm{UO}}^{*}$
and $\phi_{32}=\phi_{23}\tilde{M}_{\mathrm{UO}}^{*}/\tilde{M}_{\mathrm{LO}}^{*}$
where $\tilde{M}_{\mathrm{AT}}^{*}=587.5$, $\tilde{M}_{\mathrm{UO}}^{*}=1144$
and $\tilde{M}_{\mathrm{LO}}^{*}=18340$ are the preindustrial equilibrium
carbon concentrations in the atmosphere, upper ocean, and lower ocean
respectively; moreover, we also set $\xi_{2}=\xi_{1}\eta/\xi_{3}$
where $\xi_{3}=3$ is the equilibrium climate sensitivity and $\eta=3.8$
is the radiative forcing parameter, as in Nordhaus (2008). Thus, there
are five parameters left for the calibration: $\phi_{12}$, $\phi_{23}$,
$\xi_{1}$, $\varphi_{12}$, and $\varphi_{21}$. 

We choose parameters to match the first 500 years of the business-as-usual
paths. At first, we compute the business-as-usual data on the dynamics
of the three-dimensional carbon cycle and two-dimensional temperature
from the model of Nordhaus (2008), denoted $\mathbf{M}_{\mathrm{D}}\left(t\right)=\left(M_{\mathrm{D,AT}}\left(t\right),M_{\mathrm{D,UO}}\left(t\right),M_{\mathrm{D,LO}}\left(t\right)\right)$
and $\mathbf{T}_{\mathrm{D}}\left(t\right)=\left(T_{\mathrm{D,AT}}\left(t\right),T_{\mathrm{D,OC}}\left(t\right)\right)$
respectively, for only the values of $t$ at the decadal time nodes,
$t_{0}=0$, $t_{1}=10,$ $t_{2}=20$, ..., and $t_{50}=500$. These
serve as our data. We take the rate of emissions in the Nordhaus business-as-usual
path at decadal time, and use linear interpolation to produce data
on the business-as-usual emission path, denoted $E_{\mathrm{D}}(t)$,
for each year $t$. For each choice of the parameters, we solve the
climate system equations with emissions set at $E_{\mathrm{D}}(t)$
and find the implied climate system paths, $\mathbf{M}\left(t\right)=\left(M_{\mathrm{AT}}\left(t\right),M_{\mathrm{UO}}\left(t\right),M_{\mathrm{LO}}\left(t\right)\right)$
and $\mathbf{T}\left(t\right)=\left(T_{\mathrm{AT}}\left(t\right),T_{\mathrm{OC}}\left(t\right)\right)$.
We choose values for the parameters $\phi_{12}$, $\phi_{23}$, $\xi_{1}$,
$\varphi_{12}$, and $\varphi_{21}$ to minimize the $\mathcal{L}^{2}$
norm of the difference between the decadal time series data, $\mathbf{M}_{\mathrm{D}}\left(t_{i}\right)$
and $\mathbf{T}_{\mathrm{D}}\left(t_{i}\right)$, and our annual paths
at those decadal times, $\mathbf{M}\left(t_{i}\right)$ and $\mathbf{T}\left(t_{i}\right)$;
more precisely, we solve: 
\begin{eqnarray*}
\min_{\phi_{12},\phi_{23},\xi_{1},\varphi_{12},\varphi_{21}} &  & \sum_{i=1}^{50}\Biggl\{\left(\frac{M_{\mathrm{AT}}\left(t_{i}\right)-M_{\mathrm{D,AT}}\left(t_{i}\right)}{M_{\mathrm{D,AT}}\left(t_{i}\right)}\right)^{2}+\left(\frac{M_{\mathrm{UO}}\left(t_{i}\right)-M_{\mathrm{D,UO}}\left(t_{i}\right)}{M_{\mathrm{D,UO}}\left(t_{i}\right)}\right)^{2}+\\
 &  & \left(\frac{M_{\mathrm{LO}}\left(t_{i}\right)-M_{\mathrm{D,LO}}\left(t_{i}\right)}{M_{\mathrm{D,LO}}\left(t_{i}\right)}\right)^{2}+\left(\frac{T_{\mathrm{AT}}\left(t_{i}\right)-T_{\mathrm{D,AT}}\left(t_{i}\right)}{T_{\mathrm{D,AT}}\left(t_{i}\right)}\right)^{2}+\left(\frac{T_{\mathrm{OC}}\left(t_{i}\right)-T_{\mathrm{D,OC}}\left(t_{i}\right)}{T_{\mathrm{D,OC}}\left(t_{i}\right)}\right)^{2}\Biggr\},\\
\mathrm{s.t.}\qquad &  & \mathbf{M}(t+1)=\mathbf{\mathbf{\Phi_{\mathrm{M}}}}\mathbf{M}(t)+\left(E_{\mathrm{D}}\left(t\right),0,0\right)^{\top},\\
 &  & \mathbf{T}(t+1)=\mathbf{\mathbf{\Phi_{\mathrm{T}}}}\mathbf{T}(t)+\left(\xi_{1}\mathcal{F}_{t}\left(M_{\mathrm{AT}}(t)\right),0\right)^{\top},\\
 &  & t=0,1,2,...,500.
\end{eqnarray*}
The solution is $\phi_{12}=0.019$, $\phi_{23}=0.0054$, $\xi_{1}=0.037$,
$\varphi_{12}=0.01$ and $\varphi_{21}=0.0048$, which then imply
$\phi_{21}=\phi_{12}\tilde{M}_{\mathrm{AT}}^{*}/\tilde{M}_{\mathrm{UO}}^{*}=0.01$,
$\phi_{32}=\phi_{23}\tilde{M}_{\mathrm{UO}}^{*}/\tilde{M}_{\mathrm{LO}}^{*}=0.00034$,
and $\xi_{2}=\xi_{1}\eta/\xi_{3}=0.047$. 

\begin{figure}
\begin{centering}
\includegraphics[scale=0.46]{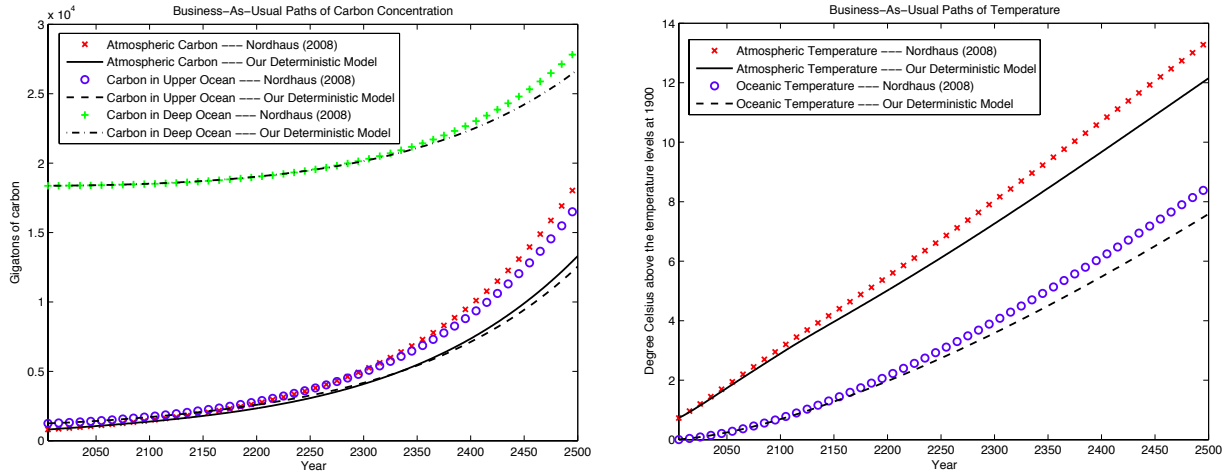}
\par\end{centering}

\protect\caption{{\small{}\label{fig:BAU-path}Business-as-usual paths for calibration}}
\end{figure}

\textcolor{black}{Figure \ref{fig:BAU-path} shows the five business-as-usual
paths from the model of Nordhaus (2008) with decadal time steps in
marks, circles, and pluses, and the corresponding fitted business-as-usual
paths from our annualized deterministic model at the calibrated parameter
values. The match is very good in the first 100 years, a horizon often
used for calibration in the climate literature. The errors in the
fit are in the later centuries where they are unlikely to significantly
affect the results for the first century, the focus of this study. }
\end{singlespace}

\begin{singlespace}

\section*{Appendix D---Calibration of the Stochastic Climate Tipping Process}
\end{singlespace}

\begin{singlespace}
We first calibrate the probability of triggering the tipping point
event. Given Equation \ref{eq:probTrigger}, we need to calibrate
$\lambda$, the hazard rate parameter, and $\underline{T_{\mathrm{AT}}}$,
the lower bound of atmospheric temperature (e.g.,\textcolor{black}{{}
today's temperature level), for which $p_{1,1,0}=0$. Ideally, climate
scientists would specify a value for the hazard rate parameter $\lambda$.
Unfortunately, no studies exist that would provide specific conditional
probabilities of triggering a tipping point event for any given location
in the state space. According to Kriegler et al. (2009) there is a
substantial lack of knowledge about the underlying physical processes
of climate tipping elements. The most reliable source of information
is reviews of expert opinion on the cumulative trigger probability
at some distant time given some degree of global warming, for example
3 degrees Celsius in 2100 (see Kriegler et al. 2009; Zickfeld et al.
2007; and Lenton 2010). The subjective probability assessments differ
significantly among experts, albeit the range of uncertainties about
the likelihood of triggering a tipping point being much lower for
some climate elements (Kriegler et al. 2009). However, the absence
of precise knowledge about the physical system is not essentially
a problem for our analysis: decisions today must be based on current
beliefs about the climate system and will reflect the imprecise nature
of those beliefs. Therefore, we make use of the assessed cumulative
trigger probabilities of tipping and infer trigger hazard rates from
them. Doing so, we assume tacitly that the subjective expert opinions
are the same as those held by the social planner. Lenton (2010) provides
for several climate elements the cumulative probability of triggering
a tipping point. We choose Lenton (2010) as the source of data because
he summarizes the findings from Kriegler et al. (2008) and other expert
elicitation studies. Since our inte}rest is in modeling a representative
tipping element we use the average of these numbers to approximate
the probability of triggering a representative climate tipping element.%
\footnote{The tipping elements used for the calibration of the benchmark tipping
element are: Arctic summer sea ice, Greenland ice sheet, Amazon rain
forests, west Antarctic ice sheet, Boreal forests, Atlantic thermohaline
circulation, El Niño Southern Oscillation, and West African monsoon. %
}

The first two rows of Table \ref{tab:Hazard Rate Calibration} depict
the cumulative probabilities of triggering a climate tipping event
until 2100 for different levels of global warming until 2100. We have
inferred these numbers from Lenton (2010).%
\footnote{We thank Timothy Lenton for helping us identifying the probability
numbers for each tipping element and their average, based on Lenton
(2010). %
} These numbers imply, for example, that given a $4$ \textcolor{black}{degrees
Celsius }increase in global warming in the year 2100, the likelihood
of a tipping point event being triggered until 2100 is 50 percent.
Our goal is to use these numbers to compute an annual conditional
trigger probability. We do this by first specifying the general relationship
between a hazard rate $h_{t}$ and the contemporaneous level of global
warming above the 2000 temperature in a simple linear form by $h_{t}=\lambda\overset{}{\tilde{T}_{\mathrm{AT},t}}$,
where $\lambda$ is a hazard rate parameter and $\tilde{T}_{\mathrm{AT},t}$
is the nominal change in temperature above the year 2000 level ($t$
represents year $t+2000$ here). We then assume that experts in general
express their subjective probabilities having in mind that temperature
linearly evolves until the specified level at the end of this century.%
\footnote{The linearity assumption comes in handy in deriving a simple formula
for the hazard rate parameter, while at the same time not being incompatible
with the recent temperature scenarios used by the Intergovernmental
Panel on Climate Change, which we have obtained from Meinshausen et
al. (2011).%
} 

We then integrate the ha\textcolor{black}{zard rate $h_{t}$ up for
100 years and obtain the cumulative hazard $H_{100}.$ The probability
of a tipping point having }\textit{\textcolor{black}{not}}\textcolor{black}{{}
occurred during this century is then $\exp\left(-H_{100}\right)$.
It follows that $P$, the cumulative probability of a tipping point
having been triggered until 2100, is $P=1-\exp\left(-H_{100}\right)$.
With the former equation we can solve for the hazard rate parameter
$\lambda$ :}

\textcolor{black}{
\begin{equation}
\lambda=-\frac{\log(1-P)}{50\left(\tilde{T}_{\mathrm{AT},100}-\tilde{T}_{\mathrm{AT},0}\right)}.
\end{equation}
}

\textcolor{black}{}
\begin{table}[H]
\noindent \begin{centering}
\textcolor{black}{\small{}}%
\begin{tabular}{c|cccccc}
\hline 
\textcolor{black}{\small{}$\tilde{T}_{\mathrm{AT},100}-\tilde{T}_{\mathrm{AT},0}$} & \textcolor{black}{\small{}1 \textdegree C} & \textcolor{black}{\small{}2 \textdegree C} & \textcolor{black}{\small{}3 \textdegree C} & \textcolor{black}{\small{}4 \textdegree C} & \textcolor{black}{\small{}5 \textdegree C} & \textcolor{black}{\small{}6 \textdegree C}\tabularnewline
\hline 
\textcolor{black}{\small{}$P$ (cumulative probability until 2100)} & \textcolor{black}{\small{}12.5\%} & \textcolor{black}{\small{}25\%} & \textcolor{black}{\small{}37.5\%} & \textcolor{black}{\small{}50\%} & \textcolor{black}{\small{}62.5\%} & \textcolor{black}{\small{}75\%}\tabularnewline
\textcolor{black}{\small{}$\lambda$ (inferred hazard rate parameter)} & \textcolor{black}{\small{}0.00267} & \textcolor{black}{\small{}0.00288} & \textcolor{black}{\small{}0.00313} & \textcolor{black}{\small{}0.00347} & \textcolor{black}{\small{}0.00392} & \textcolor{black}{\small{}0.00462}\tabularnewline
\hline 
\end{tabular}
\par\end{centering}{\small \par}

\textcolor{black}{\protect\caption{\textcolor{black}{\small{}\label{tab:Hazard Rate Calibration}Cumulative
probabilities of triggering a representative tipping element (until
2100) for different levels of global warming and the inferred hazard
rate parameter}}
}
\end{table}

\textcolor{black}{\noindent We use the above expression for the hazard
rate parameter and plug in for each column of Table \ref{tab:Hazard Rate Calibration}
the levels of $\tilde{T}_{\mathrm{AT},100}-\tilde{T}_{\mathrm{AT},0}$
and $P$. We report the computed hazard rate parameter respectively
in the bottom row} of Table \ref{tab:Hazard Rate Calibration}. For
our previous example of a 4 degrees Celsius increase in global warming
and a cumulative trigger probability of 50 percent, our calibration
implies a hazard rate parameter of roughly 0.0035, which is approximately
the central value in our calibrated range. Practically, $\lambda=0.0035$
implies that, for example, a two-degree increase in temperature (above
year 2000) until year 2090 would yield a conditional probability of
triggering the tipping point event in the year 2090 of about 0.7 percent.
We choose $\lambda=0.0035$ as the default parameter setting for our
benchmark case and set up a range for the sensitivity analysis as
$\lambda\in[0.0025,0.0045]$. Since we choose not to model any specific
tipping element explicitly, we set up the range with respect to how
different hazard rate specifications relate to the optimal social
cost of carbon. Finally, we set $\underline{T_{\mathrm{AT}}}$ = $1$
degree Celsius, which \textcolor{black}{we calibrate as today's global
warming using a climate emulator (Meinshausen et al. 2011).}

\textcolor{black}{Next, we turn to the calibration of $\overline{\mathcal{D}}$,
the expected duration of the tipping process. Climate scientists note
that tipping elements in the climate system exhibit heterogeneity
with respect to their assumed duration of the post-tipping process.
According to Lenton et al. (2008) the post-tipping process may last
less than ten years or up to several centuries. Table \ref{tab:Tipping Scales}
presents the expected transition scale for selection of the most prominent
tipping elements from Lenton et al. (2008).}

\textcolor{black}{}
\begin{table}[H]
\noindent \begin{centering}
\textcolor{black}{\small{}}%
\begin{tabular}{c|c|c|c|c|c|c|c|c}
\hline 
 &  & \textcolor{black}{\small{}Arctic} &  &  & \textcolor{black}{\small{}Atlantic} & \textcolor{black}{\small{}West} & \textcolor{black}{\small{}West} & \textcolor{black}{\small{}El Nino}\tabularnewline
\textcolor{black}{\small{}Tipping} & \textcolor{black}{\small{}Greenland} & \textcolor{black}{\small{}summer} & \textcolor{black}{\small{}Amazon} & \textcolor{black}{\small{}Boreal} & \textcolor{black}{\small{}thermohaline} & \textcolor{black}{\small{}Antarctic} & \textcolor{black}{\small{}African} & \textcolor{black}{\small{}Southern}\tabularnewline
\textcolor{black}{\small{}Element} & \textcolor{black}{\small{}ice sheet} & \textcolor{black}{\small{}sea-ice} & \textcolor{black}{\small{}rainforest} & \textcolor{black}{\small{}forests} & \textcolor{black}{\small{}circulation} & \textcolor{black}{\small{}ice sheet} & \textcolor{black}{\small{}monsoon} & \textcolor{black}{\small{}Oscillation}\tabularnewline
\hline 
\textcolor{black}{\small{}Expected} &  &  &  &  &  &  &  & \tabularnewline
\textcolor{black}{\small{}transition} & \textcolor{black}{\small{}>300} & \textcolor{black}{\small{}10} & \textcolor{black}{\small{}50} & \textcolor{black}{\small{}50} & \textcolor{black}{\small{}100} & \textcolor{black}{\small{}>300} & \textcolor{black}{\small{}10} & \textcolor{black}{\small{}100}\tabularnewline
\textcolor{black}{\small{}scale (years)} &  &  &  &  &  &  &  & \tabularnewline
\hline 
\end{tabular}
\par\end{centering}{\small \par}

\protect\caption{\textcolor{black}{\small{}\label{tab:Tipping Scales}Duration of expected
transition scales for different climate tipping elements}}
\end{table}

\textcolor{black}{Following the numbers in Table \ref{tab:Tipping Scales}
we choose $\overline{\mathcal{D}}=50$ (years) for our benchmark tipping
process. In addition, in our sensitivity analyses we also consider
$\overline{\mathcal{D}}=5$ (years) as the lower bound, $\overline{\mathcal{D}}=100$
(years), and $\overline{\mathcal{D}}=200$ (years) as the upper bound.}%
\footnote{\textcolor{black}{For $\overline{\mathcal{D}}>200$ we find very little
changes in the initial social cost of carbon than with $\overline{\mathcal{D}}=200$.}%
}\textcolor{black}{{} As mentioned in the main text, we assume that each
of our five stages occurs stochastically and the expected transition
time at each stage of the tipping process is $\overline{\mathcal{D}}/4$
years. }

In this paper, all of our examples of tipping elements assume that
there are three possible tipping damage levels in the long run if
$q>0$, and that we do not know the long-run damage level until tipping
begins. Since the variance of the long-run tipping damage level is
$q\overline{\mathcal{J}}_{\infty}^{2}$, we assume that the values
of the three possible tipping damage levels in the long run are $\left(1+\left(j-2\right)\sqrt{1.5q}\right)\overline{\mathcal{J}}_{\infty}$,
for $j=1,2,3$, with equal probability. For our benchmark case with
$\overline{\mathcal{J}}_{\infty}=5\%$ and $q=0.2$ the three possible
long-run damage levels are 2.26 percent, 5 percent, and 7.74 percent,
each with 1/3 probability. In the extreme cases of $\overline{\mathcal{J}}_{\infty}=2.5\%$
(or $\overline{\mathcal{J}}_{\infty}=10\%$) and $q=0.4$ in our sensitivity
analysis, the three possible long-run damage levels are 0.56 percent,
2.5 percent, and 4.44 percent (or 2.25 percent, 10 percent, and 17.75
percent).

We also assume that the cascading damage linearly increases over the
five post-tipping stages. That is, if the level of long-run tipping
damage is $\left(1+\left(j-2\right)\sqrt{1.5q}\right)\overline{\mathcal{J}}_{\infty}$,
then the damage at the $i$-th post-tipping stage is $\left(1+\left(j-2\right)\sqrt{1.5q}\right)\overline{\mathcal{J}}_{\infty}\cdot i/5$,
for $i=1,2,...,5$. Thus, the values of $J_{t}$ are: $\mathcal{J}_{1}=0$
and 
\begin{equation}
\mathcal{J}_{3i+j-2}=\frac{i}{5}\left(1+\left(j-2\right)\sqrt{1.5q}\right)\overline{\mathcal{J}}_{\infty},\label{eq:Tipping Damage}
\end{equation}
for $i=1,...,5$ and $j=1,2,3$. When $q=0$, there are only six values
of $J_{t}$: $\mathcal{J}_{1}=0$ and $\mathcal{J}_{i+1}=\overline{\mathcal{J}}_{\infty}\cdot i/5$
for $i=1,2,...,5$. 

Since we assume that we know the long-run damage level once the tipping
event happens and the three possible levels are equally distributed,
the probability from the pre-tipping state $\mathcal{J}_{1}$ to $\mathcal{J}_{1+j}$
at time $t$ is $p_{1,1+j,t}=\left(1-p_{1,1,t}\right)/3$ for $j=1,2,3$.
Moreover, if the current state is $\mathcal{J}_{3i+j-2}$ with $i<5$,
meaning that it is at the $i$-th post-tipping stage, then it could
either stay with the probability $\exp\left(-4/\overline{\mathcal{D}}\right)$,
or move to state $\mathcal{J}_{3(i+1)+j-2}$ (the $(i+1)$-th post-tipping
stage) with the probability $1-\exp\left(-4/\overline{\mathcal{D}}\right)$.
Therefore, the Markov chain transition matrix of the state $J_{t}$
at time $t$ is a $16\times16$ matrix with the following sparsity
shape%
\footnote{Note that the matrix could be a general upper triangular matrix, but
in our examples we assume this sparse shape for simplicity purposes.%
}: 

\[
\left[\begin{array}{ccccccc}
* & * & * & *\\
 & * &  &  & *\\
 &  & \ddots &  &  & \ddots\\
 &  &  & * &  &  & *\\
 &  &  &  & 1\\
 &  &  &  &  & 1\\
 &  &  &  &  &  & 1
\end{array}\right],
\]
where a star means one nonzero element. The $(1,1)$ element, $p_{1,1,t}$,
is given by (\ref{eq:probTrigger}), the $(1,j)$ element is $p_{1,j,t}=\left(1-p_{1,1,t}\right)/3$
for $j=2,3,4$, the $(i,i)$ element is $p_{i,i,t}=\exp\left(-4/\overline{\mathcal{D}}\right)$,
and the $(i,i+3)$ element is $p_{i,i+3,t}=1-p_{i,i,t}$ for $i=2,...,13$.
The last three diagonal elements are 1's, implying irreversibility.
When $q=0$, the matrix is reduced to a $6\times6$ bidiagonal matrix,
where the $(1,1)$ element is still given by (\ref{eq:probTrigger}),
but the $(i,i)$ element is $p_{i,i,t}=\exp\left(-4/\overline{\mathcal{D}}\right)$,
the $(i,i+1)$ element becomes $p_{i,i+1,t}=1-p_{i,i,t}$ for $i=2,3,4,5$,
and the $(6,6)$ element is 1. 
\end{singlespace}

\begin{singlespace}

\section*{Appendix E---Value Function Iteration}
\end{singlespace}

\begin{singlespace}
We summarize the numerical method we used to solve the dynamic programming
problems in this study. The value functions for our problems are continuous,
and almost everywhere differentiable in the continuous state variables.
In each case, we approximate the value function $V(x,\theta)$ using
a functional form $\hat{V}(x,\theta;{\bf b})$ with a finite number
of parameters, ${\bf b}$. In our model, the state variable $x$ represents
the six-dimensional vector $(K,\mathbf{M},\mathbf{T})$ of continuous
states, and $\theta$ represents the three-dimensional vector of discrete
states, $(\zeta,\chi,J)$, contained in a finite set $\Theta$. The
functional form $\hat{V}$ may be a linear combination of polynomials,
a rational function, a neural net, or some other parameterization
specially designed for the problem. After the functional form is set,
we focus on finding the vector of parameters, ${\bf b}$, such that
$\hat{V}(x,\theta;{\bf b})$ approximately satisfies the Bellman equation
(Bellman, 1957). Numerical dynamic programming with value function
iteration can solve the Bellman equation approximately (Judd, 1998).

A general dynamic programming model is based on the Bellman equation:
\begin{eqnarray*}
 &  & V_{t}(x,\theta)=\max_{a\in{\cal D}(x,\theta,t)}\text{ \ }u_{t}(x,a)+\beta\mathcal{H}_{t}\left(V_{t+1}(x^{+},\theta^{+})\right),\\
 &  & \text{{\rm \ \ \ \ \ \ \ \ \ \ \ \ \ \ \ \ \ s.t. \ \ }}x^{+}=F(x,\theta,a),\\
 &  & \text{{\rm \ \ \ \ \ \ \ \ \ \ \ \ \ \ \ \ \ \ \ \ \ \ \ \ }}\theta^{+}=G(x,\theta,\omega),
\end{eqnarray*}
where $V_{t}(x,\theta)$ is the value function at time $t\leq\mathcal{T}$
(the terminal value function $V_{\mathcal{T}}(x,\theta)$ is given),
$(x^{+},\theta^{+})$ is the next-stage state, ${\cal D}(x,\theta,t)$
is a feasible set of $a$, $\omega$ is a random variable vector,
$F$ and $G$ are the transition laws of $x$ and $\theta$, $\beta$
is a discount factor and $u_{t}(x,a)$ is the utility function, and
$\mathcal{H}_{t}$ is a functional operator %
\footnote{\begin{singlespace}
A typical one is $\mathbb{E}_{t}$, the expectation operator conditional
on time-$t$ information. In our model, it is 
\[
\mathcal{H}_{t}\left(V_{t+1}\right)=\left[\mathbb{E}_{t}\left\{ V_{t+1}^{\frac{1-\gamma}{1-\frac{1}{\psi}}}\right\} \right]^{\frac{1-\frac{1}{\psi}}{1-\gamma}}
\]
\end{singlespace}
} at time $t$. The following is the algorithm of parametric dynamic
programming with value function iteration for finite horizon problems.
\end{singlespace}
\begin{description}
\begin{singlespace}
\item [{Algorithm}] 1. \textit{Numerical Dynamic Programming with Value
Function Iteration for Finite Horizon Problems}
\item [{\textit{Initialization.}}] \textit{Choose the approximation nodes,
$\mathbb{X}_{t}=\{x_{i,t}:$ $1\leq i\leq m_{t}\}$ for every $t<\mathcal{T}$,
and choose a functional form for $\hat{V}(x,\theta;{\bf b})$, where
$\theta\in\Theta$. Let $\hat{V}(x,\theta;{\bf b}_{\mathcal{T}})\equiv V_{\mathcal{T}}(x,\theta)$.
Then for $t=\mathcal{T}-1,\mathcal{T}-2,\ldots,0$, iterate through
steps 1 and 2. }
\item [{\textit{Step}}] \textbf{\textit{1.}}\textit{ Maximization step.
Compute 
\begin{eqnarray*}
 &  & v_{i,j}=\max_{a\in{\cal D}(x_{i},\theta_{j},t)}\ u_{t}(x_{i},a)+\beta\mathcal{H}_{t}\left(\hat{V}\left(x^{+},\theta_{j}^{+};{\bf b}_{t+1}\right)\right)\\
 &  & \text{{\rm \ \ \ \ \ \ \ \ \ \ \ \ \ s.t. \ \ \ }}x^{+}=F(x_{i},\theta_{j},a),\\
 &  & \text{{\rm \ \ \ \ \ \ \ \ \ \ \ \ \ \ \ \ \ \ \ \ \ \ }}\theta_{j}^{+}=G(x_{i},\theta_{j},\omega),
\end{eqnarray*}
 for each $\theta_{j}\in\Theta$, $x_{i}\in\mathbb{X}_{t}$, $1\leq i\leq m_{t}.$ }
\item [{\textit{Step}}] \textbf{\textit{2.}}\textit{ Fitting step. Using
an appropriate approximation method, compute the ${\bf b}_{t}$ such
that $\hat{V}(x,\theta_{j};{\bf b}_{t})$ approximates $(x_{i},v_{i,j})$
data for each $\theta_{j}\in\Theta$. }\end{singlespace}

\end{description}
\begin{singlespace}
\textcolor{black}{We considered alternative functional forms for $\hat{V}(x,\theta;{\bf b})$,
including ordinary polynomials and polynomials in logs. We were always
able to find some parsimonious functional form for approximating the
value function. Detailed discussion of numerical dynamic programming
methods can be found in Cai (2010), Cai and Judd (2010), Judd (1998),
and Rust (2008).}

\textcolor{black}{The terminal value function is computed by assuming
that after 2600, the system is deterministic, population and productivity
growth ends, all emissions are eliminated, and the consumption-output
ratio is fixed at 0.78. Consideration of alternatives showed that
changes in the terminal value function at 2600 had no significant
impact on any results for the 21st century.}
\end{singlespace}

\begin{singlespace}

\section*{Appendix F---Code Test}
\end{singlespace}

\begin{singlespace}
\textcolor{black}{In order to have confidence in our code, we compared
how our code performed on a few cases where we could compute the true
solution using another method. The key challenge for any dynamic programming
problem, deterministic or stochastic, is the ability to accurately
approximate the value function over a large range of possible values.
Our tests were to compute solutions to deterministic versions of our
models, that is, versions where we eliminate all randomness, by two
methods: our dynamic programming code and by optimal control methods.
Since the latter can be computed with high accuracy, we can compare
those solutions with the results of the dynamic programming algorithm.
For example, among the three benchmark cases, the approximation domains
for the stochastic growth and climate tipping benchmark case are the
widest because it has the largest volatility, and then we use the
highest degree polynomials for approximation. We performed three tests,
each using the same approximation domains and polynomial degrees as
used in one of the three benchmark cases. Table \ref{tab:Accuracy}
shows the relative $\mathcal{L}^{1}$ errors over the first one hundred
years of three of the endogenous state variables ($K$, $M_{\mathrm{AT}}$,
$T_{\mathrm{AT}}$), two control variables ($C$, $\mu$), and the
social cost of carbon. It also shows the relative error in the initial
year for the control variables and the social cost of carbon. Table
\ref{tab:Accuracy} shows that all the relative errors were small. }

\textcolor{black}{\small{}}
\begin{table}[H]
\begin{centering}
\textcolor{black}{\small{}}%
\begin{tabular}{c|c|c|c|c|c|c}
\hline 
\multirow{5}{*}{\textcolor{black}{\small{}Variable}} & \multicolumn{2}{c|}{\textcolor{black}{\small{}Test case (deterministic) with }} & \multicolumn{2}{c|}{\textcolor{black}{\small{}Test case (deterministic) with }} & \multicolumn{2}{c}{\textcolor{black}{\small{}Test case (deterministic) with }}\tabularnewline
 & \multicolumn{2}{c|}{\textcolor{black}{\small{}approximation degrees and }} & \multicolumn{2}{c|}{\textcolor{black}{\small{}approximation degrees and }} & \multicolumn{2}{c}{\textcolor{black}{\small{}approximation degrees and }}\tabularnewline
 & \multicolumn{2}{c|}{\textcolor{black}{\small{}domains of the stochastic}} & \multicolumn{2}{c|}{\textcolor{black}{\small{}domains of the stochastic }} & \multicolumn{2}{c}{\textcolor{black}{\small{}domains of the stochastic growth}}\tabularnewline
 & \multicolumn{2}{c|}{\textcolor{black}{\small{} growth benchmark case}} & \multicolumn{2}{c|}{\textcolor{black}{\small{}climate benchmark case}} & \multicolumn{2}{c}{\textcolor{black}{\small{}and climate benchmark case}}\tabularnewline
\cline{2-7} 
 & \textcolor{black}{\small{}first 100 years} & \textcolor{black}{\small{}initial year} & \textcolor{black}{\small{}first 100 years} & \textcolor{black}{\small{}initial year} & \textcolor{black}{\small{}first 100 years} & \textcolor{black}{\small{}initial year}\tabularnewline
\cline{2-7} 
\textcolor{black}{\small{}$K$} & \textcolor{black}{\small{}$5.5(-3)$} & \textcolor{black}{\small{}—} & \textcolor{black}{\small{}$2.1(-4)$} & \textcolor{black}{\small{}—} & \textcolor{black}{\small{}$4.0(-3)$} & \textcolor{black}{\small{}—}\tabularnewline
\textcolor{black}{\small{}$M_{\mathrm{AT}}$} & \textcolor{black}{\small{}$1.5(-4)$} & \textcolor{black}{\small{}—} & \textcolor{black}{\small{}$1.3(-5)$} & \textcolor{black}{\small{}—} & \textcolor{black}{\small{}$1.1(-4)$} & \textcolor{black}{\small{}—}\tabularnewline
\textcolor{black}{\small{}$T_{\mathrm{AT}}$} & \textcolor{black}{\small{}$8.7(-5)$} & \textcolor{black}{\small{}—} & \textcolor{black}{\small{}$2.5(-5)$} & \textcolor{black}{\small{}—} & \textcolor{black}{\small{}$8.3(-5)$} & \textcolor{black}{\small{}—}\tabularnewline
\textcolor{black}{\small{}$C$} & \textcolor{black}{\small{}$2.0(-3)$} & \textcolor{black}{\small{}$4.6(-5)$} & \textcolor{black}{\small{}$2.4(-5)$} & \textcolor{black}{\small{}$2.6(-5)$} & \textcolor{black}{\small{}$1.6(-3)$} & \textcolor{black}{\small{}$4.0(-5)$}\tabularnewline
\textcolor{black}{\small{}$\mu$} & \textcolor{black}{\small{}$2.4(-3)$} & \textcolor{black}{\small{}$1.3(-4)$} & \textcolor{black}{\small{}$4.4(-4)$} & \textcolor{black}{\small{}$1.7(-4)$} & \textcolor{black}{\small{}$2.0(-3)$} & \textcolor{black}{\small{}$1.3(-4)$}\tabularnewline
\textcolor{black}{\small{}$\Gamma$} & \textcolor{black}{\small{}$9.2(-3)$} & \textcolor{black}{\small{}$5.4(-3)$} & \textcolor{black}{\small{}$4.1(-3)$} & \textcolor{black}{\small{}$7.2(-4)$} & \textcolor{black}{\small{}$1.0(-2)$} & \textcolor{black}{\small{}$6.2(-3)$}\tabularnewline
\hline 
\end{tabular}
\par\end{centering}{\small \par}

\textcolor{black}{\small{}\protect\caption{\textcolor{black}{\small{}Relative errors for the solution of the
deterministic case obtained by our numerical dynamic programming algorithm
with approximation domains and polynomial degrees used by the three
benchmark cases. Note---$a(-n)$ means $a\times10^{-n}$\label{tab:Accuracy}}}
}
\end{table}
{\small \par}
\end{singlespace}

\begin{singlespace}

\section*{Appendix G---Dynamics of the Economic System in the Stochastic Growth
Benchmark Case}
\end{singlespace}

\begin{figure}[H]
\noindent \begin{centering}
\includegraphics[scale=0.6]{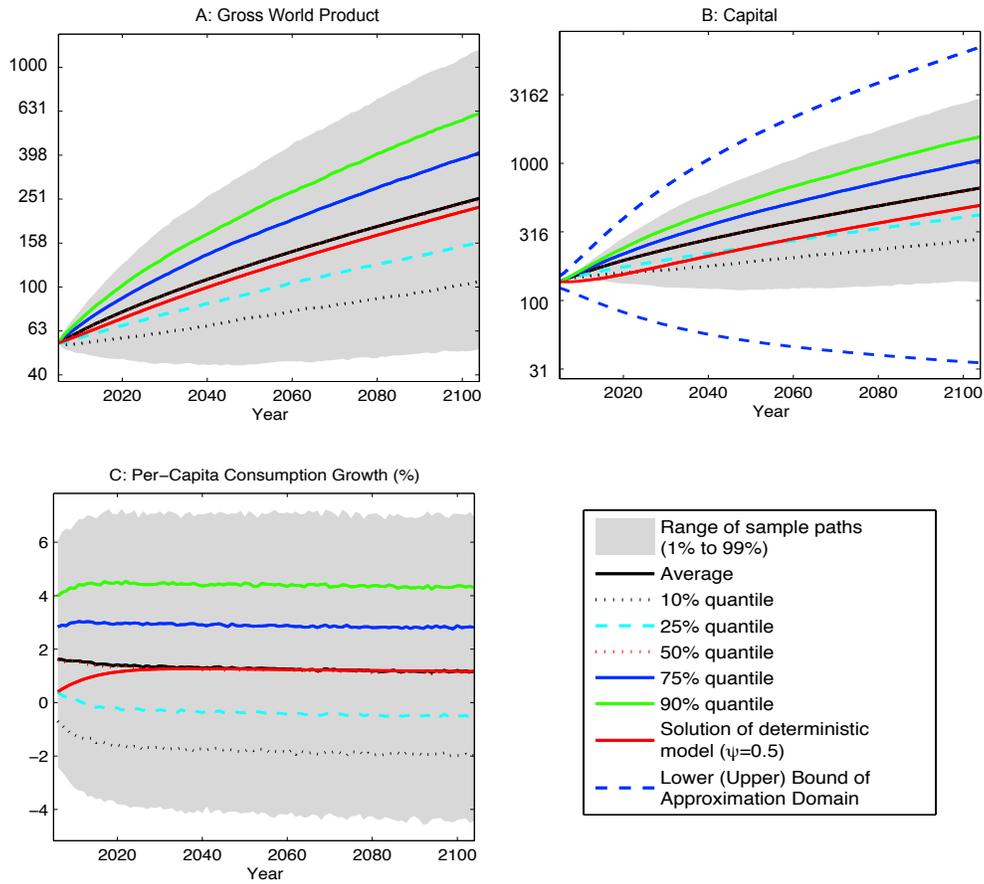}
\par\end{centering}

\protect\caption{\textcolor{black}{\small{}Simulation results of the stochastic growth
benchmark---Economics \label{fig:sim-LRR-econ}}}
\end{figure}

\begin{singlespace}
Figure \ref{fig:sim-LRR-econ} shows the \textcolor{black}{dynamics
of gross world output ($\mathcal{Y}_{t}$), total capital ($K_{t}$),
and per capita consumption growth rate. Both gross world output and
the total capital stock grow exponentially, which is not surprising
since productivity grows exponentially. The uncertainty of both is
substantial; for example the 10 percent and 90 percent quantiles of
the capital stock in 2100 are \$270 trillion and \$1,500 trillion
respectively. The growth of per capita consumption in Figure \ref{fig:sim-LRR-econ}C
ranges from $-4.4\%$ to $7.1\%$. The mean and median of the dynamic
stochastic equilibrium paths are close to the deterministic solution,
but the variation around that path is very large.}

The wide range of possible outcomes for the total capital stock displayed
in \textcolor{black}{Figure \ref{fig:sim-LRR-econ}B implied that
our approximation of the value function had to be defined over an
even larger range. The blue broken lines in Figure \ref{fig:sim-LRR-econ}B
represent, for each time $t$, the minimum and maximum level of the
capital stock (on $\log_{10}$ scale) over which the value function
is computed. The ratio of maximum to minimum capital stock exceeded
175 in 2100, and continues to grow thereafter. If we had used significantly
smaller ranges then the Bellman equation for states close to the maximum
or minimum capital stock would try to evaluate the value function
beyond the domain on which it is approximated. These ranges are the
natural outcome of the productivity process.} The large range for
capital creates difficulties for the simpler approximation methods,
such as orthogonal polynomials, but nonlinear changes in variables
allowed us to find suitable function forms for approximating the value
function.

\end{singlespace}

\end{document}